\documentclass[12pt,english]{article}
\usepackage[T1]{fontenc}
\usepackage[utf8]{inputenc}
\usepackage{geometry}
\usepackage{babel}
\usepackage{setspace}
\usepackage[authoryear]{natbib}
\usepackage{subcaption}
\usepackage[dvipsnames]{xcolor}

\usepackage[unicode=true,pdfusetitle,
bookmarks=true,bookmarksnumbered=false,bookmarksopen=false,breaklinks=false,pdfborder={0 0 1},backref=page,colorlinks=true,linkcolor=Blue, citecolor=Blue,urlcolor=blue!50!black]{hyperref}
\usepackage{breakurl}
\usepackage{comment}
\usepackage{lmodern}
\usepackage{xcolor}
\usepackage{graphicx}
\usepackage{float}
\usepackage{adjustbox}
\usepackage[labelsep=space]{caption}
\usepackage{refstyle}
\usepackage{rotfloat}
\usepackage{fixltx2e}
\usepackage{dsfont}
\usepackage{zref-xr}
\newcommand{\sym}[1]{{#1}}

\AtBeginDocument{
    
}
\usepackage{lmodern}
\usepackage[T1]{fontenc}
\usepackage[utf8]{inputenc}
\usepackage{graphicx}
\usepackage{float}
\usepackage{appendix}
\usepackage{amsmath}
\usepackage{bbm}
\usepackage{mathtools}
\usepackage{xcolor}
\usepackage{babel}
\usepackage[font=normal,labelfont={bf},tableposition=top]{caption}
\usepackage{booktabs}
\usepackage{comment}
\usepackage{appendix}
\usepackage{setspace}
\usepackage{ragged2e}
\usepackage{etoc}
\usepackage{hyperref}

\usepackage{amssymb}%
\usepackage{amsthm, amssymb}

\usepackage{pifont}%
\newcommand{\xmark}{\ding{55}}%
\makeatother

\begin{document}
\etocsettocstyle{}{} %
\sloppy

\title{\singlespacing\textbf{Financial Technologies, Labor Markets, and Wage Inequality: Evidence from Instant Payment Systems}\thanks{We thank Thorsten Beck, Susan Cherry, Tamanna Singh Dubey, Apoorv Gupta, Sean Higgins, Pulak Ghosh, Petros Katsoulis, Jeanne Lafortune, Thomas Lemieux, Pauline Liang,  Adrien Matray, Filippo Mezzanotti, Akash Raja, Janis Skrastins, Jose Tessada, Nishant Vats, Diego Vera-Cossio, Benjamín Villena Roldán, Andy Winton, and participants at SFS Cavalcade, the BSE Summer Forum, Bank of Canada Annual Economic Conference, CAFRAL-ISB Conference, University of Delaware/Philadelphia Fed Fintech Conference, Essex/Olin Corporate Finance Conference, FIRS, IDB Research Conference, IMFS Workshop on Monetary and Financial Stability, Minnesota Finance Junior Conference, MFA, NFA, WEFIDEV Workshop, Finance UC, and seminars at CAF, Central Bank of Brazil, CSOM Summer Applied Econ, Federal Reserve Bank of Minneapolis, IDB, PUC-Chile, University of Colorado Boulder, University of Minnesota, and UT San Antonio for helpful feedback. We are grateful to Kiara Lazo and Thaline do Prado for excellent research assistance. This study was funded by the Research Network of the Inter-American Development Bank (RG-K1198) and CAF.}
}
\author{Carlos Burga\thanks{Pontificia Universidad Católica de Chile. Email: \href{mailto:carlos.burga@uc.cl}{carlos.burga@uc.cl}.} \and Jacelly Cespedes\thanks{Carlson School of Management, University of Minnesota. Email: \href{mailto:cespe013@umn.edu}{cespe013@umn.edu}.} \and Carlos Parra\thanks{Pontificia Universidad Católica de Chile. Email: \href{mailto:carlosr.parra@uc.cl}{carlosr.parra@uc.cl}.} \and Bernardo Ricca\thanks{Insper. Email: \href{mailto:bernardoOGR@insper.edu.br}{bernardoOGR@insper.edu.br}.}}
\date{}
\hypersetup{
  pdftitle={Financial Technologies, Labor Markets, and Wage Inequality: Evidence from Instant Payment Systems},
  pdfauthor={Carlos Burga; Jacelly Cespedes; Carlos Parra; Bernardo Ricca}
}

\maketitle
\vspace{-0.5 cm}
\begin{abstract}
\singlespacing

\noindent While technological innovations typically increase wage inequality by favoring skilled workers, we show that instant payment systems instead reduce it. We study the labor market effects of instant payment systems in the context of Brazil's Pix rollout. Using matched employer-employee data, we implement a triple-difference design that exploits pre-Pix mobile penetration across municipalities, the differential benefits of Pix for small versus large establishments, and the timing of Pix. We find that wages in small establishments rise significantly relative to large establishments after Pix. These gains are concentrated in cash-intensive sectors such as retail and services, with no effects in wholesale or manufacturing. Crucially, wage inequality declines, driven by wage gains in the lower half of the distribution, with no effect at the top. Our evidence points to increased small-firm labor demand, consistent with lower payment frictions. These effects are amplified where low-skill labor is scarce. A calibrated monopsony model implies that uniform Pix adoption would reduce both the within- and between-municipality components of wage dispersion, amplifying the aggregate inequality reduction.

\end{abstract}

\thispagestyle{empty}\setcounter{page}{0}%

\newpage
\section{Introduction}

The relationship between technological change and wage inequality has long been a subject of intense debate. The conventional view is that production technologies increase inequality by complementing skilled labor and substituting for routine tasks typically performed by less-educated workers \citep{autor2003skill, acemoglu2011skills}. However, the labor market and distributional effects of financial technologies remain largely unexplored, despite their rapid adoption across both developed and developing economies. This raises a fundamental question: Do financial technologies affect labor markets and wage structures similarly to production technologies, or do they follow a different pattern?  %

In this paper, we contribute to this debate by examining the labor market impacts of instant payment systems, a major financial innovation adopted by over 60 countries as of 2023 \defcitealias{bis2023}{BIS, 2023}(\citetalias{bis2023}). Instant payment systems reduce transaction costs such as merchant fees and cash-handling expenses, enable immediate settlement, and generate digital transaction records that can expand financial access. These features alleviate frictions that are particularly burdensome for small, cash-intensive businesses.

However, the impact of these technologies on wage inequality is theoretically ambiguous. Inequality could increase if instant payments complement high-skill labor, as production technologies do. In contrast, inequality could decline if they complement low-skill workers or favor low-skill-intensive firms such as small businesses and retailers. Furthermore, even if low-skill-intensive firms benefit disproportionately, the pass-through to wages may be muted under monopsony. As these firms expand, their growing labor market shares increase wage markdowns, potentially leaving inequality unchanged or even widening it. Thus, whether instant payment systems reduce wage inequality is an empirical question.

We analyze Brazil's Pix system, an instant payment platform launched by the central bank in November 2020, which reduced transaction fees to approximately one-tenth those of credit cards while achieving unprecedented adoption—67 percent of Brazil's adult population within its first year.\footnote{Pix is an instant payment system developed by Brazil's central bank to reduce cash usage and promote financial inclusion. It facilitates real-time transfers through simplified identifiers such as email addresses or phone numbers. Key institutional features include mandatory participation by large banks, zero transaction fees for individuals, and substantially reduced fees for merchants—approximately 0.22 percent per transaction compared to roughly 2.2 percent for credit cards. By early 2024, Pix had over 152 million individual users and 15 million registered businesses.} Contrary to conventional patterns of skill-biased technical change, we find that instant payment systems reduce wage inequality by raising wages in small establishments within cash-intensive sectors such as retail and services, with gains concentrated among workers in the lower half of the wage distribution and amplified in areas where low-skill labor is scarce. %

We develop a conceptual framework that highlights two channels through which instant payment systems affect wage inequality: a competition channel and a supply elasticity channel. The framework features heterogeneous firms with monopsony power and a self-employment outside option. Through the competition channel, a reduction in transaction costs that disproportionately benefits small businesses boosts their labor demand and reduces labor market concentration, narrowing markdowns and compressing wage dispersion. This effect is stronger when the benefited firms employ predominantly low‑skill workers.

Through the supply elasticity channel, the same reduction in transaction costs raises the value of self‑employment, particularly for low‑skill workers, increasing the labor supply elasticity that firms face, narrowing markdowns, and further reducing inequality. The decline in inequality is weaker, and can even reverse, if the payment technology is complementary to high-skill labor or substitutes for low‑skill labor. %

We then estimate the labor market effects of Pix using administrative data on formal workers and establishments from 2016 to 2024. Our identification strategy is an intention-to-treat triple difference-in-differences (DiD) design exploiting Pix's nationwide launch in November 2020, pre-existing mobile penetration across municipalities, and the differential exposure of small and large establishments to payment frictions. We compare wage changes between small and large establishments within the same municipality before and after Pix. We interact this contrast with pre-Pix mobile penetration, which strongly predicts Pix adoption intensity, and control for local economic conditions and industry-specific trends. Small establishments face higher payment frictions than large establishments, especially in cash-intensive sectors, and therefore have more scope to benefit from Pix's instant and low-cost payments.\footnote{For example, \citet{egan2026pays} show that large U.S. grocery and merchandise merchants pay interchange fees about 50 basis points lower than smaller merchants in the same sectors, consistent with volume discounts and private negotiations.}

Our baseline triple DiD design assumes that, without Pix, the wage
differential between small and large establishments would have evolved
similarly across municipalities with different pre-Pix mobile
penetration, conditional on our fixed effects. This assumption does
not require mobile penetration to be randomly assigned across
municipalities. Our fixed effects narrow the residual threat by
absorbing time-invariant establishment differences, time-varying local
shocks, and national size and industry trends. Event-study estimates
show no differential pre-trends before Pix, supporting this
assumption.

We find that following Pix's introduction, small establishments in municipalities with higher mobile penetration experienced significant wage increases relative to large establishments within the same municipalities. Specifically, our estimates indicate that a one-standard-deviation increase in mobile penetration leads to a 0.4 percent increase in wages for small establishments (those with fewer than 20 employees).\footnote{This wage response, combined with the 1.7 percent employment increase for micro establishments, implies a firm-level labor supply elasticity of approximately 4, consistent with recent estimates derived from firm-specific labor demand shocks \citep{sokolova2021monopsony}.} Our results remain robust to alternative definitions of small establishments (fewer than 5 or 10 employees) and are notably stronger in cash-intensive sectors, specifically retail (0.8 percent) and services (1.4 percent), which rely heavily on small-value, consumer-facing transactions. In contrast, we find no significant wage effects in less cash-intensive sectors such as wholesale or manufacturing.

A key question is whether our estimates capture the effect of Pix
adoption or faster non-Pix digital adaptation during the COVID-19
period. Because Pix launched nationwide, our design identifies an
intention-to-treat effect of Pix exposure on the small-versus-large
wage gap, using pre-Pix mobile penetration as the exposure proxy. Our
preferred fixed effects absorb municipality-by-industry-by-year shocks
and national size-by-industry trends, so a digital-adaptation
explanation would require size-specific pandemic impacts correlated
with pre-Pix mobile penetration. Augmenting the specification with
direct measures of pandemic intensity (e.g., per-capita emergency
transfers, wage-subsidy beneficiaries, stay-at-home orders, and COVID
cases) interacted with $\textit{Small} \times \textit{Post}$ leaves
the coefficient of interest largely unchanged. We also find
heterogeneity that is difficult to reconcile with broad non-Pix
digital adaptation: effects concentrate in cash-intensive
consumer-facing sectors and are larger in physical services than in
delivery-capable services including restaurants. Finally, the dynamic
effects grow over time rather than attenuate, consistent with
continued merchant adoption of Pix.

We next examine the impact of Pix adoption on wage inequality.
We find that municipalities with higher mobile penetration
experienced a significant decline in wage inequality following
Pix's introduction. A one-standard-deviation increase in mobile
penetration is associated with approximately a 1 percent reduction
in the Gini coefficient relative to the sample mean. This reduction
is driven by wage gains among workers in the lower half of the
wage distribution, with no significant effects at the top.\footnote{Separately, we estimate wage responses for low- and high-skill workers and find similar gains (0.4\% and 0.5\%, respectively), suggesting that the wage effects are not strongly skill-biased.}

Our evidence points to increased labor demand from small businesses as the key mechanism behind these inequality-reducing effects. These findings are consistent with our conceptual framework, which predicts that lowering transaction costs increases firm profitability and stimulates employment growth among small, cash-dependent establishments. Because these firms rely heavily on low-skill labor, the resulting increase in labor demand raises wages at the bottom of the distribution and reduces inequality.

We find three key pieces of evidence supporting this mechanism.
First, municipalities with higher mobile penetration experienced
significant employment growth in small establishments (a 1.7
percent increase) after Pix's introduction. Second, higher mobile
penetration led to increased small-business entry in retail
sectors, with no corresponding effects in manufacturing, indicating
that Pix's impact is concentrated in cash-intensive businesses.
Third, the inequality-reducing effects of Pix are notably stronger
in areas characterized by tighter low-skill labor markets, with no
significant effect in regions where low-skill workers are abundant.
This heterogeneity aligns with a mechanism in which increased labor
demand generates greater wage pressure in tighter local labor
markets, amplifying the reductions in wage inequality.

We evaluate two channels through which Pix could increase small-firm expansion. First, Pix may reduce transaction costs. It substitutes for cash, reducing the need to store, transport, and safeguard physical currency. It also lowers payment costs for merchants, especially small firms with less bargaining power over card fees. Consistent with this channel, we document declines in bank-branch cash inventories in higher-exposure municipalities after Pix and stronger wage effects in high-crime settings where cash holding is costlier. We also find larger wage gains for small establishments embedded in small firms, which are likely to face higher merchant fees.

Second, Pix could relax financing frictions. Digital transaction records may reduce information frictions between lenders and small firms, allowing firms to finance expansion. We find no measurable increase in aggregate business lending after Pix adoption, so the evidence does not support credit expansion as the primary driver. However, our data do not allow us to rule out changes in loan terms, borrower composition, or specific types of credit not captured by aggregate lending volumes. In any case, this channel would operate through the same mechanism as our main interpretation: Pix relaxes constraints on small firms, enabling expansion and increasing labor demand.

While our findings point to increased labor demand as the main
driver, we evaluate alternative explanations. Formalization or wage
reporting alone cannot fully explain the wage gains. Formal
employment is costly in Brazil, and prior reductions in
formalization costs have generated limited and short-lived changes
in registration \citep{cavalcante2020encargos,
rocha2018taxinformality}. Moreover, incentives to formalize due to
fear of tax enforcement are limited, as Pix transactions remain
subject to banking secrecy and tax authorities do not observe
transaction-level payer or payee identifiers. Wage increases also
extend to small establishments with up to 20 employees, which
typically already comply with formal labor regulations. We find no
differential wage effects based on municipalities' pre-Pix share
of informal workers, and worker composition is unchanged with null
effects on gender, skill, tenure, and age. Effects are also
stronger where low-skill labor is scarce, which is difficult to
reconcile with a pure compliance or reporting channel. %

Motivated by our empirical results, we extend our conceptual
framework to quantify the aggregate gains from Pix and to
simulate uniform adoption across municipalities. While our
empirical analysis finds that Pix reduced wage inequality within
high-adoption municipalities, the aggregate effect is ambiguous. Cities that adopted Pix more intensively also tend to
have higher average wages, so uneven adoption could increase
between-city inequality even as it reduces within-city inequality.
We calibrate the model to match key moments of the Brazilian labor
market and simulate Pix as a transaction-cost reduction for small,
cash-intensive firms. The calibrated model reproduces the
reduced-form wage and inequality patterns and reveals two sources
of welfare gains not captured by the empirical design. First, as
small firms gain market share, the employment HHI declines and
average wage markdowns fall. This creates a spillover effect that increases wages among large firms, a segment that is not directly benefited by Pix.\footnote{Thus, our model allows us to deal with the missing intercept problem present in reduced-form designs \citep{NS_2014, Wolf_2021}.} Second, the decline of the within-city
component of wage variance is partially offset by a rise in the
between-city component. A counterfactual that closes the adoption
gap amplifies these effects: unlike the baseline, uniform adoption
reduces both the within- and between-city components of wage
variance. The uneven rollout of Pix increased the total wage-bill
by 0.04\%, while uniform adoption would generate a 0.21\%
increase, relative to a world without Pix.

\paragraph{Related Literature.} Our paper contributes to the
literature on financial technology and digital payments. Prior work
has studied the effects of digital payments on risk-sharing and
poverty reduction \citep{jack2014risk, suri2016long}, financial
inclusion \citep{ouyang2021cashless}, economic growth
\citep{dubey2023can}, and small business outcomes
\citep{agarwal2019mobile, agarwal2022real, klapper2023digital,
dalton2024electronic, higgins2024}, as well as credit access via
cashless payment records \citep{ghosh2024fintech}, but the effects
of instant payment systems on wages and wage inequality remain
largely unexplored. We also contribute to the payments literature
studying adoption and banking outcomes, including cash disruptions
and payment-technology adoption dynamics
\citep{chodorow2020cash, crouzet2023}, deposit competition
\citep{sarkisyan2023instant}, payment technology
complementarities \citep{sampaio2024payment}, and the role of
physical bank infrastructure in digital payment adoption
\citep{mariani2023banks}.

To our knowledge, our paper is the first to examine the effects of an instant payment system on wages and wage inequality using matched employer-employee data. We show that Pix raises wages in small, cash-intensive establishments, compressing the wage distribution from below. Our findings are consistent with a novel mechanism: by lowering transaction costs, instant payments increase labor demand among small, cash-intensive firms, with larger wage effects where low-skill workers are scarce. This contrasts with the conventional view that technological innovation primarily benefits skilled labor. We complement these reduced-form results with a calibrated monopsony
model that quantifies their aggregate implications and shows that
uniform adoption would reduce wage dispersion and raise the aggregate
wage bill.\footnote{Related work links faster payments or broader financial access to labor market outcomes. \cite{BARROT} show that the U.S.\ QuickPay reform raised employment for treated small contractors, while \cite{FONSECA2024103854} find that expanding bank branch coverage in Brazil increased wages. We complement these studies by analyzing a nationwide instant payment platform that reduces payment frictions and delivers concentrated wage gains and inequality compression among cash-intensive businesses.}

We also contribute to the large literature on technological change and wage inequality. The conventional view, supported by evidence from computerization and automation, posits that new technologies generally raise inequality by complementing skilled labor and substituting for routine tasks \citep{autor2003skill, acemoglu2011skills}. This includes both skill-biased technological change \citep{goldin1998origins, bound1992changes, krusell2000capital} and routine-biased technological change \citep{goos2014explaining, autor2013polarization, michaels2014has}. In contrast, we show that financial technologies exhibit markedly different distributional effects. Our results suggest that the impact of technology on wage inequality critically depends on the intersection of three factors: which specific market frictions it alleviates, the types of businesses it disproportionately affects, and the skill composition of their workforces.

\section{Conceptual Framework}\label{sect:model}

We develop a stylized model to illustrate how a reduction in transaction costs generated by instant payment systems passes through to wages and reshapes wage inequality. The economy consists of heterogeneous firms competing à la Cournot, two skill groups, and a self-employment outside option. Wages respond through a direct effect on the marginal revenue product of labor, a competition channel operating through firm market shares, and a supply-elasticity channel from the outside option. The distributional consequences depend on a skill-composition effect, namely whether the treated firms employ predominantly low- or high-skill workers, and a skill-bias effect from potential complementarity between the new payment infrastructure and high-skill labor.

\subsection{Workers}
The labor force consists of two skill types, low- ($L$) and high-skill ($H$), with fixed masses $\lambda_L$ and $\lambda_H = 1 - \lambda_L$. Workers of each skill type $k$ choose among $M$ firms and a self-employment option indexed by $i=0$. A worker $\omega$ of skill $k$ employed at firm $i$ obtains utility
\begin{equation*}
U_{ik\omega} = \ln w_{ik} + \xi_{ik\omega}
\end{equation*}
where $w_{ik}$ is the wage paid by option $i$ to skill-$k$ workers and $\xi_{ik\omega}$ is a match-specific preference shock. The self-employment payoff is $w_{0k} = (1 - \tau_0)\, \tilde{w}_{0k}$, net of transaction costs $\tau_0$. Preference shocks follow a nested generalized extreme value distribution with parameters $\eta$ and $\delta$ governing, respectively, the substitutability across firms and between the firm sector and self-employment. We assume $\eta > \delta$, so workers view firms as closer substitutes with one another than with self-employment, and we take these parameters to be symmetric across skill groups, $\eta_k=\eta$ and $\delta_k=\delta$.

Standard aggregation of the logit choice probabilities (Appendix~\ref{app:elasticity-derivation}) yields the fraction of skill-$k$ workers who choose self-employment:
\begin{equation}
\pi_k = \frac{w_{0k}^{\,\delta}}{w_{0k}^{\,\delta} + W_k^{\,\delta}}
\label{eq:self_emp_share}
\end{equation}
where $W_k$ is the CES wage index of the firm sector. Total employment of skill $k$ at firm $i$ is $n_{ik} = \lambda_k\,(1 - \pi_k)\, s_{ik}$, where $s_{ik}$ is the firm's payroll share within the firm-sector and skill group $k$. The labor supply elasticity that firm $i$ faces for skill-$k$ workers is
\begin{equation}
\frac{1}{\varepsilon_{ik}}
\;=\; \frac{1-s_{ik}}{\eta} \;+\; \frac{s_{ik}}{\delta\,\pi_k}
\label{eq:elasticity}
\end{equation}
which is increasing in the self-employment share and decreasing in the firm's payroll share.

\subsection{Firms}
There is a fixed set of $M$ heterogeneous firms indexed by $i=1,\dots,M$. Firm $i$ has idiosyncratic productivity $z_i$ and produces a homogeneous good, sold in a competitive market at a price normalized to one, using low-skill and high-skill labor with a Cobb-Douglas technology:
\begin{equation}
y_i = (1 - \tau_i)\, z_i\, n_{iL}^{\alpha}\, n_{iH}^{1-\alpha}
\label{eq:production}
\end{equation}
where $\tau_i$ is a firm-specific transaction cost and $\alpha \in (0,1)$ is the low-skill labor share. Firms compete à la Cournot, choosing employment $n_{iL}$ and $n_{iH}$ taking competitors' employment as given. The first-order condition yields:
\begin{equation}
w_{ik} = \mu_{ik}\times \mathrm{MRPL}_{ik}
\label{eq:markdown}
\end{equation}
where $\mathrm{MRPL}_{ik}$ is the marginal revenue product of skill-$k$ labor and the markdown $\mu_{ik} = \varepsilon_{ik} / (\varepsilon_{ik}+1)$ is increasing in the elasticity $\varepsilon_{ik}$ defined by equation~(\ref{eq:elasticity}).

\subsection{Comparative Statics}

We illustrate how labor markets and wage inequality respond to instant payment systems. We consider the initial transaction cost to be inversely related to firm productivity, and model instant payments as a decline of $\tau_i$ and $\tau_0$ to zero.

\paragraph{Wages.} Equation~(\ref{eq:markdown}) shows that a fall in $\tau_i$ raises wages through three channels. First, a \emph{direct effect} that raises $\mathrm{MRPL}_{ik}$, mediated by the markdown $\mu_{ik}$. Second, a \emph{competition channel}: as small firms expand and large, high-share firms lose payroll share, the markdowns of the firms with the most market power compress. Third, a \emph{supply-elasticity channel} generated by a lower $\tau_0$ that improves the outside option, increasing $\varepsilon_{ik}$ and $\mu_{ik}$. The latter two are general-equilibrium forces that amplify the direct effect.

\paragraph{Wage inequality.} Whether these gains compress the wage distribution depends on who the benefited workers are. Because small firms are the most benefited by lower transaction costs, the gains fall disproportionately on their workers, so the distributional impact hinges on their skill composition, governed by $\alpha$. Figure~\ref{fig:cf_inequality} plots the change in the Gini coefficient against $\alpha$, showing that inequality falls, and the decline is stronger when the benefited firms are more low-skill intensive, so that the gains accrue predominantly to low-skill workers.

This skill-composition force can be offset by a skill-bias channel. If the digital infrastructure accompanying instant payments complements high-skill labor, it raises their marginal product and widens the skill premium. We formalize this in an extension with capital-skill complementarity in Appendix~\ref{app:capital}, and show that when the complementarity is strong it can partly offset the skill-composition force, so the sign of the employment response is a question we take to the data

\paragraph{Employment.} The effect on firm-sector employment is ambiguous and varies with firm size. A lower $\tau_i$ raises labor demand, while a lower $\tau_0$ draws workers into self-employment; the net change depends on which force dominates. In the smallest, most affected by transaction costs before Pix, the demand gain is large and employment rises. In medium-sized firms, which benefit less and compete with a more attractive outside option, the two forces may cancel leaving employment unaffected. Whether instant payments raise firm-sector employment is therefore an empirical question, which we take up in the next sections.

\section{Institutional Background}\label{sect:setting}

On November 16, 2020, the Central Bank of Brazil launched Pix, an instant payment system. Pix enables immediate, 24/7 transfers between individuals, businesses, and government entities. Users initiate transfers through banks' mobile apps or websites using aliases (e.g., tax IDs, phone numbers) or QR codes. As of early 2024, over 152 million individuals and 15 million firms had registered for Pix.

Two design choices are relevant for adoption and for our empirical strategy. First, participation is mandatory for large banks (those with more than 500,000 accounts), which covered more than 99 percent of bank accounts in 2020. This ensured widespread availability and interoperability. Second, Pix transfers are free for individuals. Since many small firms use the owner's personal bank account for business transactions, they pay zero or near-zero fees. Even when firms use Pix through business accounts, fees are substantially lower than those of other methods: merchants pay an average fee of 0.22 percent per Pix transaction, compared to 2.2 percent for credit card payments \citep{duarte2022bulletin}. Pix's pricing structure is particularly relevant given the inverse relationship between firm size and merchant discount rates, whereby smaller firms face systematically higher fees for credit and debit card payments than larger firms (Appendix Figure \ref{appendix_fig:MDR}), and therefore stand to benefit disproportionately from the lower costs of Pix.\footnote{The inverse relationship between firm size and merchant discount rate is also observed in the United States \citep{egan2026pays}.}

Pix quickly became a major payment method. About a year after its launch, the number of Pix transactions surpassed those of credit and debit cards as well as TED (the electronic wire transfer system used before Pix). In terms of transaction value, Pix exceeded credit and debit cards two quarters after its introduction (Appendix Figure \ref{appendix_fig:differ_payment}). Pix initially grew through transfers between individual-linked accounts and then expanded to transfers from individual-linked to business-linked accounts (Appendix Figure \ref{appendix_fig:desc_p2p_p2m}).\footnote{Bank accounts are associated with either an individual tax ID or a business tax ID. Some accounts tied to individuals may also be used for business purposes, especially by small businesses and informal firms.}

The rise in Pix coincides with a 35 percent drop in the number of cash withdrawals (Appendix Figure \ref{appendix_fig:withdrawal}), consistent with reduced reliance on cash.\footnote{Central Bank of Brazil surveys show the decline: cash was identified as the most common payment method by 60.2 percent of respondents in 2018, 41.7 percent in 2021, and 22 percent in 2024 (\emph{O brasileiro e sua relação com o dinheiro}).}

Pix adoption is concentrated in cash-intensive sectors. Between 2021 and 2023, retail accounted for 32.8 percent of Pix transactions and services for 31.6 percent, while manufacturing accounted for only 3.1 percent (Appendix Table \ref{appendix_tab:desc_pix_sector}). This concentration motivates our sectoral heterogeneity tests in the empirical analysis.

\section{Data and Research Design}\label{sect:data}

\subsection{Data}

We use data from multiple sources. First, we draw on data from the Central Bank of Brazil, which provides information on the number and value of Pix transactions initiated or received by firms and individuals, broken down by municipality and month.\footnote{Brazil has 5,570 municipalities, grouped into 26 states and one federal district.} Second, we use matched employer-establishment data from the Brazilian Ministry of Labor (\emph{Relação Anual de Informações Sociais}, RAIS). This dataset includes information on workers (e.g., education), job contracts (e.g., wages), and employers (e.g., size, industry) for all Brazilian formal employment records.\footnote{RAIS data cover only formal employment relationships, omitting independent contractors, self-employment, and informal work.} Third, we use data on firm creation from the \emph{Cadastro Nacional da Pessoa Jurídica}, a firm registry maintained by the Brazilian Federal Tax Authority. This dataset allows us to compute the monthly creation of micro-firms, small businesses, and large enterprises at the municipality-industry level.  Finally, we collect data on municipality characteristics, such as the age profile of the population, the number of bank branches, homicide rate, GDP per capita, and mobile penetration from the Brazilian Institute of Geography and Statistics (IBGE), the Central Bank of Brazil, and the Telecommunications Agency (ANATEL). We restrict our analysis to the 2016–2024 period. %

Table \ref{tab:summ_stat} provides summary statistics of the main variables. The average municipality has a GDP per capita of 24,500 BRL (median 18,200), 4 branches (median 1), and 36,460 inhabitants (median 11,065). The average municipality has a Gini coefficient of 0.29 (median 0.28). The share of the population with access to 3G+ is 54 percent  (median 53 percent).

\subsection{Research Design}

We study how instant payment systems affect labor markets and wage
inequality. These technologies reduce payment frictions: cash-handling
costs, high processing fees, and settlement delays. We exploit
Brazil's nationwide introduction of Pix in November 2020. Before Pix,
77 percent of retail transactions used cash, credit card fees averaged
2.2 percent, and settlements took multiple days. Pix reduced
cash-handling costs, lowered merchant fees to 0.22 percent, and
enabled instant settlement. These cost reductions matter most for
small establishments in cash-intensive sectors like retail, where
payment frictions consume a larger share of revenue.

We employ a triple difference-in-differences design leveraging variation across three dimensions. First, we exploit Pix's nationwide introduction in November 2020. Second, we use pre-existing mobile penetration across Brazilian municipalities, measured as the 2019 ratio of 3G-capable devices to population, which strongly predicts Pix adoption intensity. Third, we contrast small versus large establishments within the same municipality, hypothesizing that small establishments benefit disproportionately from reduced transaction costs and accelerated cash flow enabled by instant payments. This design identifies intention-to-treat effects of instant payment availability on establishment-level wages.

Formally, we define pre-Pix mobile penetration at the municipality level \( c \) as:

\begin{equation}
\text{Mobile Penetration}_c = \frac{\text{Number of cellphones with 3G or above}_c}{\text{Population}_c}
\end{equation}

Figure~\ref{fig:mobile_penetration} shows the spatial distribution of
mobile penetration across municipalities, measured as the ratio of
3G-capable devices to population in 2019. Mobile penetration varies
substantially across Brazilian municipalities. Because mobile
penetration is measured before Pix and strongly predicts subsequent Pix
adoption intensity, we use it to capture cross-municipality differences
in exposure to Pix. Section~\ref{sec:first_stage} presents the
first-stage evidence.

Our triple difference-in-differences specification is:
\begin{equation}
Y_{kjct} = \beta \times (\text{Mobile Penetration}_c \times \text{Small}_k \times \text{Post}_t) + \delta_{kjc} + \delta_{ct} + \delta_{kt} + \delta_{jt} + u_{kjct}
\end{equation}

where $Y_{kjct}$ is the log average wage for establishment size group $k$ in industry $j$, municipality $c$, and year $t$.\footnote{Size bins are constructed based on the number of employees (up to 4, 9, 19, 49, 99, 249, 499, 999, and 1000 or more workers).} \textit{Small}$_{k}$ equals one for establishments with fewer than 20 employees (or fewer than 10 or 5 employees) in year $t$. \textit{Post}$_t$ equals one for $t \ge 2021$ (post-Pix) and zero otherwise.\footnote{Pix launched on November 16, 2020. Because we measure outcomes at the annual level, we code 2020 as the last pre-treatment year.} %

Our granular data allow a rich set of fixed effects to address potential confounders. Municipality-by-size-by-industry fixed effects ($\delta_{kjc}$) absorb time-invariant differences across establishment types within each municipality. Municipality-by-year fixed effects ($\delta_{ct}$) control for time-varying local economic shocks affecting all establishments in a municipality. Size-by-year fixed effects ($\delta_{kt}$) capture national trends that differentially affect establishments of different sizes. Industry-by-year fixed effects ($\delta_{jt}$) account for sector-specific shocks and trends at the national level. While this is our baseline specification, we also present results using more demanding fixed-effect structures. Standard errors are clustered at the municipality level to account for serial correlation.

The coefficient $\beta$ captures the intention-to-treat effect of Pix
on wages at small establishments relative to large establishments.
Because the baseline specification pools all sectors, $\beta$ averages
across sectors with different pre-Pix exposure to payment frictions.
We therefore examine heterogeneity across cash-intensive sectors, such
as retail, and less cash-intensive sectors, such as wholesale and
manufacturing.

The key identifying assumption underlying our triple difference-in-differences design is that, in the absence of Pix, wage trends for small and large establishments would have evolved similarly across municipalities with different pre-Pix mobile penetration, conditional on our fixed effects. While this assumption cannot be tested directly, we provide supporting evidence by examining pre-trends.

This assumption does not require mobile penetration to be randomly
assigned across municipalities. It requires only parallel trends in
the wage differential between small and large establishments,
conditional on the fixed
effects.\footnote{High-mobile-penetration municipalities are
unconditionally more urban with higher wages.
Figure~\ref{fig:covar_balance} shows that these differences shrink
materially within population size and agricultural exposure deciles.}
Municipality-by-year fixed effects absorb time-varying shocks common
to all establishments within a municipality, while size-by-year and
industry-by-year fixed effects control for national trends. Our
triple-difference estimator thus relies solely on within-municipality
wage differentials between small and large establishments across
varying pre-Pix mobile penetration.

Because pre-Pix mobile penetration is a proxy for exposure to Pix
rather than a direct measure of adoption, our estimate should be
interpreted as an intention-to-treat effect. A remaining concern is
that mobile penetration also captures non-Pix digital adaptation
during the COVID-19 period that differentially benefited small
establishments. Three pieces of evidence weigh against this
interpretation. First, augmenting the specification with direct
measures of pandemic intensity interacted with
$\textit{Small} \times \textit{Post}$ leaves the coefficient of
interest largely unchanged. Second, effects concentrate in
cash-intensive sectors and are absent in manufacturing and wholesale,
a pattern difficult to reconcile with broad digital adaptation. Within
services, effects are larger in in-person services (e.g., barber
shops, auto repair) than in delivery-capable services (e.g.,
restaurants), weighing against delivery platforms as the primary
driver. Third, the dynamic effects grow over time rather than
attenuate, more consistent with continued merchant adoption of Pix
than with a transitory pandemic shock.
Section~\ref{sec:alternative} presents these tests in detail.

\section{Main Results}\label{sect:main}

\subsection{Pix Adoption}\label{sec:first_stage}

Figure \ref{fig:pix_first_stage} reports coefficient estimates from regressions analyzing Pix adoption and cash usage following Pix's introduction, controlling for municipality fixed effects, time fixed effects, and municipality characteristics (e.g., GDP per capita, percent of young population, and importance of the agriculture sector) interacted with time.

Panel A plots the per capita value of Pix transactions received by firms (business-linked accounts), households (individual-linked accounts), and in total, showing a substantial increase in usage over time. For example, by the third quarter of 2023, the value of Pix transactions received by firms had increased by approximately BRL 750 per capita in response to a one standard deviation increase in mobile penetration. The growing magnitude of the coefficients suggests that pre-existing differences in mobile infrastructure significantly influenced the intensity of Pix adoption across municipalities. Appendix Table \ref{appendix_tab:penetration_pix_adoption} quantifies this relationship, showing that a one-standard-deviation increase in mobile penetration is associated with 5.8 additional transactions per capita and approximately BRL 2,073 higher transaction value per capita following Pix introduction.

Panel B presents the relationship between our measure of predetermined cash intensity and ex-post Pix usage across industries. We define cash intensity as the industry share of sales to households, based on the 2019 Brazilian input-output table. We normalize the value of Pix transactions by the total wage bill of each industry. Cash intensity explains 30 percent of the variability of Pix adoption in the cross-section of industries.

Both panels reveal that municipalities with higher initial mobile penetration experienced significantly greater Pix adoption, and industries with higher cash-dependence use Pix more intensively. These results highlight the strong role of pre-existing mobile infrastructure and consumer-facing transactions in shaping the intensity of Pix usage across municipalities and industries. Overall, these results support our approach of using mobile penetration as a variable to capture differential exposure to Pix adoption.

\subsection{Wage Effects}

Table \ref{tab:triple_diff_size} reports triple difference-in-differences estimates of Pix's wage effects by establishment size. The dependent variable is the log of average monthly wages. The triple interaction term (Mobile Penetration × Small × Post) captures Pix's wage effect by comparing small and large establishments across municipalities with differing mobile penetration before and after Pix's rollout. Small establishments are defined as those with fewer than 20 employees.

Panel A of Table \ref{tab:triple_diff_size} presents estimates across alternative sets of fixed effects to address potential confounders. Column (1) includes municipality-by-size-bin-by-industry fixed effects to absorb all time-invariant differences between firms, as well as municipality-by-year, size-by-year, and industry-by-year fixed effects. This specification controls for any general time-varying local shocks (e.g., a city-wide lockdown or demand shock) and national-level trends affecting all small firms or all firms within an industry. The triple interaction coefficient is 0.003, indicating that a one-standard-deviation increase in mobile penetration leads to a 0.3 percent (approximately BRL 8.08, equivalent to USD 2.05 in 2019 terms) monthly wage increase for small establishments relative to large ones post-Pix.

In Column (2), we replace the separate municipality-by-year and industry-by-year fixed effects with the more demanding municipality-by-industry-by-year fixed effects. These new fixed effects absorb any unobserved, time-varying shocks that are specific to an industry within a given municipality, such as localized COVID-19 restrictions on retail, or a local demand shock that disproportionately affects the service sector. This rigorous control ensures the comparison is not driven by these localized, industry-specific trends. The estimate remains similar (0.4 percent) and highly significant.

Finally, Column (3), our preferred specification, replaces the
size-by-year control with size-by-industry-by-year fixed effects. This
fully saturated model additionally controls for national-level trends,
policies, or economic shocks that affect firms differently based on
both their size and industry (e.g., a federal relief program targeted
only at small service-sector businesses). The estimate remains similar
(0.4 percent) and highly significant. More importantly, the
coefficient changes little as we move from the less saturated
specification in Column (1) to the more saturated specifications in
Columns (2) and (3), from 0.003 to 0.004, which narrows concerns that
omitted local or sector-specific shocks drive the result.

Panel B further decomposes these effects by worker skill level. For each skill category (high- or low-skill), we restrict the data, aggregate observations to municipality × industry × establishment-size × year cells, and apply our main specification. The positive wage effects of Pix adoption are significant for both low-skill (0.4 percent) and high-skill workers (0.5 percent), consistent with both skill groups benefiting from reduced transaction frictions in cash‑intensive sectors.\footnote{Note that column 3 has fewer observations because some cells lack high-skill workers; this also occurs occasionally for low-skill workers.}

Figure \ref{fig:triple_diff_industry} plots dynamic effects from our triple difference specification based on establishment size, with coefficients normalized to 2020 as the baseline. Small establishments are defined as those with fewer than 20 employees. Pre-Pix coefficients (2016–2019) are small, statistically insignificant, and exhibit no evidence of differential pre-trends. A joint test fails to reject that all four pre-Pix coefficients are equal to zero ($F=1.51$, $p=0.196$),  supporting the parallel trends assumption. After Pix's introduction, wages in small establishments within municipalities characterized by higher mobile penetration increase steadily, reaching approximately 0.8 percent by 2024. These results indicate that Pix adoption significantly increased wages in smaller establishments, particularly in municipalities with higher mobile penetration.

Finally, we conduct four robustness tests on the wage estimates. First, Table \ref{tab:triple_diff_size_robust} examines robustness across alternative definitions of small establishments (fewer than 5, 10, 20, and 50 employees). The results remain consistently stable, significant, and robust for establishments defined with fewer than 5, 10, and 20 employees, but are insignificant for establishments with fewer than 50 employees. All significant results have coefficients of 0.004. These results validate that Pix's positive impact on wages concentrates in smaller establishments, which are the type of businesses that benefit the most from this technology. Second, Appendix Table \ref{tab:triple_diff_size_noceo} excludes CEOs as a proxy for establishment owners. The preferred specification yields a coefficient of 0.005, consistent with wage gains accruing to workers rather than reflecting changes in owner compensation. Third, in Appendix Table \ref{tab:continuing_estabs}  we address compositional churn. One concern is that changes in the composition of establishments, rather than wage growth within firms, drive the baseline estimate. Entry of high-wage establishments or exit of low-wage ones could raise average wages without any within-firm gains. We re-estimate \ref{tab:triple_diff_size} using a balanced panel. The coefficient ranges from 0.015 to 0.020 across different specifications. Because survival may respond to treatment, we use this sample only to assess composition and retain the full sample for the baseline. Finally, Appendix Table \ref{tab:commuting_zone} addresses treatment leakage across municipal boundaries. Because Pix reduces transaction costs independently of physical borders, commuting and cross-border shopping could expose establishments in neighboring municipalities to each other's treatment. We re-estimate equation (6) aggregating exposure and outcomes to commuting zones. The preferred specification yields a coefficient of 0.004, indistinguishable from the baseline, consistent with boundary spillovers not driving the municipal estimates.

\subsubsection{Industry Heterogeneity Analysis}

Our central hypothesis is that Pix boosts wages by reducing transaction frictions ($\tau_c(z)$) that are particularly costly for small, cash-intensive firms. A key implication of this mechanism is that the wage effect should not be uniform across the economy. Instead, wage effects should be concentrated primarily in industries heavily reliant on in-person, small-value, business-to-consumer (B2C) transactions. Conversely, we should observe minimal or no effects in sectors that are not cash-intensive, such as capital-intensive or business-to-business (B2B) industries.

To test this critical prediction, we disaggregate our preferred triple-difference specification (Panel A of Table \ref{tab:triple_diff_size}, Column (3)) and re-estimate it separately for different industries. This provides a robust test against alternative explanations, such as a general non-Pix-related technology shock, which might affect all industries broadly rather than specifically targeting the cash-intensive sectors identified by our mechanism.

Table \ref{tab:triple_diff_size_new} presents these results, which support our proposed transaction-cost mechanism. Column (1) of Panel A presents pooled results for all industries, closely replicating our main findings. Column (2) shows results specifically for Retail, where we find a large and statistically significant coefficient (0.008), double the magnitude of the pooled effect. In contrast, in Wholesale (Column (3)) and Manufacturing (Column (4))—both primarily B2B sectors—the estimated effects are small and statistically insignificant. This null finding is particularly important. While Pix can also facilitate B2B transfers, its marginal benefit is largest when replacing high-friction payment methods like physical cash or high-fee credit cards (common in B2C retail). For B2B firms, Pix primarily replaces existing, low-cost bank transfers (like TED), offering a comparatively smaller efficiency gain. The absence of wage effects in these B2B sectors thus suggests our results are not driven by general efficiency improvements, but rather by the reduction of high-friction, cash-based transactions.

Similarly, within service industries (Panel B), we observe the largest effect in ``Other Services'' (e.g., barber shops, Column (4)), with a coefficient of 0.014, while more formal service sectors like Hotels (Column (2)) show no significant effect. This pronounced heterogeneity—large, positive effects in cash-intensive consumer-facing sectors and null effects in B2B or formal sectors—is consistent with wage gains reflecting reduced payment frictions faced by small businesses that previously relied heavily on cash transactions.

\subsection{Alternative Interpretations: COVID and Digital Readiness}\label{sec:alternative}

Pre-trend evidence supports the parallel trends assumption of our triple-difference design (Figure~\ref{fig:triple_diff_industry}). Because our treatment variable is pre-Pix mobile penetration, our estimate is an intent-to-treat effect of Pix exposure rather than a direct measure of Pix adoption. Therefore, the concern is that mobile penetration proxies for broader non-Pix digital adaptation during COVID-19 that disproportionately benefited small establishments. We evaluate this alternative below.

\noindent\textbf{COVID intensity.}
Our preferred specification includes municipality-by-industry-by-year fixed effects, which absorb any time-varying shock that affects all establishments in an industry within a municipality (e.g., local COVID severity, local demand changes). The remaining concern is size-specific COVID shocks correlated with pre-Pix mobile penetration. We therefore augment the specification with direct measures of pandemic intensity interacted with \textit{Small} and \textit{Post} (Table~\ref{tab:robustness_covid}). Column (1) adds per-capita emergency aid transfers to individuals (Aux'{i}lio Emergencial), Column (2) adds per-capita beneficiaries of the federal wage-subsidy program BEm (Programa Emergencial de Manuten\c{c}~{a}o do Emprego e da Renda), which compensated workers at firms that reduced working hours or suspended employment contracts during the pandemic, Column (3) adds municipal stay-at-home orders, and Column (4) adds cumulative COVID cases per capita. In all four columns, the triple interaction of interest remains stable at 0.004--0.005. The BEm, lockdown, and case interactions are small and insignificant. The transfer interaction is positive and significant, but the triple interaction remains unchanged. Following \cite{altonji2005selection}, this stability suggests that selection on unobservables would need to be large relative to selection on observed COVID proxies to explain our estimate. As an additional check, Appendix Table~\ref{tab:triple_diff_size_donut} drops the year 2020 entirely and redefines the pre-period as 2016--2019 and the post-period as 2021--2024. The preferred specification yields a coefficient of 0.005, consistent with the baseline estimate.

\medskip
\noindent\textbf{Sectoral and within-services heterogeneity.}
If Pix operates through reduced payment frictions, effects should concentrate
in cash-intensive sectors where Pix substitutes for costly payment methods.
A digital-readiness alternative would predict positive wage effects across a
broader set of sectors. Table~\ref{tab:triple_diff_size_new} shows a
significant effect in retail (0.8\%), but no significant effects in wholesale
or manufacturing. More directly, we split industries by their pre-Pix cash
intensity, measured as the share of sales to households. The wage effect is
close to zero in the lowest cash-intensity tercile and rises monotonically
across terciles, as shown in Section~\ref{sec:profit}.

A more specific alternative is that delivery platforms (e.g., iFood, Rappi) raised wages in small consumer-facing establishments. If delivery platforms drive our results, the effects should be strongest in delivery-oriented services such as restaurants. Under the Pix mechanism, effects should be relatively larger in in-person services where payment frictions were more salient. Appendix Table~\ref{tab:delivery_split} shows a small and statistically insignificant estimate for delivery-capable services such as restaurants (0.2\%), while physical services (e.g., barber shops, auto repair) show a significant 1.9\% increase. This within-services pattern is harder to reconcile with delivery platforms as the primary driver.\footnote{Delivery couriers in Brazil operate as independent contractors (typically MEIs) and are not recorded in RAIS, which covers only formal employment contracts (CLT).}

\medskip
\noindent\textbf{Post-period dynamics.}
A COVID digital-adaptation channel would plausibly attenuate as acute disruptions fade. Under the Pix mechanism, effects should grow with merchant adoption. Figure~\ref{fig:triple_diff_industry} shows that the dynamic triple-interaction coefficients rise from approximately 0.3\% in 2021 to 0.8\% by 2024, with no sign of attenuation. This time path is more consistent with the continued expansion of person-to-merchant Pix transactions (Appendix Figure~\ref{appendix_fig:desc_p2p_p2m}) than with a transitory pandemic shock.

\medskip

\medskip

Taken together, the evidence aligns more closely with Pix operating through payment frictions than with a broad post-2020 digital-readiness channel. 

\subsection{Effects on Wage Inequality}

Section \ref{sect:model} implies that a reduction in transaction frictions compresses wages when the firms with the largest wage gains employ relatively more low-skill labor, with the magnitude shaped by the skill–productivity complementarities across sectors and the elasticity of labor supply that governs wage markdowns under monopsony. 

Appendix Figure \ref{fig:skill_composition} shows that this condition holds most strongly in retail.  Small retail establishments are particularly intensive in low-skill labor, allocating over 91 percent of their total payroll to low-skill workers, compared to only 58 percent in large retail establishments. This contrasts with small manufacturing firms, which have a notably lower low-skill payroll share (around 75 percent).

Motivated by this composition, we test whether wage inequality declines more after Pix in municipalities with higher pre-Pix mobile penetration. We employ a municipality-level difference-in-differences specification, regressing our measure of inequality, the Gini coefficient, on standardized mobile penetration while controlling for municipality fixed effects, region-by-year fixed effects, and municipality characteristics interacted with year.

Figure \ref{fig:gini} supports our model's prediction using the Gini coefficient of wages, calculated from the distribution of individual worker wages within each municipality-year and multiplied by 100.\footnote{We exclude government employees from this calculation.} This approach isolates the effect of Pix adoption on wage inequality while accounting for time-invariant municipality characteristics and local time-varying shocks. The estimates show no differential trends prior to Pix's introduction, consistent with our identifying assumption. Following Pix adoption, we observe a significant decline in the Gini coefficient in municipalities with higher mobile penetration relative to those with lower mobile penetration. This reduction in wage inequality emerges gradually, starting in 2021, and becomes more pronounced through the end of our sample.

Column (1) in Panel A of Table \ref{tab:gini_scarcity} quantifies this effect, showing that a one-standard-deviation increase in mobile penetration is associated with a statistically significant 0.270-point decline in the Gini coefficient (scaled by 100) following Pix adoption. Relative to the sample mean of 29 (corresponding to an unscaled Gini of 0.29), this represents a 0.9 percent reduction in wage inequality.

Figure \ref{fig:wage_distribution} further unpacks these results by examining wage dynamics separately for different segments of the wage distribution. Panel A of Figure \ref{fig:wage_distribution} reveals a noticeable increase in wages for workers in the bottom 50\% of the distribution following Pix adoption. In contrast, Panel B shows no significant wage changes for workers in the top 50\% of the distribution. The regression results in Table \ref{tab:gini_scarcity} (Columns (2) and (3)) show that the bottom 50\% experiences a significant 0.6 percent increase in wages, while the top 50\% experiences no significant effect. This decomposition indicates that the observed reduction in wage inequality is driven by wage growth at the bottom of the distribution, rather than wage compression at the top.

The decline in wage inequality is particularly notable given the canonical literature on skill-biased technical change, which suggests that new technologies typically increase wage inequality by raising the productivity and wages of skilled workers \citep{autor1998computing,acemoglu2011skills}. Our findings show that financial technologies can have markedly different distributional effects. By reducing frictions in cash-intensive sectors that disproportionately employ low-skill workers, digital payment technologies can generate more inclusive patterns of wage growth.

\section{Mechanisms}\label{sect:mechanism}

We examine why wages rise after Pix. First, we estimate effects on employment and new firm creation. Next, we test if wage effects are larger where labor markets are tighter. Finally, we examine the sources of small-firm profitability gains and evaluate alternative explanations.

\subsection{Labor Demand Effects}

Our main results indicate that higher mobile penetration increases wages in small establishments, especially within retail, and compresses wage inequality. According to our model, reductions in transaction frictions increase firm profitability and stimulate labor demand, particularly in cash-intensive sectors that heavily rely on low-skilled labor.

We test this hypothesis by examining employment growth using a triple difference-in-differences approach comparing small versus large establishments. Figure \ref{fig:net_job_creation_small} illustrates the dynamic effects on job growth for the smallest establishments (fewer than 5 employees). The estimates show no significant pre-trends and a clear increase after Pix's introduction in November 2020. Table \ref{tab:triple_diff_active_small} quantifies these effects, using the logarithm of active jobs as the dependent variable. We find a positive and significant effect of 0.017 for the smallest firms (N<5, Column 1). Slightly larger definitions of ``small'' (N<10, Column 2; N<20, Column 3) yield smaller, insignificant estimates.%

Given that employment growth could result either from existing firms expanding or from new firm entry, we next analyze firm creation directly. Appendix Figure \ref{fig:firm_entry} presents entry patterns by sector and firm size. After Pix’s introduction, municipalities with higher mobile penetration show a significant increase in small retail firm entry (Panel A). In contrast, entry rates for small manufacturing firms remain unchanged (Panel B), consistent with our earlier findings. Table~\ref{tab:small_entry} quantifies these results, showing that a one-standard-deviation increase in mobile penetration raises small retail firm entry per 1,000 residents by 0.004, with no significant effect in small manufacturing.

We further examine entrepreneurship using municipal MEI-ICMS data. Our conceptual framework shows that the reduction in transaction costs brought by Pix enhances self-employment incentives, amplifying wage gains and reinforcing the decline in inequality through the supply elasticity channel. The Micro-Empreendedor Individual (MEI) tax charges a fixed \text{R\$}1 monthly
ICMS fee regardless of sales volume, making annual per-capita MEI-ICMS receipts proportional to the number of active MEIs.

Appendix Figure \ref{fig:taxes} plots the dynamic effects of mobile penetration on annual MEI--ICMS receipts per 1,000 residents. The estimates show no
significant pre-trends, then rise after Pix's introduction in November 2020 for high-mobile-penetration municipalities. Appendix Table \ref{tab:mei_entry}
quantifies this effect: a one-standard-deviation increase in mobile penetration raises annual MEI--ICMS receipts by BRL 1.417 per 1,000 residents.

This MEI expansion following Pix adoption is consistent with reduced transaction costs facilitating self-employment registry (or formalization). Combined with our evidence on small retail firm entry, these results indicate that Pix expands entrepreneurship at the extensive margin, particularly among cash-intensive sole proprietors. Employment growth concentrates in micro-firms (N<5) and new entrants, while wage effects extend to all small establishments (N<20). This pattern is consistent with Figure~\ref{fig:cf_employees} in our conceptual framework. The expansion of self-employment raises outside options for workers, pushing up wages at slightly larger establishments even without headcount growth.

\subsection{Labor Market Tightness and Wage Amplification}

The evidence above shows that Pix increases labor demand for
low-skill workers through employment growth and new firm entry. Whether this demand shift translates into wage gains
depends on local labor supply conditions. If low-skill labor is
abundant, firms can expand without raising wages. If low-skill
labor is scarce, the same demand shift bids up wages and
compresses inequality.

Table~\ref{tab:gini_scarcity}, Panel B, tests this prediction by interacting the treatment with an indicator for tight labor markets, defined as municipalities where the share of low-skill workers is below the population-weighted median. Column (1) shows that a one-standard-deviation increase in mobile penetration reduces the Gini coefficient by an additional 0.323 points in tight labor markets. When examining wage components, the differential effect on Bottom 50\% wages in tight markets (Column 2) is positive but imprecisely estimated. These results indicate that Pix's inequality-reducing effects concentrate where low-skill labor is relatively scarce, consistent with a labor demand channel amplified by local labor market tightness.

\subsection{Small Firm Profitability}\label{sec:profit}

We examine three channels through which Pix may increase small-firm profitability: reductions in cash-handling costs, variation in merchant payment fees, and higher local demand. The first two also speak to the digital-readiness alternative discussed in Section~\ref{sec:alternative}.

\paragraph{Payment Frictions and Cash-Handling Costs.}

If Pix substitutes for cash rather than expanding overall payment volume, it
can reduce the costs firms incur from storing, transporting, and safeguarding
physical currency. We estimate the effect of Pix on bank branch cash
inventories. Figure~\ref{fig:cash_inventories} shows that municipalities with
higher mobile penetration experienced a statistically significant decline in
cash balances held by bank branches. The reduction intensifies over time,
suggesting that Pix adoption substituted for cash transactions.

We estimate the wage response for industries with different levels of
cash-intensity, measured by their share of sales to households
(Table~\ref{tab:triple_diff_cash_intensity}). Wage effects are null in
low-cash industries (Column~1) and rise monotonically with cash-intensity
(Columns~2--3).

Cash dependence is more costly in high-crime environments because firms must
store, transport, and safeguard physical currency, increasing losses from
theft and security expenditures. We split municipalities by their 2019
homicide rate into above- and below-median groups and re-estimate the
baseline specification for each (Figure~\ref{fig:ddd_wage_homicide},
Table~\ref{tab:ddd_wage_homicide}). In high-homicide municipalities, wages in
small establishments rise by approximately 0.6 percent per
one-standard-deviation increase in mobile penetration. In low-homicide
municipalities, effects are small and imprecise.

\paragraph{Merchant Fees and Bargaining Power.}

When customers pay by card, merchants pay a merchant discount rate (MDR), the
share of the transaction value retained by the acquiring chain. MDR comprises
interchange fees set by card networks, scheme fees, and the acquirer's
processing margin. Only part of this bundle is negotiable: larger merchants
with greater volume and bargaining power can typically secure lower processing
margins and better contract terms, even if interchange and scheme fees are
uniform. Small merchants face higher effective MDR because they are less able
to negotiate discounts. In Brazil, average merchant costs on Pix are
substantially lower than the average MDR faced by merchants on card
transactions.

If negotiated MDR differs by firm size, switching to Pix should lower payment
costs more for small establishments in small firms than for small
establishments in larger firms. We estimate a quadruple difference-in-difference
 specification. Figure~\ref{fig:dddd_wage_small} plots the event-study coefficients of the interaction $\text{Mobile Penetration} \times
\text{Small Establishment} \times \text{Small Firm} \times \text{Post}$.
The coefficients are close to zero prior to Pix and become positive after
2020.

\paragraph{Local Demand.}

Pix may also boost demand by providing a safer and more convenient
alternative to cash. We split industries into tradable and non-tradable
categories based on payroll geographical concentration, following
\cite{mian2014explains}.\footnote{Industries with payroll concentrated
in few locations are classified as tradable; those dispersed nationally
as non-tradable. We exclude low-cash sectors, which were unaffected by
Pix.} If local demand were a key mechanism, non-tradable industries in
high-adoption municipalities should show larger wage gains. Wage
responses are similar across the two groups (Appendix
Table~\ref{tab:triple_diff_T_vs_NT}). Local demand does not appear to
be the primary driver.

These results are consistent with cash-handling costs being an important component of Pix’s gains for small firms.

\subsection{Additional Channels for Pix's Effect on Wages}\label{sec:add_channels}

The preceding sections present evidence consistent with Pix reducing
transaction costs for small, cash-intensive businesses, increasing
labor demand and encouraging entry. Wage effects are larger where
low-skill labor is scarce. We now evaluate two additional channels
through which Pix may affect wages: formalization and credit access.

\paragraph{Formalization and wage reporting.}

Pix may have increased formalization or improved wage reporting,
raising measured wages without a corresponding change in labor market
conditions. This matters because our main wage outcomes come from
administrative data on formal employment. A formalization-driven
channel predicts that wage effects should be larger where baseline
informality is higher, and that the observed composition of formal
workers should shift toward worker types that are overrepresented in
informal employment.

Our findings indicate that Pix spurred some formalization, particularly the registration of solo entrepreneurs (MEIs). Several institutional
and empirical patterns, however, make formalization unlikely to be the main driver of observed wage increases.

First, total mandated employer costs in Brazil exceed the contractual salary by more than 50 percent \citep{cavalcante2020encargos}. Even
large, targeted reductions in formalization costs have produced only limited and transitory effects on firm registration,\footnote{The MEI program halved monthly taxes for micro-firms but increased registration by only 1.9 percentage points from a 20 percent baseline, with the effect dissipating after six
months \citep{rocha2018taxinformality}.} making a formalization-only explanation of the wage results unlikely.

Second, Pix transactions remain covered by bank secrecy. Tax authorities do not observe transaction-level origin or destination details, limiting the scope for Pix to increase enforcement exposure and trigger formalization.

Third, wage effects remain stable across different size thresholds, with identical effects for firms with fewer than 5, 10, and 20 employees.
Since firms approaching 20 employees are substantially more likely to already comply with labor regulations, the persistence of wage gains in
this group suggests generalized labor demand increases rather than formalization-driven compliance effects.\footnote{\citet{ulyssea2018firms} documents that informality declines sharply with firm size in Brazil, reaching near-zero levels well below 20 employees.}

Fourth, pure compliance effects would generate uniform wage increases toward the regulatory minimum, independent of local labor market
conditions. Instead, we find wage effects concentrated where low-skill labor is scarce. This pattern is consistent with increased labor demand
interacting with constrained labor supply. 

Fifth, if formalization drives our results, wage effects should be larger in municipalities with higher baseline informality, where there
is greater scope to formalize previously informal employment. We split municipalities by a pre-Pix informality measure constructed from the
2010 Census and estimate our baseline triple-difference specification separately for each group. Appendix Table~\ref{tab:ddd_wage_informal}
shows that wage effects are identical across the two groups. This pattern is inconsistent with a compliance or reporting channel, which
predicts larger wage increases where initial informality was higher.

Sixth, we test directly for shifts in the composition of formal workers. If wage gains reflect formalization, we should see significant changes
in the characteristics of formal employees after Pix. In Appendix Table~\ref{tab:composition}, we re-estimate our baseline specification
using worker observables as outcomes: gender, skill, tenure, and age. These estimates are uniformly close to zero. The null results are
informative because informal workers differ sharply from formal workers along these dimensions in the 2010 Census (Appendix
Table~\ref{tab:formal_informal}). A substantial inflow from informal employment into formal jobs would produce detectable shifts in
administrative data.

Worker tenure provides a further test. If wage gains reflected formalization, they would concentrate among newly hired workers entering the formal sector at the bottom of the tenure distribution. Appendix Table \ref{tab:tenure_monotonicity} interacts the treatment with tenure-group indicators, using workers with fewer than 12 months of tenure as the omitted group. Wage gains are 1.7 percent larger for workers with 12 to 24 months of tenure and 3.7 percent larger for workers with at least 24 months, and the two coefficients differ significantly. Wage gains rise with tenure, consistent with higher pay for existing formal workers rather than a change in the composition of who is recorded as formal.

Finally, we use PNADC, IBGE's household survey, to assess whether the wage gains reflect previously informal employees being registered as formal workers. If Pix raised measured wages by inducing firms to register employees previously paid off the books, rather than by raising pay within existing formal jobs, then treated areas should show a decline in the share of private-sector employees without a signed work contract. We aggregate the microdata to 146 survey-stratum-by-year cells and compare post-2020 outcomes across strata with different pre-Pix mobile penetration. Appendix Table \ref{tab:pnad_informality} shows no such decline. A one-standard-deviation increase in pre-Pix mobile penetration is associated with a statistically insignificant 0.26 percentage point increase in this share. These results weigh against an explanation in which the formalization of previously informal employees drives the RAIS wage gains.

Overall, Pix may have contributed to some formalization, but the available evidence does not support a formalization-only explanation
for the wage results. We interpret formalization as a complementary channel that can amplify measured wage gains, while the primary
mechanism operates through increased labor demand in cash-intensive businesses interacting with local labor-market frictions.

\paragraph{Credit Access.} Pix could also support small-firm expansion by improving access to external finance. Digital transaction records may reduce information asymmetries between lenders and small firms \citep{ouyang2021cashless, alok2024open, ghosh2024fintech, cramer2024shadow}. We test this channel using municipality-level business lending data. Appendix Figure~\ref{fig:business_loans} plots the dynamic effects of mobile penetration on log per-capita business loans. We find no differential trends before Pix and no significant increase afterward.\footnote{In Brazil, third-party lenders can access Pix transaction histories only with explicit customer authorization under the Open Finance framework, which launched after Pix and was still scaling during our sample period. Evidence from India shows that digital payment histories can expand credit access when data-sharing infrastructure is more mature \citep{alok2024open, ghosh2024fintech}.} This evidence does not support aggregate credit expansion as the primary driver of the wage effects we document. However, our test captures lending volumes and may not detect changes in loan terms, borrower composition, or specific types of credit. If present, these credit-margin effects would reinforce the same mechanism: Pix would relax constraints on small firms, enabling expansion and increasing labor demand. We do not rule out credit effects at other margins or over a longer horizon.

\medskip

Taken together, the evidence points to reductions in transaction
costs as the main channel through which Pix raises wages in small
establishments. Formalization contributes as a complementary
channel but it cannot account for the concentration of wage effects
where low-skill labor is scarce. We do not detect aggregate lending expansion in our data, although we
cannot rule out credit effects at other margins. %

\section{Quantitative Analysis}\label{sect:quantitative}

We consider a set of $c = 1, \dots, C$ cities, each containing $M_c$ firms. A fixed share $\gamma_k$ of firms operates in sector $k$, which can be retail, wholesale, or manufacturing. Workers are grouped into four skill categories $s = \{1,\dots,4\}$, each with mass $H_s$, where $\sum_s H_s = 1$.

A worker $i$ of skill $s$ derives the following utility from working at firm $j$ located in city $c$:
\begin{equation*}
    u_{ijc} = \log(w_{sjc}) + \xi_{ijc}
\end{equation*}

where $w_{sjc}$ is the wage offered by firm $j$ in city $c$ to workers of skill $s$, and $\xi_{ijc}$ is a worker-specific taste shock for firm $j$. These shocks are i.i.d. across individuals and follow the nested logit distribution:
\[
F(\xi_{1,1}, \dots, \xi_{M_c,C}) = 
\exp\left[ -\sum_{c=1}^{C} \left( \sum_{j=1}^{M_c} e^{-(1+\eta)\xi_{jc}} \right)^{\frac{1+\theta}{1+\eta}} \right],
\]
where $\eta$ and $\theta$ govern within-city and between-city substitution elasticities, respectively.

Given this preference structure, the labor supply to firm $j$ in city $c$ for skill $s$ is given by:\footnote{The aggregate indices are defined as:
$N_s = \left[ \sum_{c=1}^C N_{sc}^{\frac{\theta+1}{\theta}} \right]^{\frac{\theta}{\theta+1}}$,
$W_s = \left[ \sum_{c=1}^C W_{sc}^{\theta+1} \right]^{\frac{1}{\theta+1}}$,
$N_{sc} = \left[ \sum_{j=1}^{M_c} n_{jsc}^{\frac{\eta+1}{\eta}} \right]^{\frac{\eta}{\eta+1}}$, and
$W_{sc} = \left[ \sum_{j=1}^{M_c} w_{jsc}^{\eta+1} \right]^{\frac{1}{\eta+1}}$.}
\begin{equation}
\label{eq:labor_supply_full}
n_{sjc} = \left( \frac{w_{sjc}}{W_{sc}} \right)^{\eta} 
          \left( \frac{W_{sc}}{W_s} \right)^{\theta} N_s.
\end{equation}

Firm $j$ in industry $k$ and city $c$ uses a decreasing-returns technology that combines the four types of labor additively. Its profit function is given by:
\begin{equation*}
    \Pi_{jkc} = \sum_s ( 1 - \tau_{jc} ) A_s A_c z_j f_{ks}(z_j) n_{jsc}^{\alpha} - \sum_s w_s\left(n_{jsc},N_{sc}(n_{jsc},n_{(-j)sc}),N_s,W_s\right) n_{jsc}
\end{equation*}

where $A_s$ and $A_c$ are skill- and city-specific productivity terms, and $\tau_{jc}$ is a transaction cost. The function $f_{ks}(z_j)$ captures sector-specific skill complementarities and is defined as
\begin{equation*}
f_{ks}(z) = \frac{z^{\lambda_{ks}}}{\sum_{q\in S} z^{\lambda_{kq}}},
\qquad \sum_{q\in S} \lambda_{kq}=0.
\end{equation*}

The wage function $w_s\bigl(n_{jsc},N_{sc},N_s,W_s\bigr)$ is the inverse of the labor-supply relation given in equation \ref{eq:labor_supply_full}. For each skill group $s$, firms internalize their impact on $N_{sc}$, but take as given the country level values $N_{s}$ and $W_s$, as well as their competitors' strategies. Profit maximization yields the optimal wage for skill $s$ offered by firm $j$ operating in city $c$:
\begin{equation} \label{eq:labor_demand_full}
w_{jsc} = \frac{1}{1 + \bigl[ s_{jsc}/\theta + (1-s_{jsc})/\eta \bigr]} 
\; \alpha (1-\tau_{jc}) A_s A_c z_j f_{ks}(z_j) n_{jsc}^{\alpha-1},
\end{equation}

where $s_{jsc}$ denotes the firm’s wage-bill market share in city $c$ for skill group $s$. %
Thus, a firm with a larger market share $s_{jsc}$ sets a larger markdown, i.e., pays a lower wage relative to marginal product, if and only if $\theta < \eta$.

We solve the model through the following steps. First, following \citet{berger2022labor}, market shares satisfy the fixed-point condition

\begin{equation*}
s_{js} = \frac{\mu_{js} \, s_{js}^{\frac{\alpha \eta}{1 + \eta}} \, (1 - \tau_j) \, z_j \, f_{k(j)s}(z_j)}
{\sum_\ell \mu_{\ell s} \, s_{\ell s}^{\frac{\alpha \eta}{1 + \eta}} \, (1 - \tau_\ell) \, z_\ell \, f_{k(\ell)s}(z_\ell)},
\end{equation*}

where $\mu_{jsc} = 1 \big/ \big( 1 + [ s_{jsc}/\theta + (1-s_{jsc})/\eta ] \big)$ is the wage markdown. This system of equations determines the equilibrium vector of firm market shares.

Second, given the resulting markdowns $\mu_{jsc}$, equilibrium wages and labor supply are computed directly from equations \ref{eq:labor_supply_full} and \ref{eq:labor_demand_full}. Finally, equilibrium requires that the market clears for each skill group $\sum_{j,c} n_{jsc} = H_s$ for all $s$.

\subsection{Calibration}

Our calibration distinguishes between two sets of parameters: those that are predetermined, based either on external evidence or on observable moments that are perfectly matched, and those that are calibrated internally to match specific moments of the data.

\textbf{Predetermined Parameters.}
We standardize the number of cities to $C = 100$, skill groups to $S = 4$, and sectors to $K = 3$. We group Brazilian cities evenly into 100 bins. The number of firms $M_c$ is such that the smallest bin has $x_m = 695$ firms, corresponding to the empirical minimum; and for the remaining bins, the number of firms is drawn from a Pareto distribution with shape parameter $\alpha_c = 1.24$. This parameterization ensures that the largest 10\% of cities (the top 10 bins) account for 64\% of all firms, matching the empirical concentration of firms across cities. Skill groups are defined by educational attainment: primary education, secondary education, some college, and college graduates. Their population shares $H_s$ are set to match the empirical distribution of workers. The three sectors--manufacturing, retail, and wholesale--have shares $\gamma_k$ calibrated to the observed number of firms in each sector.

We set the within-city and between-city labor supply elasticities to $\eta = 7$ and $\theta = 0.5$, respectively, following \citet{berger2022labor}. The decreasing-returns parameter is $\alpha = 0.7$. Transaction costs $\tau$ are set to 5\% for small retailers--defined as firms with productivity below the 90th percentile of the productivity distribution, which in our data have less than 19 employees--and zero otherwise.\footnote{This structure can be rationalized by a technology adoption decision: firms draw a potential productivity and decide whether to pay a fixed cost to adopt a technology that avoids transaction costs \citet{ulyssea2018firms}. Then, bigger firms are more likely to adopt the technology. Pix eliminates this fixed cost, thereby reducing transaction costs to zero for all firms.} All predetermined parameters are summarized in Appendix Table \ref{tab:pre_parameters}.

\textbf{Calibrated parameters.}
A second set of parameters is calibrated to match key moments of the Brazilian labor market. The sector-skill complementarity parameters $\lambda_{ks}$ target the skill composition of employment in each sector $k$. The Pareto shape parameters $\alpha_k$ for the firm productivity distribution are chosen to match the employment share of the top 10\% of firms in each sector. Finally, skill-specific productivity levels $A_s$ are calibrated to replicate the observed skill premium relative to the least educated group ($s = 1$). Appendix Table \ref{tab:cal_parameters} lists these 15 calibrated parameters and their corresponding empirical targets.

Our calibrated model successfully reproduces key features of the Brazilian labor market. As illustrated in Appendix Table \ref{tab:model_fit}, the model captures fairly well the skill composition of each sector. Retail is the least skill-intensive sector, with around 10 percent of its workers holding a bachelor's degree. In contrast, manufacturing and wholesale are more skill-intensive, with college graduate shares of 15 and 22 percent, respectively. Appendix Table \ref{tab:model_fit} shows that the model also matches the aggregate employment distribution across sectors. Retail and manufacturing each account for approximately 40 percent of total employment, while wholesale employs about 10 percent of the workforce. Finally, the bottom panel of Appendix Table \ref{tab:model_fit} confirms that the skill-specific average wages generated by the model align closely with their empirical counterparts.

\subsection{Counterfactual Analysis}

We use the calibrated model to quantify the aggregate effects of Pix and to simulate a counterfactual in which Pix adoption is uniform across municipalities.

\textbf{Pix Introduction.} We study Pix's implementation through the lens of our calibrated model. Consistent with our empirical findings, we model Pix as a reduction in transaction costs for small firms in the retail sector as follows:
\begin{equation*}
    \tau^{\text{new}} = \tau^{\text{old}} \times \big(1 - \omega(\text{MP}_c)\big),
\end{equation*}
where $\omega: [0,1] \to [0,1]$ is a strictly increasing function and MP$_c$ denotes mobile penetration. %
Because the model does not include mobile penetration directly, we impute it using fitted values from a linear projection of observed mobile penetration on city-level aggregates:
$\text{MP}_c = \hat{\beta}_0 + \hat{\boldsymbol{\beta}}' \mathbf{X}_c$,
where $\mathbf{X}_c$ collects city-level aggregates (for example, log employment and log wage bill). %
Cities with higher imputed mobile penetration experience a larger reduction in transaction costs. We define $\omega$ as a linear function of $\text{MP}_c$, rescaled to lie in $[0,1]$. Thus, we choose $\tau^{\text{old}} = 5\%$ as described above to match the effect of one standard deviation higher mobile penetration on average wages of small relative to large firms in the retail sector.
We first quantify the wage response and report our results in Appendix Figures \ref{fig:wage_change_firmsize} and \ref{fig:wage_change_skill}. In the model, average wages increase by 0.23\% in small firms and by 0.04\% in large firms, with the differential reflecting the direct effect of lower transaction costs on small-firm labor demand. Because small firms predominantly employ low-skill workers, the wage gains differ substantially across skill groups. Workers in skill groups 1 and 2 experience average wage increases of 0.12\% and 0.22\%, respectively, while workers in groups 3 and 4 see changes of 0.01\% and nearly zero, respectively.

To assess whether Pix increased competition, we compute changes in the wage-bill share of small firms and the Herfindahl–Hirschman Index (HHI). As shown in Figure~\ref{fig:share_hhi}, the wage-bill share of small firms rises by 0.09 percentage points, while HHI falls by 0.16\%. Both findings are consistent with Pix fostering greater competition in local labor markets.

Given that wage gains are concentrated below the median, one might expect wage inequality to decline after Pix. However, the transaction cost reduction is stronger in larger cities. If pre-Pix wages were already higher in these cities, overall inequality could remain unchanged or even increase. Appendix Figure \ref{fig:city_wage_num_firms} shows that average wages are higher in cities with more firms. %

To quantify the net effect, we decompose the change in wage inequality into within-city and between-city components using the variance of log wages. %
We find a net decline of 0.04\% in overall wage variance. As shown in Panel A of Figure \ref{fig:var_deco_Pix}, this result is driven entirely by a contraction in within-city inequality, which is partially offset by a rise in between-city inequality. %

\textbf{Uniform Adoption.}
We next compute the effects of uniform Pix adoption across all municipalities. In this scenario, Pix reduces transaction costs to zero for all small firms nationwide, corresponding to $\omega(\text{MP}_c) = 1$ for every city $c$. The impact of Pix in this case differs from the baseline simulation in two key dimensions.

First, the effects on labor market competition are stronger. The wage-bill share of small firms increases by 0.7 percentage points, leading to a 1.2\% decline in the Herfindahl–Hirschman Index (HHI). Consequently, the average wage increase is larger, with gains concentrated among low-skill workers: skill groups 1 and 2 experience average wage increases of 0.8\% and 1.5\%, respectively.

Second, the effect on wage inequality is more pronounced. As reported in Panel B of Figure \ref{fig:var_deco_Pix}, the variance of log wages declines by 0.6\%. In contrast to the baseline scenario, this reduction is explained by contributions from both within-city and between-city components. This indicates that uniform adoption intensifies within-city wage compression and also mitigates the spatial inequality that arose from the uneven initial uptake of Pix. %

We compute the aggregate wage and profit changes from uniform adoption under Pix. The initial implementation of Pix under the observed, uneven adoption increases the total wage bill by 0.04\% and total firm profits by 0.02\%. However, uniform adoption would generate larger gains: a 0.21\% increase in the total wage bill and a 0.03\% increase in profits, relative to a baseline scenario without Pix.

These wage-bill gains are highly concentrated among low-skill workers. The bottom skill groups (1 and 2) see their total wage-bill rise by only 0.19\% under the observed rollout, but by a substantial 1.3\% under uniform adoption. %

Overall, our quantitative analysis shows that, by reducing transaction costs, Pix increases competition in local labor markets and raises wages, particularly for low-skill workers. The policy also reduces overall wage inequality, primarily by compressing within-city wage inequality. We find that uniform adoption would amplify the wage and competition effects and reduce inequality more equally across regions and skill groups. %

\section{Conclusion}\label{sect:conclusion}

We study how instant payment adoption affects labor markets and wage inequality. Using Brazil's rollout of Pix, we find that wages in small, cash-intensive establishments increase relative to large establishments. Wage inequality falls, driven by gains below the median. The patterns we document are consistent with lower payment frictions raising labor demand among small firms, with larger effects on inequality where low-skill labor is scarce. These findings show that, in contrast to production technologies that prior work associates with rising inequality, financial technology can compress it by alleviating frictions in sectors that disproportionately employ low-skill labor.

Cash-intensive firms rely heavily on low-skill labor, so the wage gains we document concentrate in sectors that employ lower-wage workers. Our quantitative model implies that under the observed, uneven rollout, the within-city component of wage dispersion falls but the between-city component rises, partially offsetting the aggregate compression. Under uniform adoption, both components fall, so more even access amplifies the inequality-reducing potential of instant payments.

\newpage
\bibliography{literature}
\bibliographystyle{jf}
\index{Bibliography@\emph{Bibliography}}%

\setcounter{figure}{0}
\renewcommand{\thefigure}{\arabic{figure}}
\setcounter{table}{0}
\renewcommand{\thetable}{\arabic{table}}
\singlespacing
\begin{figure}[H]

\centering
\includegraphics[width=0.8\textwidth]{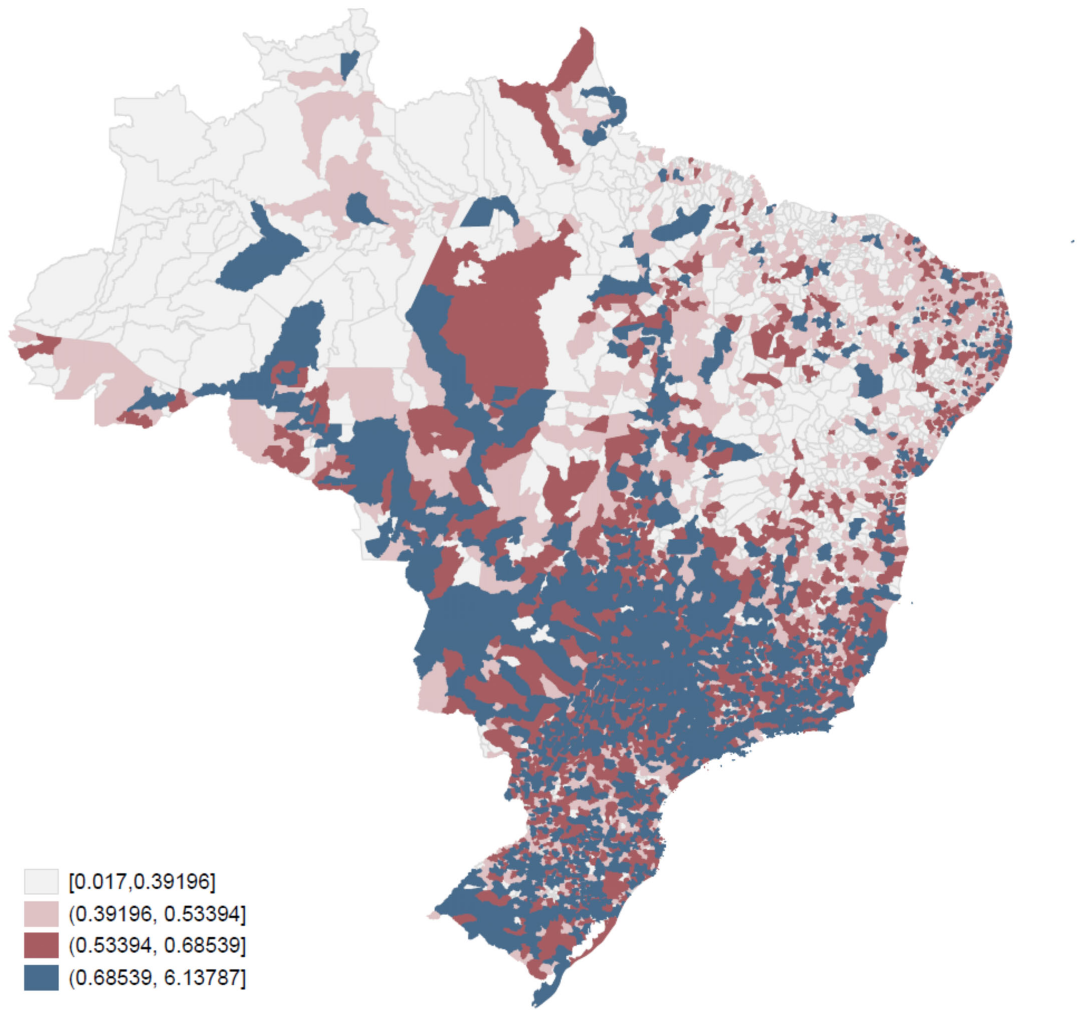}
\caption{Geographic Distribution of Mobile Technology Penetration}
\label{fig:mobile_penetration}

\justify

{\footnotesize
This figure displays the spatial variation in mobile technology penetration across Brazilian municipalities as of 2019. Mobile penetration is defined as the ratio of mobile devices with 3G or higher capability to the total population in each municipality. The map uses four color categories to represent different levels of penetration. These ranges correspond to quartiles of the mobile penetration distribution across municipalities. The data combine telecommunications infrastructure information from ANATEL (Agência Nacional de Telecomunicações) with population estimates from IBGE (Instituto Brasileiro de Geografia e Estatística).

}
\end{figure}

\newpage
\begin{figure}[H]
\begin{center}
    \begin{tabular}{c}
    \includegraphics[width=0.6\textwidth]{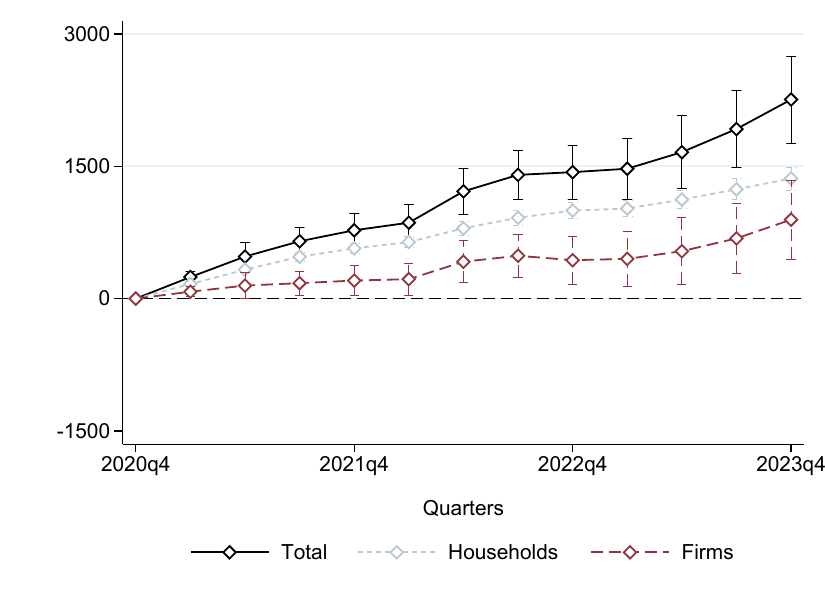} \\
    \multicolumn{1}{c}{Panel A. Pix Value Per Capita} \\[1cm]
    \includegraphics[width=0.6\textwidth]{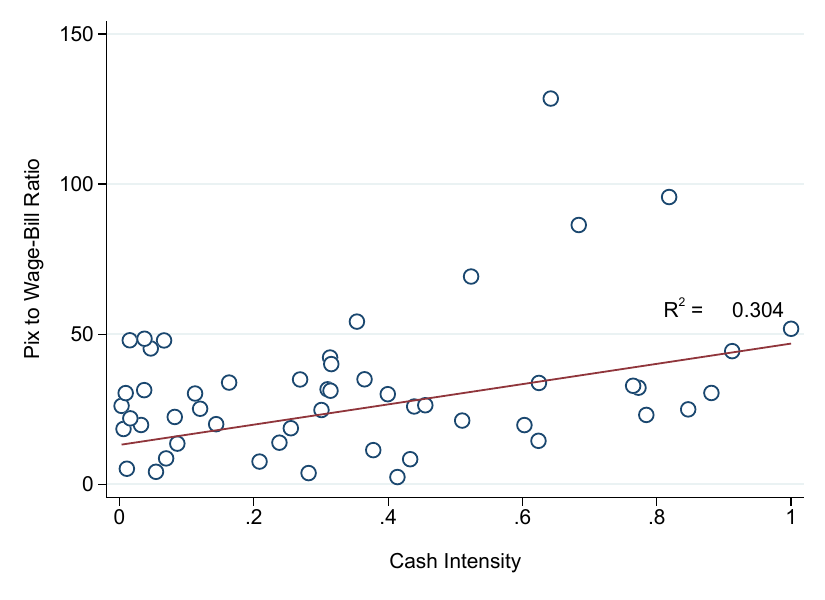} \\
    \multicolumn{1}{c}{Panel B. Pix Usage and Cash Intensity}
    \end{tabular}
\end{center}

\caption{Effect of Mobile Penetration and Cash Intensity on Pix Value}
\label{fig:pix_first_stage}

\justify
{\footnotesize
Panel (a) presents the estimated coefficient from municipality-level regressions examining the effect of standardized mobile penetration on Pix transaction value around the introduction of Pix in November 2020. The dependent variable is the value of Pix transactions per capita received by firms and households, measured using the municipality's 2019 population. Panel B plots the relationship between industry Pix usage divided by wage-bill as of 2022 and cash-intensity measured by the share of sales to households in the input-output table of 2019. Panel (a) includes municipality fixed effects, time fixed effects, and municipality characteristics interacted with time fixed effects. Municipality characteristics include deciles of GDP per capita, percentage of young population, and percentage of value added in agriculture. The dashed lines represent 95\% confidence intervals based on robust standard errors clustered at the municipality level. Mobile penetration is defined as the ratio of mobile devices with 3G or higher capability to the total municipal population, standardized to have unit variance and winsorized at the 1\% level. Data are from the Central Bank of Brazil.

}
\end{figure}

\newpage
\begin{figure}[H]

\centering
\includegraphics[width=0.8\textwidth]%
{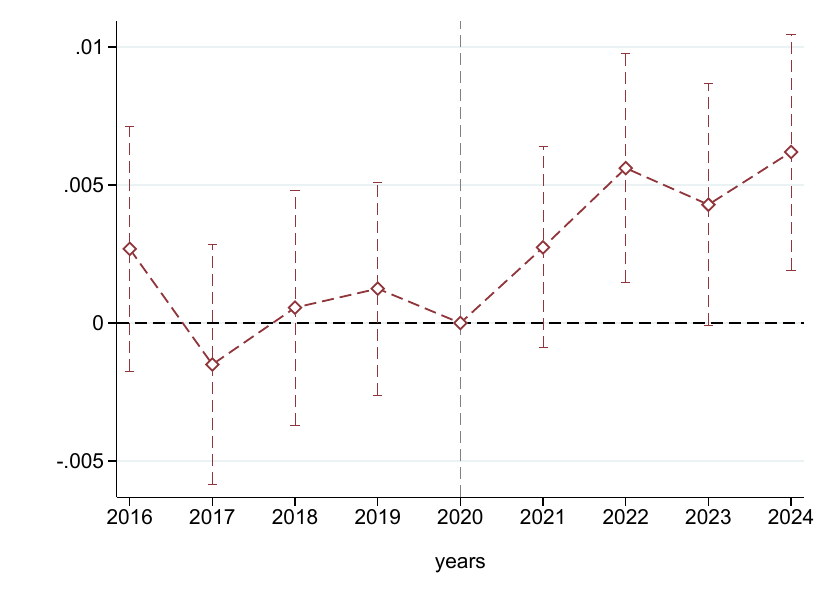} 
\caption{Effect of Mobile Penetration on Wages by Small Establishments}\label{fig:triple_diff_industry}

\justify
{\footnotesize
This figure plots the estimated yearly coefficients from a triple difference-in-differences specification that measures the differential effect of mobile penetration on the logarithm of average wages between small and large establishments around the introduction of Pix in late 2020. Small establishments are defined as those with fewer than 20 employees. The dependent variable is the logarithm of average wages. The regression specification includes fixed effects for municipality-by-industry-by-year, municipality-by-size-by-industry, and size-by-industry-by-year. The dashed lines represent 95\% confidence intervals constructed using robust standard errors clustered at the municipality level. A vertical dashed line marks the introduction of Pix in 2020. Mobile penetration is defined as the ratio of mobile devices with 3G or higher capability to total municipal population, standardized to have unit variance and winsorized at the 1\% level.
}
\end{figure}

\newpage
\begin{figure}[H]
\centering
\begin{tabular}{c}
    \includegraphics[width=0.6\textwidth]{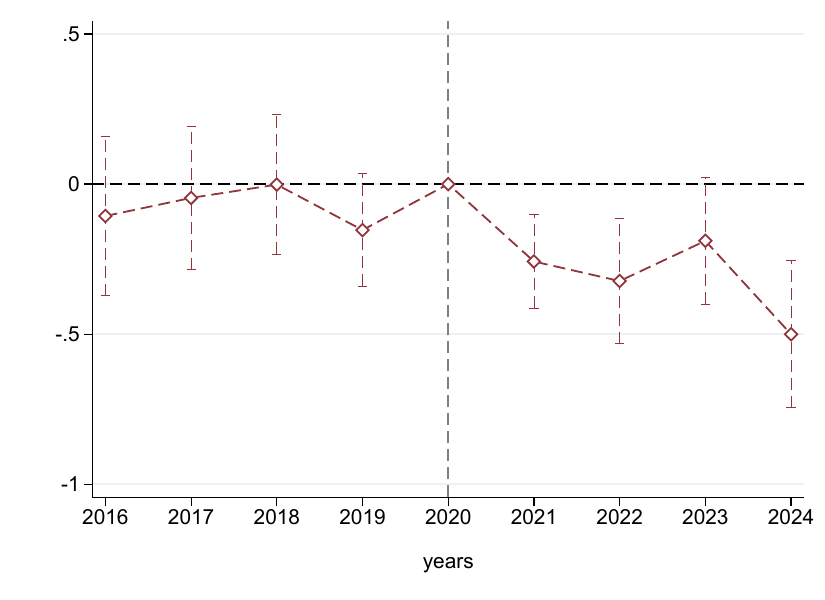}
\end{tabular}
\caption{Effect of Mobile Penetration on Wage Inequality}
\label{fig:gini}
\justify
{\footnotesize
This figure presents the estimated yearly coefficients from municipality-level difference-in-differences regressions examining the effect of standardized mobile penetration on the Gini coefficient of wages around the introduction of Pix in November 2020. The dependent variable is the Gini coefficient calculated from the distribution of individual worker wages within each municipality-year, multiplied by 100. The regression specification includes municipality fixed effects, region-by-year fixed effects, and municipality characteristics interacted with year fixed effects. Municipality characteristics include deciles of GDP per capita, percentage of young population, and percentage of value added in agriculture. The dashed lines represent 95\% confidence intervals constructed using robust standard errors clustered at the municipality level. A vertical dashed line marks the introduction of Pix in 2020. Mobile penetration is defined as the ratio of mobile devices with 3G or higher capability to total municipal population, standardized to have unit variance and winsorized at the 1\% level.
}

\end{figure}

\newpage
\begin{figure}[H]
\centering
\begin{tabular}{c}
    \includegraphics[width=0.58\textwidth]{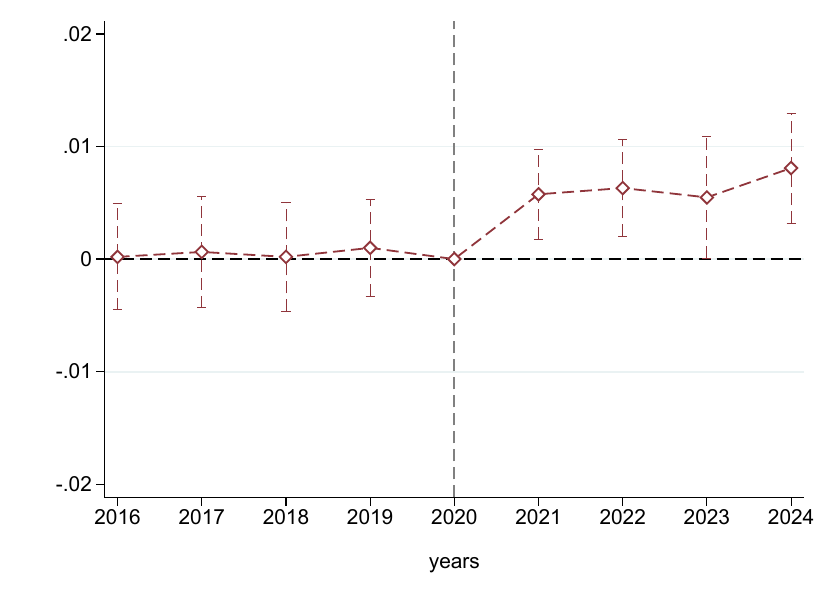} \\
    Panel A. Workers' Wages Bottom 50\% \\[0.8cm]
    \includegraphics[width=0.58\textwidth]{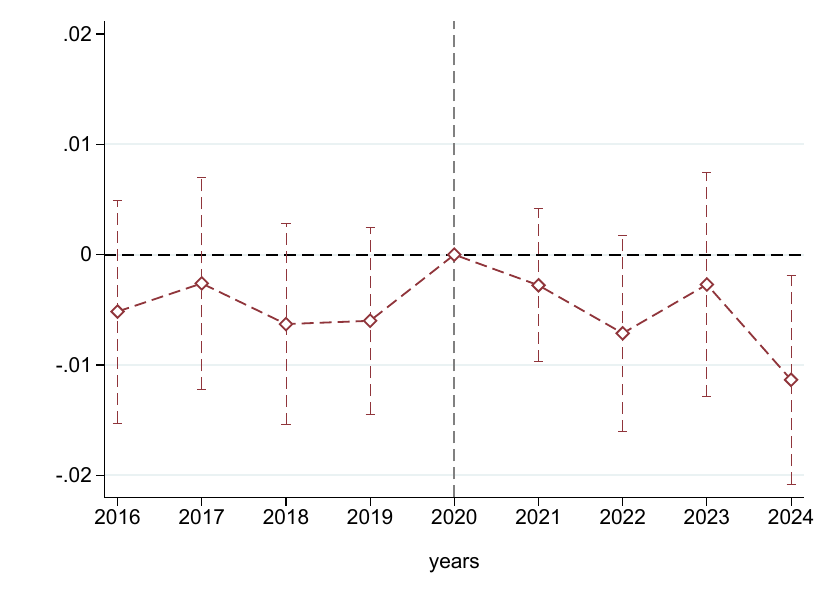} \\
    Panel B. Workers' Wages Top 50\%
\end{tabular}
\caption{Effect of Mobile Penetration on Wages Across the Wage Distribution}
\label{fig:wage_distribution}
\justify
{\footnotesize
This figure presents the estimated yearly coefficients from municipality-level difference-in-differences regressions examining the effect of standardized mobile penetration on the logarithm of average wages across different segments of the wage distribution around the introduction of Pix in November 2020. Panel A plots coefficient estimates for workers in the bottom 50\% of the wage distribution within each municipality-year. Panel B plots coefficient estimates for workers in the top 50\% of the wage distribution. The wage distribution is calculated separately for each municipality-year based on individual worker wages. The regression specification includes municipality fixed effects, region-by-year fixed effects, and municipality characteristics interacted with year fixed effects. Municipality characteristics include deciles of GDP per capita, percentage of young population, and percentage of value added in agriculture. The dashed lines represent 95\% confidence intervals constructed using robust standard errors clustered at the municipality level. A vertical dashed line marks the introduction of Pix in 2020. Mobile penetration is defined as the ratio of mobile devices with 3G or higher capability to total municipal population, standardized to have unit variance and winsorized at the 1\% level.
}
\end{figure}

\newpage
\begin{figure}[H]
\centering
\includegraphics[width=0.6\textwidth]{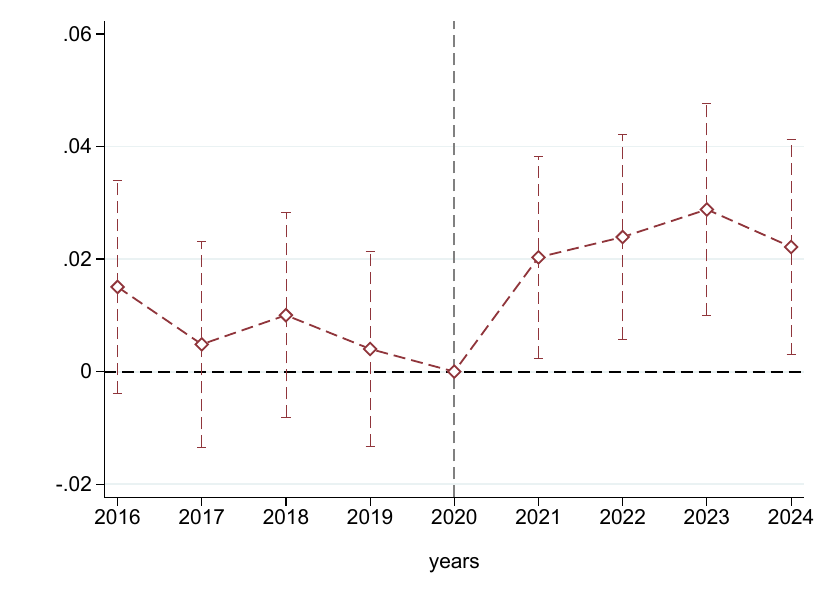}
 
\caption{Effect of Mobile Penetration on Employment Growth in Small Establishments}\label{fig:net_job_creation_small}
\justify
{\footnotesize
This figure presents the estimated yearly coefficients from a triple difference-in-differences specification examining the differential effect of standardized mobile penetration on the logarithm of the number of active jobs in small establishments relative to large establishments around the introduction of Pix in November 2020. Small establishments are defined as those with fewer than 5 employees. The dependent variable is the logarithm of active jobs. The regression specification includes  fixed effects for municipality-by-industry-by-year, municipality-by-size-by-industry, and size-by-industry-by-year. The dashed lines represent 95\% confidence intervals constructed using robust standard errors clustered at the municipality level. A vertical dashed line marks the introduction of Pix in 2020. Mobile penetration is defined as the ratio of mobile devices with 3G or higher capability to total municipal population, standardized to have unit variance and winsorized at the 1\% level.
}
\end{figure}

\newpage
\begin{figure}[H]
\begin{center}
    \begin{tabular}{c}    \includegraphics[width=0.6\textwidth]{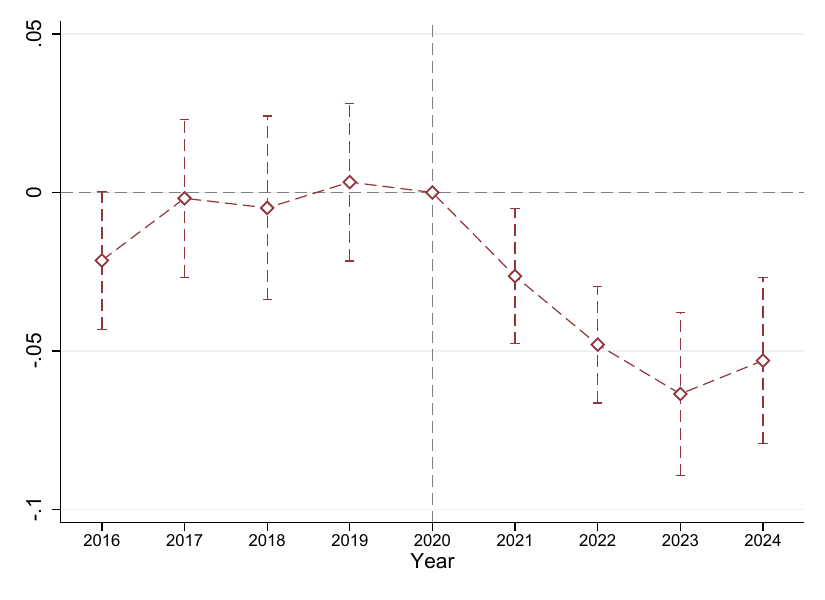} \\
    \multicolumn{1}{c}{Log Cash}
    \end{tabular}
\end{center}

\caption{Effect of Mobile Penetration on Bank Branch Cash Inventories}
\label{fig:cash_inventories}

\justify
{\footnotesize This figure presents the estimated coefficient from municipality-level regressions examining the effect of standardized mobile penetration on cash usage around the introduction of Pix in November 2020. The dependent variable is the logarithm of cash, measured as the stock of cash held by bank branches in each municipality and year. Our specification includes municipality fixed effects, time fixed effects, and municipality characteristics interacted with time fixed effects. Municipality characteristics include deciles of GDP per capita, percentage of young population, and percentage of value added in agriculture. The dashed lines represent 95\% confidence intervals constructed using robust standard errors. Mobile penetration is defined as the ratio of mobile devices with 3G or higher capability to total municipal population, standardized to have unit variance and winsorized at the 1\% level. Data are from the Central Bank of Brazil.

}
\end{figure}

\newpage
\begin{figure}[H]

\begin{center}
    \begin{tabular}{c}
    \includegraphics[width=0.6\textwidth]{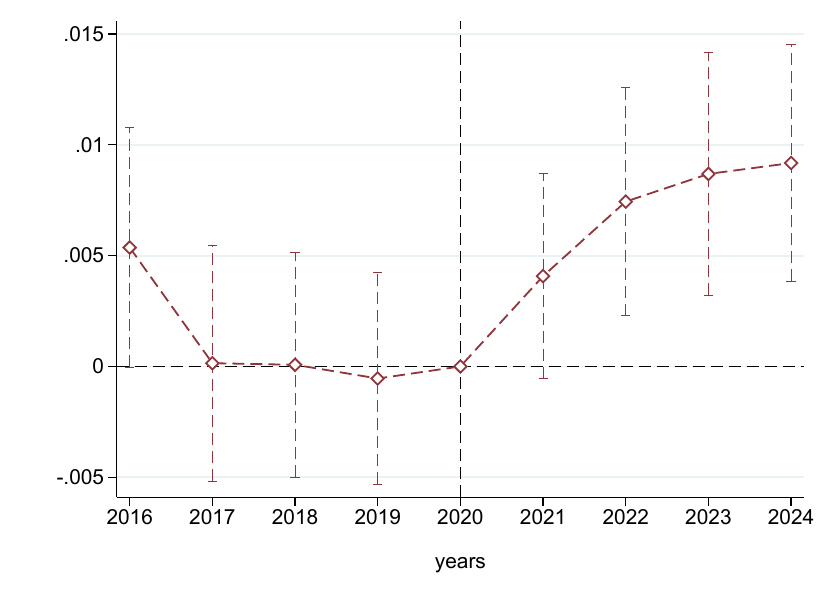} \\
    \multicolumn{1}{c}{Panel A. Above-Median Homicide Rate} \\[1cm]
    \includegraphics[width=0.6\textwidth]{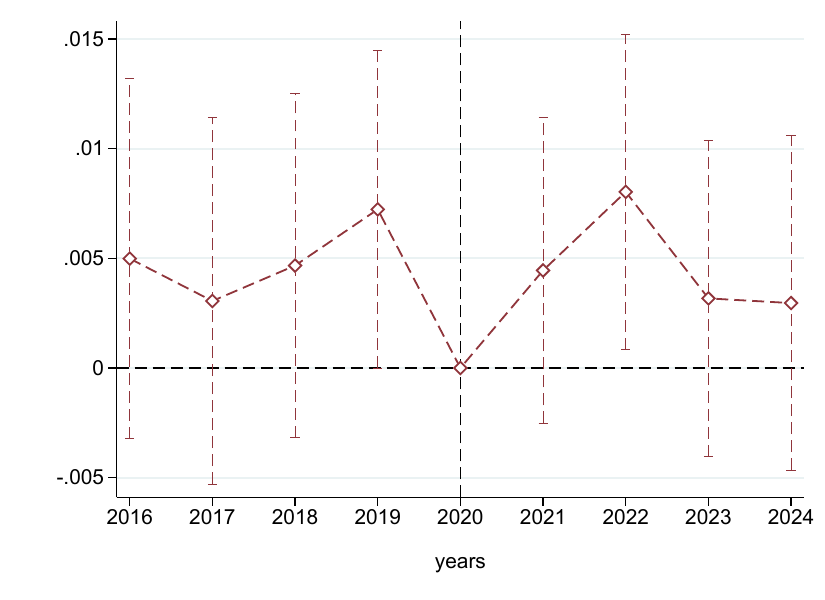} \\
    \multicolumn{1}{c}{Panel B. Below-Median Homicide Rate}
    \end{tabular}
\end{center}

\caption{Effect of Mobile Penetration on Wages by Small Size and Homicide Rate}
\label{fig:ddd_wage_homicide}

\justify
{\footnotesize
This figure plots the estimated yearly coefficients from a triple difference-in-differences specification that measures the differential effect of mobile penetration on the logarithm of average wages between small and large establishments around the introduction of Pix in late 2020, splitting the sample by pre-Pix municipal homicide rate (measured in 2019). Panel A restricts the sample to municipalities with above-median homicide rates and Panel B to below-median. Small establishments are defined as those with fewer than 20 employees. The regression specification includes fixed effects for municipality-by-size-by-industry, municipality-by-industry-by-year, and size-by-industry-by-year. Dashed lines represent 95\% confidence intervals from standard errors clustered at the municipality level. Mobile penetration is defined as the ratio of mobile devices with 3G or higher capability to total municipal population, standardized to have unit variance and winsorized at the 1\% level.
}
\end{figure}

\newpage
\begin{figure}[H]
\centering
\includegraphics[width=0.6\textwidth]{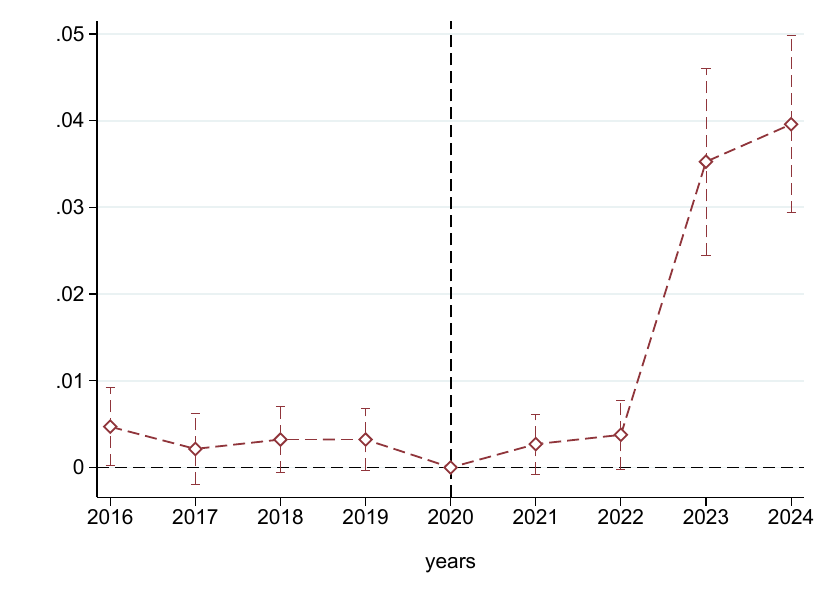}
\caption{Effect of Mobile Penetration on Wages by Establishment and Firm Size}
\label{fig:dddd_wage_small}
\justify
{\footnotesize
This figure plots the estimated yearly coefficients from a quadruple difference-in-differences specification that measures the differential effect of mobile penetration on the logarithm of average wages across establishment size and firm size around the introduction of Pix in late 2020. Small establishments and small firms are defined as those with fewer than 20 employees. The dependent variable is the logarithm of average monthly wages. The regression specification includes fixed effects for municipality-by-industry-by-year, municipality-by-firm-by-industry, and firm-by-industry-by-year. The dashed lines represent 95\% confidence intervals constructed using robust standard errors clustered at the municipality level. A vertical dashed line marks the introduction of Pix in 2020. Mobile penetration is defined as the ratio of mobile devices with 3G or higher capability to total municipal population, standardized to have unit variance and winsorized at the 1\% level.
}
\end{figure}

\newpage

\begin{figure}[H]
\begin{center}
\includegraphics[width=0.7\textwidth]{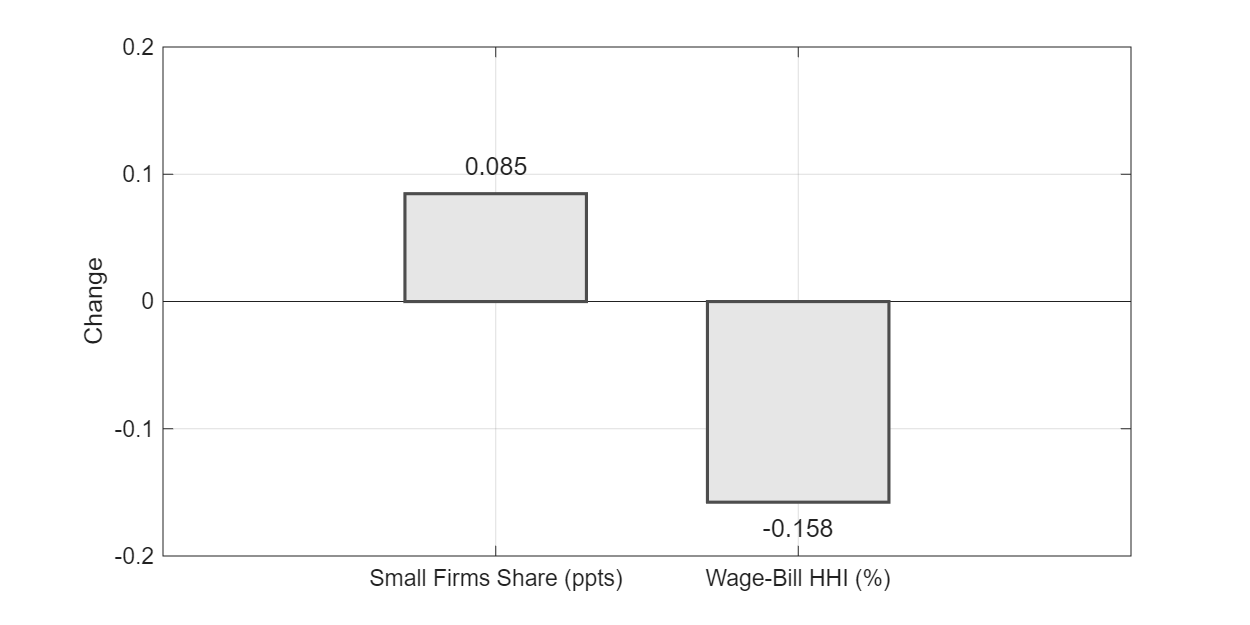}
\end{center}
\caption{Wage-Bill: Small Firm Share and HHI}
\label{fig:share_hhi}
\justify
{\footnotesize
This figure shows the change in the wage-bill share of small firms (left bar) and the percentage change in the wage-bill Herfindahl–Hirschman Index (HHI) for local labor markets (right bar) following the implementation of Pix.
}
\end{figure}

\newpage
\begin{figure}[H]
\centering

\begin{subfigure}[t]{\textwidth}
  \centering
  \includegraphics[width=0.75\textwidth]{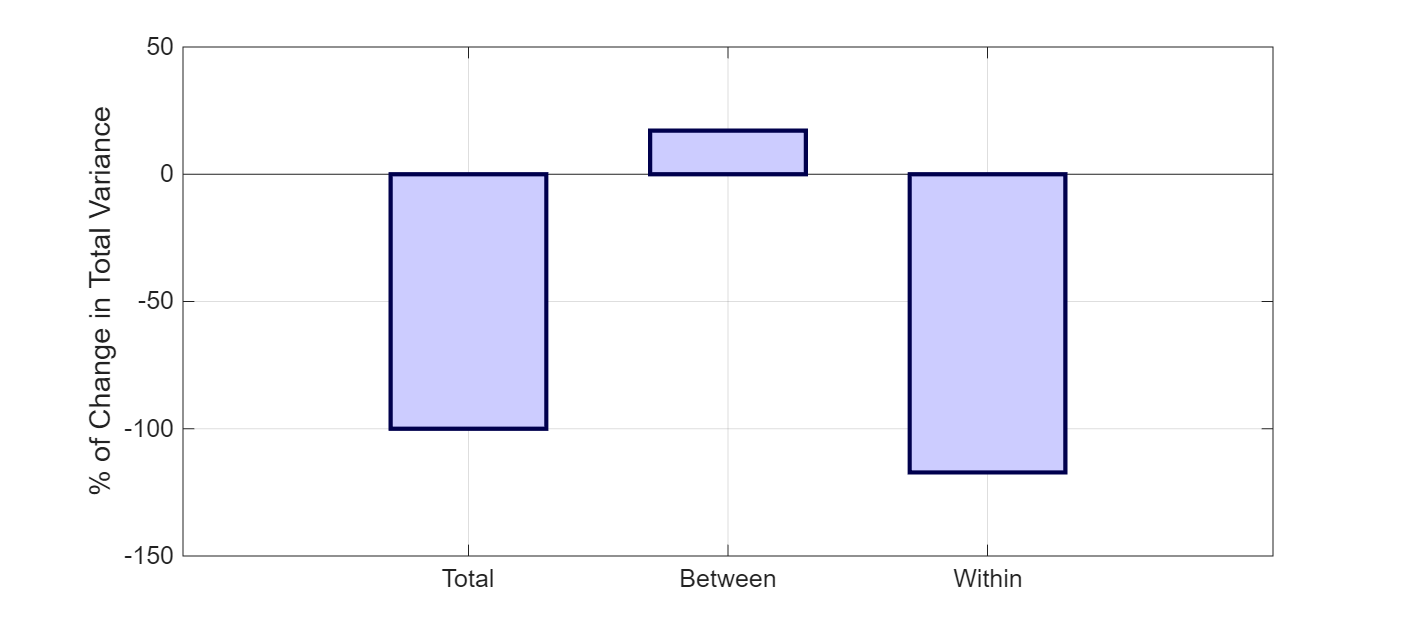} \\
  Panel A. Baseline Pix Adoption
  \label{fig:var_deco_Pix_A}
\end{subfigure}

\vspace{0.8em}

\begin{subfigure}[t]{\textwidth}
  \centering
  \includegraphics[width=0.75\textwidth]{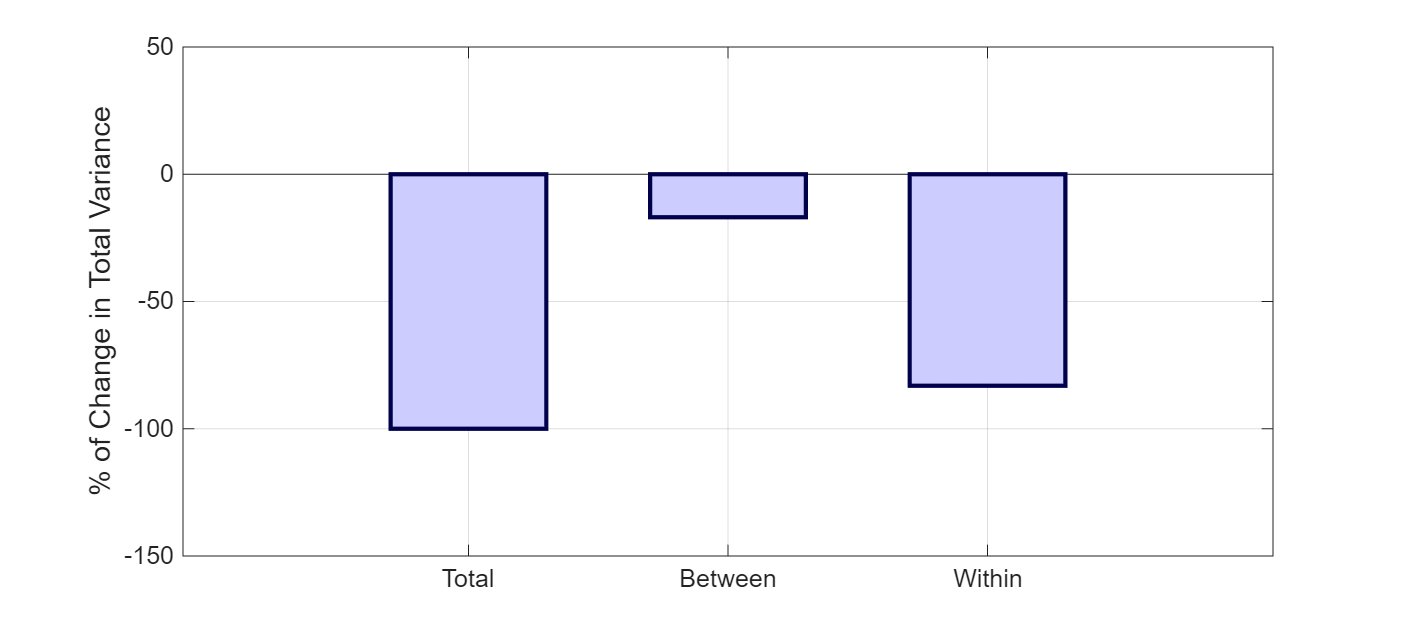} \\
  Panel B. Uniform Pix Adoption
  \label{fig:var_deco_Pix_B}
\end{subfigure}

\caption{Decomposition of Change in Variance of Wages}
\label{fig:var_deco_Pix}
\justify
{\footnotesize
This figure decomposes the decline in the total variance of log wages into contributions from changes within cities and between cities. Panel A reports results under observed Pix adoption. Panel B reports results under counterfactual uniform adoption, in which mobile penetration gaps are fully closed. In each panel, the left bar corresponds to the total decline in variance (100\%), the middle bar corresponds to the between-city component change divided by the change in total variance, and the right bar corresponds to the within-city component change divided by the total variance change.
}
\end{figure}

\newpage
\begin{table}[H]
\setlength{\tabcolsep}{0pt}
\caption{Summary Statistics}
\label{tab:summ_stat}
\justify
{\footnotesize 
This table presents summary statistics for average municipality-level characteristics in 2019. Wage values are measured in Brazilian reais (BRL). The number of bank branches is sourced from the ESTBAN database. Wages and skill composition measures are derived from the RAIS employer–employee dataset. Local GDP per capita, population, and the shares of agriculture and services in local value added are obtained from the Brazilian Institute of Geography and Statistics (IBGE). Mobile penetration (Share Mobile) is defined as the ratio of mobile devices with 3G or higher capability to the total municipal population. Low-skill workers are those without a college degree, while high-skill workers hold at least a college degree. The average wage is the mean monthly wage across all workers in the municipality.}
\vspace{0.5cm}
\noindent
{\small
\makebox[\textwidth]{
\begin{tabular*}{\textwidth}{@{\extracolsep{\fill}}l*{4}{c}@{}}
\toprule
 & (1) & (2) & (3) & (4) \\ 
 & Mean & Median & SD & N \\
\midrule
GDP per capita (in 1,000 BRL) & 24.5 & 18.2 & 25.6 & 5,570 \\ 
Share services & 0.35 &  0.32 & 0.14 & 5,570 \\ 
Share agriculture & 0.18 & 0.14 & 0.15 & 5,570 \\ 
Num. branches & 4 & 1 & 35 & 5,570 \\ 
Population & 36,460 & 11,065 & 206,519 & 5,570 \\ 
Share mobile & 0.54 & 0.53 & 0.20 & 5,570 \\ 
Share young & 0.43 & 0.43 & 0.03 & 5,570 \\
Avg. monthly wage & 2,020 & 1,946 & 443 & 5,570 \\
Low-skill avg. wage & 1,697 & 1,649 & 355 & 5,570 \\
High-skill avg. wage & 3,312 & 3,148 & 1,009 & 5,570 \\
Gini coefficient & 0.29 & 0.28 & 0.06 & 5,570 \\
Wage Top 50\% & 2,607.19 & 2,383.79 & 980.29 & 5,524 \\
Wage Bottom 50\% & 1,109.32 & 1,074.68 & 171.85 & 5,524 \\
\bottomrule
\end{tabular*}}
}
\end{table}

\newpage
\begin{table}[H]
\caption{Effect of Mobile Penetration on Average Wages in Small Establishments}
\label{tab:triple_diff_size}
\begin{singlespace}
{\footnotesize
This table presents triple difference-in-differences estimates examining the differential effect of standardized mobile penetration on average wages in small establishments relative to large establishments following the introduction of Pix in November 2020. The dependent variable is the logarithm of average monthly wages. Small establishments are defined as those with fewer than 20 employees. The key coefficient of interest, Mobile Penetration $\times$ Small $\times$ Post, measures how the effect of mobile penetration on wages varies with establishment size after Pix introduction. Panel A examines robustness across different fixed effects specifications. Panel B examines the effects separately for all workers, low-skill workers, and high-skill workers, using the specification from Panel A Column (3). Low-skill workers are those without a college degree and high-skill workers are those with at least a college degree. Standard errors clustered at the municipality level are reported in parentheses. Mobile penetration is defined as the ratio of mobile devices with 3G or higher capability to total municipal population, standardized to have unit variance and winsorized at the 1\% level. ***p$<$0.01, **p$<$0.05, *p$<$0.1.
}
\end{singlespace}

\vspace{0.5cm}
\textbf{Panel A: Alternative Fixed Effects Specifications}\\
\vspace{0.3cm}
{\small
\begin{tabular*}{\linewidth}{l@{\extracolsep{\fill}}*{3}{c}}
\toprule
& (1) & (2) & (3) \\ 
\midrule
Mobile Penetration $\times$ Small $\times$ Post & 0.003*** & 0.004*** & 0.004*** \\
 & (0.001) & (0.001) & (0.001) \\
\midrule
Observations & 1,529,548 & 1,529,548 & 1,529,548 \\
\midrule
Municipality $\times$ Size $\times$ Industry & $\checkmark$ & $\checkmark$ & $\checkmark$ \\
Municipality $\times$ Industry $\times$ Year & \xmark & $\checkmark$ & $\checkmark$ \\
Size $\times$ Year & $\checkmark$ & $\checkmark$ & \xmark \\
Industry $\times$ Year & $\checkmark$ & \xmark & \xmark \\
Municipality $\times$ Year & $\checkmark$ & \xmark & \xmark \\
Size $\times$ Industry $\times$ Year & \xmark & \xmark & $\checkmark$ \\
\bottomrule
\end{tabular*}
}

\textbf{Panel B: Effects by Worker Skill Level}\\
\vspace{0.3cm}
{\small
\begin{tabular*}{\linewidth}{l@{\extracolsep{\fill}}*{3}{c}}
\toprule
 & (1) & (2) & (3) \\ 
 \midrule
  & All & Low-Skill & High-Skill \\
\midrule
Mobile Penetration $\times$ Small $\times$ Post & 0.004*** & 0.004*** & 0.005* \\
 & (0.001) & (0.001) & (0.003) \\
\midrule
Observations & 1,586,407 & 1,569,418 & 1,005,200 \\ 
\midrule
Municipality $\times$ Size $\times$ Industry & \checkmark & \checkmark & \checkmark \\
Municipality $\times$ Industry $\times$ Year & \checkmark & \checkmark & \checkmark \\
Size $\times$ Industry $\times$ Year & \checkmark & \checkmark & \checkmark \\
\bottomrule
\end{tabular*}
}
\end{table}

\newpage
\begin{table}[H]
\caption{Effect of Mobile Penetration on Average Wages in Small Establishments: Alternative Size Definitions}
\label{tab:triple_diff_size_robust}
\begin{singlespace}
{\footnotesize
This table presents triple difference-in-differences estimates examining the differential effect of standardized mobile penetration on average wages in small establishments relative to large establishments following the introduction of Pix in November 2020, with robustness checks across alternative definitions of small establishments. The dependent variable is the logarithm of average monthly wages. The key coefficient of interest, Mobile Penetration $\times$ Small $\times$ Post, measures how the effect of mobile penetration on wages varies with establishment size after Pix introduction. Columns vary the threshold for small establishments: fewer than 5, 10, 20, and 50 employees. All specifications include municipality-by-size-by-industry fixed effects, municipality-by-industry-by-year fixed effects, and size-by-industry-by-year fixed effects. Standard errors clustered at the municipality level are reported in parentheses. Mobile penetration is defined as the ratio of mobile devices with 3G or higher capability to total municipal population, standardized to have unit variance and winsorized at the 1\% level. ***p$<$0.01, **p$<$0.05, *p$<$0.1.
}
\end{singlespace}
\vspace{0.5cm}
{\small
\begin{tabular*}{\linewidth}{l@{\extracolsep{\fill}}*{4}{c}}
\toprule
& (1) & (2) & (3) & (4) \\ 
\midrule
 & $N < 5$ & $N < 10$ & $N < 20$ & $N < 50$ \\
\midrule
Mobile Penetration $\times$ Small $\times$ Post & 0.004*** & 0.004*** & 0.004*** & 0.001 \\
 & (0.001) & (0.001) & (0.001) & (0.002) \\
\midrule
Observations & 1,529,548 & 1,529,548 & 1,529,548 & 1,529,548 \\ 
\midrule
Municipality $\times$ Size $\times$ Industry & \checkmark & \checkmark & \checkmark & \checkmark \\
Municipality $\times$ Industry $\times$ Year & \checkmark & \checkmark & \checkmark & \checkmark \\
Size $\times$ Industry $\times$ Year & \checkmark & \checkmark & \checkmark & \checkmark \\
\bottomrule
\end{tabular*}
}
\end{table}

\newpage
\begin{table}[H]
\caption{\label{tab:triple_diff_size_new} Effect of Mobile Penetration on Average Wages in Small Establishments by Industry}
\begin{singlespace}
{\footnotesize
This table presents triple difference-in-differences estimates examining the differential effect of standardized mobile penetration on average wages in small establishments relative to large establishments across different industries following the introduction of Pix in November 2020. The dependent variable is the logarithm of average monthly wages. Small establishments are defined as those with fewer than 20 employees. The key coefficient of interest, Mobile Penetration $\times$ Small $\times$ Post, measures how the effect of mobile penetration on wages varies with establishment size after Pix introduction, estimated separately for each industry. Panel A presents results disaggregated by industry. Column (1) presents the pooled estimate across all industries, replicating Column (3) from Panel A of Table \ref{tab:triple_diff_size}. Columns (2)-(5) present results for retail, wholesale, manufacturing, and all other industries. Panel B presents results for service industry subsectors: restaurants, hotels, household services, and other services. All specifications include municipality-by-size-by-industry fixed effects, municipality-by-industry-by-year fixed effects, and size-by-industry-by-year fixed effects. Standard errors clustered at the municipality level are reported in parentheses. Mobile penetration is defined as the ratio of mobile devices with 3G or higher capability to total municipal population, standardized to have unit variance and winsorized at the 1\% level. ***p$<$0.01, **p$<$0.05, *p$<$0.1.
}
\end{singlespace}

\vspace{0.5cm}
\textbf{Panel A: All Industries, Retail, Wholesale, Manufacturing, and Other}\\
\vspace{0.3cm}
{\small
\begin{tabular*}{\linewidth}{l@{\extracolsep{\fill}}*{5}{c}}
\toprule
 & (1) & (2) & (3) & (4) & (5) \\
\midrule
 & All & Retail & Wholesale & Manuf. & Other Ind. \\
\midrule
Mobile Penetration $\times$ Small $\times$ Post & 0.004*** & 0.008*** & -0.009 & -0.000 & 0.001 \\
 & (0.001) & (0.002) & (0.006) & (0.002) & (0.003) \\
\midrule
Observations & 1,529,548 & 122,242 & 71,142 & 476,797 & 289,488 \\ 
\midrule
Municipality $\times$ Size $\times$ Industry & \checkmark & \checkmark & \checkmark & \checkmark & \checkmark \\
Municipality $\times$ Industry $\times$ Year & \checkmark & \checkmark & \checkmark & \checkmark & \checkmark \\
Size $\times$ Industry $\times$ Year & \checkmark & \checkmark & \checkmark & \checkmark & \checkmark \\
\bottomrule
\end{tabular*}
}

\vspace{0.8cm}
\textbf{Panel B: Service Industry Subsectors}\\
\vspace{0.3cm}
{\small
\begin{tabular*}{\linewidth}{l@{\extracolsep{\fill}}*{4}{c}}
\toprule
 & (1) & (2) & (3) & (4) \\
\midrule
 & Restaurants & Hotel & House Serv. & Other Serv. \\
\midrule
Mobile Penetration $\times$ Small $\times$ Post & 0.002 & -0.002 & 0.001 & 0.014*** \\
 & (0.004) & (0.005) & (0.011) & (0.003) \\
\midrule
Observations & 64,406 & 41,582 & 45,554 & 456,773 \\ 
\midrule
Municipality $\times$ Size $\times$ Industry & \checkmark & \checkmark & \checkmark & \checkmark \\
Municipality $\times$ Industry $\times$ Year & \checkmark & \checkmark & \checkmark & \checkmark \\
Size $\times$ Industry $\times$ Year & \checkmark & \checkmark & \checkmark & \checkmark \\
\bottomrule
\end{tabular*}
}
\end{table}
\newpage

\newpage
\begin{table}[H]
\caption{\label{tab:gini_scarcity} Effect of Mobile Penetration on Wages Across the Wage Distribution}
\begin{singlespace}
{\footnotesize
This table presents difference-in-differences estimates examining the effect of standardized mobile penetration on wage inequality and wages across the wage distribution following the introduction of Pix in November 2020. The dependent variable in Column (1) is the Gini coefficient computed from the distribution of individual worker wages within each municipality-year, multiplied by 100. Columns (2) and (3) use the log of average wages for workers in the bottom and top 50\% of that within-municipality-year wage distribution. Panel A presents baseline estimates. Panel B interacts the treatment with an indicator for tight low-skill labor markets, defined as municipalities where the ratio of low-skill workers to total workers is below the population-weighted median. Mobile Penetration × Post measures the average effect, and Mobile Penetration × Post × Tight measures the differential effect in tight markets. All specifications include municipality fixed effects, region-by-year fixed effects, and municipality characteristics interacted with year fixed effects. Panel B additionally includes Tight-by-year fixed effects. Standard errors clustered at the municipality level are reported in parentheses. Mobile penetration is standardized to unit variance and winsorized at the 1\% level. ***p<0.01, **p<0.05, *p<0.1.
}
\end{singlespace}

\vspace{0.5cm}
\textbf{Panel A: Average Effect Across Wage Distribution} \\
\vspace{0.3cm}
{\footnotesize
\begin{tabular*}{\linewidth}{l@{\extracolsep{\fill}}*{3}{c}}
\toprule
 & (1) & (2) & (3) \\
\midrule
 & Gini & Bottom 50\% & Top 50\% \\
\midrule
Mobile Penetration $\times$ Post & -0.270** & 0.006** & -0.002 \\
 & (0.085) & (0.002) & (0.004) \\
\midrule
Observations & 23,331 & 23,331 & 23,331 \\ 
\midrule
Municipality & $\checkmark$ & $\checkmark$ & $\checkmark$ \\
Region $\times$ Year & $\checkmark$ & $\checkmark$ & $\checkmark$ \\
Controls $\times$ Year & $\checkmark$ & $\checkmark$ & $\checkmark$ \\
\bottomrule
\end{tabular*}
}

\vspace{0.8cm}
\textbf{Panel B: Heterogeneous Effects by Low-Skill Labor Market Scarcity} \\
\vspace{0.3cm}
{\footnotesize
\begin{tabular*}{\linewidth}{l@{\extracolsep{\fill}}*{3}{c}}
\toprule
 & (1) & (2) & (3) \\
\midrule
 & Gini & Bottom 50\% & Top 50\% \\
\midrule
Mobile Penetration $\times$ Post & -0.142 & 0.005*** & 0.001 \\
 & (0.089) & (0.002) & (0.004) \\
Mobile Penetration $\times$ Post $\times$ Tight & -0.323*** & 0.003 & -0.007 \\
 & (0.122) & (0.002) & (0.005) \\
\midrule
Observations & 23,331 & 23,331 & 23,331 \\ 
\midrule
Municipality & $\checkmark$ & $\checkmark$ & $\checkmark$ \\
Region $\times$ Year & $\checkmark$ & $\checkmark$ & $\checkmark$ \\
Controls $\times$ Year & $\checkmark$ & $\checkmark$ & $\checkmark$ \\
Tight $\times$ Time & $\checkmark$ & $\checkmark$ & $\checkmark$ \\
\bottomrule
\end{tabular*}

}
\end{table}

\newpage

\begin{table}[H]
\caption{Effect of Mobile Penetration on Average Wages in Small Establishments: Robustness to COVID-19}
\label{tab:robustness_covid}
\begin{singlespace}
{\footnotesize
This table presents triple difference-in-differences estimates examining the differential effect of standardized mobile penetration on average wages in small establishments relative to large establishments following the introduction of Pix in November 2020, with additional robustness controls for the effect of COVID-19. The dependent variable is the logarithm of average monthly wages. The key coefficient of interest, Mobile Penetration $\times$ Small $\times$ Post, measures how the effect of mobile penetration on wages varies with establishment size after Pix introduction. Small establishments are defined as those with fewer than 20 employees. Column (1) includes controls for government transfers per capita (Aux\'{i}lio Emergencial) interacted with small and post-period indicators. Column (2) controls for the federal Emergency Employment and Income Preservation Program (BEm), measured as beneficiaries per capita and interacted with small and post-period indicators. Column (3) includes controls for COVID-19 isolation policies using municipal-level isolation policy indicators. Column (4) controls for COVID-19 severity by including cumulative COVID cases per capita at the municipality level, interacted with small and post-period indicators. All specifications include municipality-by-size-by-industry fixed effects, municipality-by-industry-by-year fixed effects, and size-by-industry-by-year fixed effects. Standard errors clustered at the municipality level are reported in parentheses. Mobile penetration is defined as the ratio of mobile devices with 3G or higher capability to total municipal population, standardized to have unit variance and winsorized at the 1\% level. ***p$<$0.01, **p$<$0.05, *p$<$0.1.

}
\end{singlespace}

\vspace{0.5cm}
{\small
\begin{tabular*}{\linewidth}{l@{\extracolsep{\fill}}cccc}
\toprule
& (1) & (2) & (3) & (4) \\
\midrule
Mobile Penetration $\times$ Small $\times$ Post & 0.005*** & 0.005*** & 0.004*** & 0.004*** \\
 & (0.001) & (0.002) & (0.001) & (0.001) \\
Transfers $\times$ Small $\times$ Post & 0.004*** &  &  &  \\
 & (0.001) &  &  &  \\
BEm per capita $\times$ Small $\times$ Post &  & 0.008 &  &  \\
 &  & (0.019) &  &  \\
Isolation $\times$ Small $\times$ Post &  &  & -0.008 &  \\
 &  &  & (0.009) &  \\
COVID cases per capita $\times$ Small $\times$ Post &  &  &  & -0.001 \\
 &  &  &  & (0.001) \\
\midrule
Observations & 1,529,548 & 1,529,548 & 1,529,548 & 1,529,548 \\
\midrule
Municipality $\times$ Size $\times$ Industry & \checkmark & \checkmark & \checkmark & \checkmark \\
Municipality $\times$ Industry $\times$ Year & \checkmark & \checkmark & \checkmark & \checkmark \\
Size $\times$ Industry $\times$ Year & \checkmark & \checkmark & \checkmark & \checkmark \\
\bottomrule
\end{tabular*}
}
\end{table}

\newpage
\begin{table}[H]
\caption{Effect of Mobile Penetration on Active Jobs in Small Establishments} \label{tab:triple_diff_active_small}
\begin{singlespace}

{\footnotesize
This table presents triple difference-in-differences estimates examining the differential effect of standardized mobile penetration on the logarithm of the number of active jobs in small establishments relative to large establishments following the introduction of Pix in November 2020. The dependent variable is the logarithm of the number of active jobs. The key coefficient of interest, Mobile Penetration $\times$ Small $\times$ Post, measures how the effect of mobile penetration on active jobs varies with establishment size after Pix introduction, estimated separately for alternative definitions of small establishments. Small establishments are defined with varying thresholds across columns: fewer than 5, 10, and 20 employees. All specifications include municipality-by-size-by-industry fixed effects, municipality-by-industry-by-year fixed effects, and size-by-industry-by-year fixed effects. Standard errors clustered at the municipality level are reported in parentheses. Mobile penetration is defined as the ratio of mobile devices with 3G or higher capability to total municipal population, standardized to have unit variance and winsorized at the 1\% level. ***p$<$0.01, **p$<$0.05, *p$<$0.1.
}
\end{singlespace}
\vspace{0.5cm}
{\small
\begin{tabular*}{\linewidth}{l@{\extracolsep{\fill}}*{3}{c}}
\toprule
 & (1) & (2) & (3) \\
\midrule
 & $N < 5$ & $N < 10$ & $N < 20$ \\
\midrule
Mobile Penetration $\times$ Small $\times$ Post & 0.017*** & 0.007 & -0.005 \\
 & (0.006) & (0.006) & (0.007) \\
\midrule
Observations & 122,199 & 122,199 & 122,199 \\ 
\midrule
Municipality $\times$ Size $\times$ Industry & $\checkmark$ & $\checkmark$ & $\checkmark$ \\
Municipality $\times$ Industry $\times$ Year & $\checkmark$ & $\checkmark$ & $\checkmark$ \\
Size $\times$ Industry $\times$ Year & $\checkmark$ & $\checkmark$ & $\checkmark$ \\
\bottomrule
\end{tabular*}
}
\end{table}

\newpage
\begin{table}[H]
\caption{Effect of Mobile Penetration on Entry of Small Establishments by Industry}
\label{tab:small_entry}
\begin{singlespace}
{\footnotesize
This table presents difference-in-differences estimates examining the effect of standardized mobile penetration on the number of small firm entrants per 1,000 population following the introduction of Pix in November 2020, estimated separately for small retail firms and small manufacturing firms. Small firms are defined as those with annual sales between USD 70,000 and USD 970,000. The dependent variable is the number of firm entrants per 1,000 population. The key coefficient of interest, Mobile Penetration $\times$ Post, measures how mobile penetration affects small firm entry after Pix introduction. Column (1) presents estimates for small retail firms, while Column (2) presents estimates for small manufacturing firms. All specifications include municipality fixed effects, region-by-year fixed effects, and municipality characteristics interacted with year fixed effects. Standard errors clustered at the municipality level are reported in parentheses. Mobile penetration is defined as the ratio of mobile devices with 3G or higher capability to total municipal population, standardized to have unit variance and winsorized at the 1\% level. ***p$<$0.01, **p$<$0.05, *p$<$0.1.
}
\end{singlespace}
\vspace{0.5cm}
{\small
\begin{tabular*}{\linewidth}{l@{\extracolsep{\fill}}*{2}{c}}
\toprule
 & (1) & (2) \\ 
\midrule
 & Small Retail & Small Manufacturing \\
\midrule
Mobile Penetration $\times$ Post & 0.004*** & -0.001 \\
 & (0.001) & (0.001) \\
\midrule
Observations & 18,445 & 18,445 \\
\midrule
Municipality & $\checkmark$ & $\checkmark$ \\
Region $\times$ Year & $\checkmark$ & $\checkmark$ \\
Controls $\times$ Year & $\checkmark$ & $\checkmark$ \\
\bottomrule
\end{tabular*}
}
\end{table}

\newpage

\begin{table}[H]
\caption{Effect of Mobile Penetration on Wages in Small Establishments by Industry Cash-Intensity} \label{tab:triple_diff_cash_intensity}
\begin{singlespace}

{\footnotesize
This table presents triple difference-in-differences estimates examining the differential effect of standardized mobile penetration on average wages in small establishments relative to large establishments across different groups of industries following the introduction of Pix in November 2020. We define an industry-specific measure of cash-intensity equal to the share of sales to households. Low-cash industries are in the bottom tercile of the cash-intensity distribution, medium-cash are in the middle tercile, and high-cash are at the top tercile. The dependent variable is the logarithm of average monthly wages. Small establishments are defined as those with fewer than 20 employees. The key coefficient of interest, Mobile Penetration $\times$ Small $\times$ Post, measures how the effect of mobile penetration on wages varies with establishment size after Pix introduction, estimated separately for each group of industries. Column (1) presents the estimate for low-cash industries, column (2) for medium-cash, and column (3) for high-cash. All specifications include municipality-by-size-by-industry fixed effects, municipality-by-industry-by-year fixed effects, and size-by-industry-by-year fixed effects. Standard errors clustered at the municipality level are reported in parentheses. Mobile penetration is defined as the ratio of mobile devices with 3G or higher capability to total municipal population, standardized to have unit variance and winsorized at the 1\% level. ***p$<$0.01, **p$<$0.05, *p$<$0.1.
}
\end{singlespace}
\vspace{0.5cm}
{\small
\begin{tabular*}{\linewidth}{l@{\extracolsep{\fill}}*{3}{c}}
\toprule
 & (1) & (2) & (3) \\
\midrule
 & Low-Cash & Medium-Cash & High-Cash \\
\midrule
Mobile Penetration $\times$ Small $\times$ Post & 0.001 & 0.006** & 0.008*** \\
 & (0.003) & (0.003) & (0.002) \\
\midrule
Observations & 438,395 & 474,332 & 374,534 \\ 
\midrule
Municipality $\times$ Size $\times$ Industry & $\checkmark$ & $\checkmark$ & $\checkmark$ \\
Municipality $\times$ Industry $\times$ Year & $\checkmark$ & $\checkmark$ & $\checkmark$ \\
Size $\times$ Industry $\times$ Year & $\checkmark$ & $\checkmark$ & $\checkmark$ \\
\bottomrule
\end{tabular*}
}
\end{table}

\newpage

\begin{table}[H]
\setlength{\tabcolsep}{0pt}
\caption{\label{tab:ddd_wage_homicide} Effect of Mobile Penetration on Average Wages in Small Establishments by Municipal Homicide Rate}

\justify
{\footnotesize
This table presents triple difference-in-differences estimates examining the differential effect of standardized mobile penetration on average wages in small establishments relative to large establishments following the introduction of Pix in November 2020, separately for municipalities with above-median and below-median homicide rates. The dependent variable is the logarithm of average monthly wages. Small establishments are defined as those with fewer than 20 employees. The key coefficient of interest, Mobile Penetration $\times$ Small $\times$ Post, measures how the effect of mobile penetration on wages varies with establishment size after Pix introduction. Panel A restricts the sample to municipalities with above-median homicide rates and Panel B to municipalities with below-median homicide rates. Both panels examine robustness across alternative fixed effects specifications. Standard errors clustered at the municipality level are reported in parentheses. Mobile penetration is defined as the ratio of mobile devices with 3G or higher capability to total municipal population, standardized to have unit variance and winsorized at the 1\% level. ***p$<$0.01, **p$<$0.05, *p$<$0.1.
}

\vspace{0.5cm}

\noindent
{\small
\makebox[\textwidth]{
\begin{tabular*}{\textwidth}{@{\extracolsep{\fill}}l*{3}{c}@{}}

\multicolumn{4}{l}{\textbf{Panel A: Above-Median Homicide Rate}} \\
\toprule
& (1) & (2) & (3) \\
\midrule
Mobile Penetration $\times$ Small $\times$ Post & 0.005*** & 0.005*** & 0.006*** \\
 & (0.002) & (0.002) & (0.002) \\
\midrule
Observations & 880,123 & 880,123 & 880,123 \\
\midrule
Municipality $\times$ Size $\times$ Industry & $\checkmark$ & $\checkmark$ & $\checkmark$ \\
Municipality $\times$ Industry $\times$ Year & \xmark & $\checkmark$ & $\checkmark$ \\
Size $\times$ Year & $\checkmark$ & $\checkmark$ & \xmark \\
Industry $\times$ Year & $\checkmark$ & \xmark & \xmark \\
Municipality $\times$ Year & $\checkmark$ & \xmark & \xmark \\
Size $\times$ Industry $\times$ Year & \xmark & \xmark & $\checkmark$\\
\midrule
\\
\multicolumn{4}{l}{\textbf{Panel B: Below-Median Homicide Rate}} \\
\midrule
& (1) & (2) & (3) \\
\midrule
Mobile Penetration $\times$ Small $\times$ Post & $-$0.002 & $-$0.002 & 0.001 \\
 & (0.002) & (0.002) & (0.002) \\
\midrule
Observations & 495,065 & 495,065 & 495,065 \\
\midrule
Municipality $\times$ Size $\times$ Industry & $\checkmark$ & $\checkmark$ & $\checkmark$ \\
Municipality $\times$ Industry $\times$ Year & \xmark & $\checkmark$ & $\checkmark$ \\
Size $\times$ Year & $\checkmark$ & $\checkmark$ & \xmark \\
Industry $\times$ Year & $\checkmark$ & \xmark & \xmark \\
Municipality $\times$ Year & $\checkmark$ & \xmark & \xmark \\
Size $\times$ Industry $\times$ Year & \xmark & \xmark & $\checkmark$\\
\bottomrule
\end{tabular*}}
}
\end{table}

\clearpage
\setcounter{page}{1}
\begin{appendices}

\setcounter{table}{0}
\renewcommand{\thetable}{A.\arabic{table}}
\setcounter{figure}{0}
\renewcommand{\thefigure}{A.\arabic{figure}}

\begin{center}
  \large\textbf{Online Appendix}  
\end{center}

{
  \hypersetup{linkcolor=black}
}

\clearpage
\setcounter{subsection}{0}
\renewcommand{\thesubsection}{A\arabic{subsection}}
\setcounter{figure}{0}
\renewcommand{\thefigure}{A.\arabic{figure}}

\section{Conceptual Framework}

\subsection{Derivation of the labor supply elasticity}
\label{app:elasticity-derivation}

This appendix derives equation~\ref{eq:elasticity} under Cournot competition.
Write the sector‑wide employment as $N_k = \sum_j n_{jk}$. From the nested logit
structure, aggregate employment and the wage index are linked by
$1-\pi_k = W_k^{\theta_k}/(w_{0k}^{\theta_k}+W_k^{\theta_k})$ and
$N_k = \lambda_k (1-\pi_k)$. Inverting gives the sector‑level inverse supply
curve
\begin{equation*}
W_k = w_{0k}\, \Bigl(\frac{N_k}{\lambda_k - N_k}\Bigr)^{1/\theta_k}
\end{equation*}
Within the sector, workers allocate themselves across firms according to
$n_{ik}/N_k = (w_{ik}/W_k)^{\eta_k}$, so the wage needed to attract $n_{ik}$
workers is
\begin{equation*}
w_{ik} = W_k\, \Bigl(\frac{n_{ik}}{N_k}\Bigr)^{1/\eta_k}
= w_{0k}\,
\Bigl(\frac{N_k}{\lambda_k - N_k}\Bigr)^{1/\theta_k}
\Bigl(\frac{n_{ik}}{N_k}\Bigr)^{1/\eta_k}
\end{equation*}
Under Cournot, firm $i$ takes $N_{-ik} = \sum_{j\neq i} n_{jk}$ as given, so
$\mathrm d N_k / \mathrm d n_{ik} = 1$. Differentiating $\ln w_{ik}$ with
respect to $\ln n_{ik}$ yields
\begin{align*}
\frac{\partial \ln w_{ik}}{\partial \ln n_{ik}}
&= \frac{1}{\eta_k}
+ \Bigl(\frac{1}{\theta_k} - \frac{1}{\eta_k}\Bigr)\frac{n_{ik}}{N_k}
+ \frac{1}{\theta_k}\,\frac{n_{ik}}{\lambda_k - N_k} \\[4pt]
&= \frac{\eta_k s^{\mathrm{wit}}_{ik} + \theta_k\pi_k\big(1-s^{\mathrm{wit}}_{ik}\big)}
{\eta_k\theta_k\pi_k}
\end{align*}
The structural labour supply elasticity $\varepsilon_{ik}$ is the reciprocal,
hence
\begin{equation*}
\varepsilon_{ik} = \frac{\eta_k\,\theta_k\,\pi_k}
{\theta_k\pi_k\big(1-s^{\mathrm{wit}}_{ik}\big) + \eta_k s^{\mathrm{wit}}_{ik}}
\end{equation*}
which is equation~\ref{eq:elasticity}. \qed

\subsection{Proportional hiring under symmetric elasticities}
\label{app:proportional-hiring}

This appendix proves that under symmetric labor supply elasticities ($\eta_L = \eta_H = \eta$, $\theta_L = \theta_H = \theta$) and equal self-employment shares ($\pi_L = \pi_H = \pi$), every firm in the sector hires low- and high-skill labor in the same ratio: $\rho_i = n_{iH}/n_{iL} = \rho$ for all $i$.

\paragraph{Step 1: First-order conditions imply a level-curve restriction.} From  equation~\ref{eq:markdown} and the marginal revenue productivity of each labor type, the ratio of first-order conditions for $L$ and $H$ at firm $i$ gives
\begin{equation*}
\frac{w_{iL} \, n_{iL}}{w_{iH} \, n_{iH}}
\;=\; \frac{\mu_{iL}}{\mu_{iH}} \cdot \frac{\alpha}{1-\alpha}
\end{equation*}
Multiplying numerator and denominator by appropriate sums and recognizing $s^{\mathrm{wit}}_{ik} = w_{ik} n_{ik} / \sum_j w_{jk} n_{jk}$,
\begin{equation}
\frac{s^{\mathrm{wit}}_{iL}}{s^{\mathrm{wit}}_{iH}}
\;=\; \Gamma \cdot \frac{\mu_{iL}}{\mu_{iH}}
\label{eq:app-fixed-curve}
\end{equation}
where $\Gamma$ is a sector-level constant independent of $i$.

\paragraph{Step 2: Markdowns are a common function of the payroll share.} Under $\eta_L = \eta_H$ and $\theta_L = \theta_H$ and $\pi_L=\pi_H=\pi$, the markdown formula~\ref{eq:markdown} together with~\ref{eq:elasticity} implies
\begin{equation*}
\mu_{ik}
\;=\;
m\big(s^{\mathrm{wit}}_{ik}\big)
\;\equiv\;
\frac{\eta\theta\pi}
{\eta\theta\pi + \theta\pi(1-s^{\mathrm{wit}}_{ik}) + \eta s^{\mathrm{wit}}_{ik}}
\end{equation*}
The same function $m(\cdot)$ governs $\mu_{iL}$ and $\mu_{iH}$. Substituting into~\ref{eq:app-fixed-curve},
\begin{equation*}
\frac{s^{\mathrm{wit}}_{iL}}{m(s^{\mathrm{wit}}_{iL})}
\;=\;
\Gamma \cdot \frac{s^{\mathrm{wit}}_{iH}}{m(s^{\mathrm{wit}}_{iH})}
\end{equation*}
The function $g(s) \equiv s/m(s) = s + \frac{\theta\pi(1-s) + \eta s}{\eta\theta\pi}\,s$ is strictly increasing on $[0,1]$. Hence every firm lies on the same level curve in $(s^{\mathrm{wit}}_{iL}, s^{\mathrm{wit}}_{iH})$ space, parameterized by $\Gamma$.

\paragraph{Step 3: Adding-up forces $\Gamma = 1$.} Suppose for contradiction $\Gamma > 1$. Since $g$ is strictly increasing, the level-curve relation forces $s^{\mathrm{wit}}_{iL} > s^{\mathrm{wit}}_{iH}$ for every firm $i$. Summing,
\begin{equation*}
\sum_{i=1}^{M} s^{\mathrm{wit}}_{iL} \;>\; \sum_{i=1}^{M} s^{\mathrm{wit}}_{iH}
\end{equation*}
But both sums equal one, a contradiction. Symmetrically, $\Gamma < 1$ leads to $\sum_i s^{\mathrm{wit}}_{iL} < \sum_i s^{\mathrm{wit}}_{iH}$, also a contradiction. Hence $\Gamma = 1$ and, by strict monotonicity of $g$, $s^{\mathrm{wit}}_{iL} = s^{\mathrm{wit}}_{iH}$ for every firm $i$.

\paragraph{Step 4: Equal payroll shares imply equal employment ratios.} From payroll shares, we have:
\begin{equation*}
\frac{w_{iL}^{\eta}}{W_L^{\eta}} \;=\; \frac{w_{iH}^{\eta}}{W_H^{\eta}}
\end{equation*}
Combined with the constancy of $w_{iH}/w_{iL}$ across firms, this implies $n_{iH}/n_{iL} = \rho$ is firm-invariant. \qed

\subsection{Cournot fixed point on payroll shares under symmetry}
\label{app:cournot-fixed}

When the preference parameters are symmetric across skills and the self‑employment shares equalize, the equilibrium forces every firm to hire the two skill types in the same proportion $\rho$. As a result, the payroll share $s^{\mathrm{wit}}_i$ and the markdown $\mu_i$ are common across skill groups within a firm. Using the fact that $\mathrm{MRPL}_{ik}$ is proportional to $(1-\tau_i)\,z_i$ (up to a factor that is identical for both skills), we obtain
\begin{equation}
s^{\mathrm{wit}}_i = \frac{\big(\mu_i z_i\big)^{\eta}}{\sum_{j=1}^{M} \big(\mu_j z_j\big)^{\eta}},
\qquad
\mu_i = \frac{\varepsilon_i}{\varepsilon_i + 1},
\qquad
\varepsilon_i = \frac{\eta\,\theta\,\pi}
{\theta\pi\big(1 - s^{\mathrm{wit}}_i\big) + \eta\,s^{\mathrm{wit}}_i}
\label{eq:FP_app}
\end{equation}
Note that the term $(1-\tau_i)$ is common to all firms after the shock and cancels out of the relative shares in equilibrium, so it does not appear explicitly in the share equation. This system jointly determines the distribution of payroll shares and markdowns, and it must be solved numerically together with the aggregate conditions for $\pi$ and $W$.

\subsection{Capital input with capital-skill complementarity}
\label{app:capital}

The baseline model abstracts from capital. The digital-payments
interpretation of the productivity shock, however, naturally introduces a
capital-like input: firms that adopt instant payment infrastructure
acquire digital equipment whose complementarity with skill is itself an
empirical question. This appendix extends the baseline to a CES
production structure with a third input $K_i$ that has flexible
complementarity with high-skill labor \citep{krusell2000capital}.

\paragraph{Production technology.} Firm $i$ produces with the nested CES
\begin{equation}
y_i \;=\; \tilde A \, z_i \, Q_i,
\qquad
Q_i
\;=\;
\left[
\alpha \, n_{iL}^{\,(\sigma_L-1)/\sigma_L}
\;+\;
(1-\alpha) \, G_i^{\,(\sigma_L-1)/\sigma_L}
\right]^{\sigma_L/(\sigma_L-1)}
\label{eq:Q_appendix}
\end{equation}
\begin{equation}
G_i
\;=\;
\left[
\beta \, K_i^{\,(\sigma_K-1)/\sigma_K}
\;+\;
(1-\beta) \, n_{iH}^{\,(\sigma_K-1)/\sigma_K}
\right]^{\sigma_K/(\sigma_K-1)}
\label{eq:G_appendix}
\end{equation}
where $K_i$ is firm-level capital (treated as exogenous), and
$\sigma_L > 0$ and $\sigma_K > 0$ are the elasticities of substitution at
the outer and inner CES nests, respectively. The empirical literature finds $\sigma_L$ above one (substitutes)
and $\sigma_K$ below one (complements), so capital tends to complement
high-skill labor more strongly than it complements low-skill labor.

Marginal revenue products are as follows:
\begin{equation*}
\mathrm{MRPL}_{iL}
\;=\;
\tilde A z_i \, \alpha \, \big( Q_i / n_{iL} \big)^{1/\sigma_L},
\end{equation*}
\begin{equation*}
\mathrm{MRPL}_{iH}
\;=\;
\tilde A z_i \, (1-\alpha)(1-\beta) \, \big( Q_i / G_i \big)^{1/\sigma_L}
\big( G_i / n_{iH} \big)^{1/\sigma_K}
\end{equation*}
The wage equation~\ref{eq:markdown} continues to apply with the same
markdowns.

Instant-payments now reduce transaction costs, increase the value of self-employment, and expand capital. The first two channels operate exactly as in the baseline. The
transaction-cost shock activates the competition channel and the self-employment shock activates
the elasticity channel, both benefiting mainly low-skill workers. The capital shock pushes in the opposite direction when capital and high-skill labor are
complements: by raising $\mathrm{MRPL}_{iH}$ disproportionately, it
widens the skill premium.

The strength of the
capital channel depends on $\sigma_K$. For low $\sigma_K$, capital is strongly complementary to high-skill labor and then, a capital expansion pushes the skill premium up, partly
offsetting the inequality compression delivered by the other two channels. For $\sigma_K > 1$, the capital shock raises the low-skill marginal product more than the high-skill marginal product, reinforcing inequality compression.

\newpage

\begin{figure}[H]

\begin{center}
\includegraphics[width=0.7\textwidth]{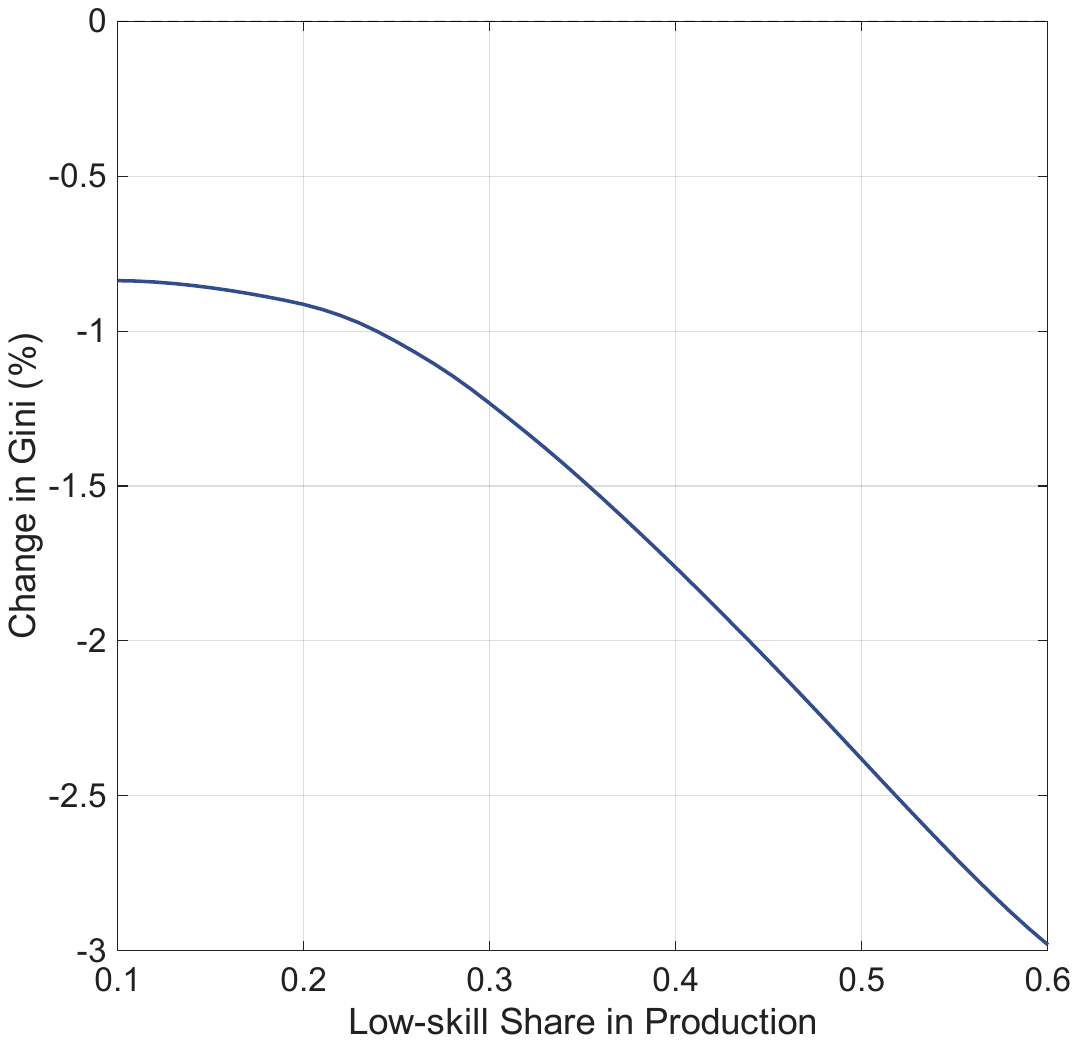}
\end{center}
\vspace{-2cm}

\caption{Effect of Lowering Transaction Cost on Wage Inequality}
\label{fig:cf_inequality}

\justify
{\footnotesize
This figure displays the percentage change in inequality resulting from reducing transaction costs to zero, using the Gini coefficient as our inequality measure.
}
\end{figure}

\newpage

\begin{figure}[H]

\begin{center}
\includegraphics[width=0.7\textwidth]{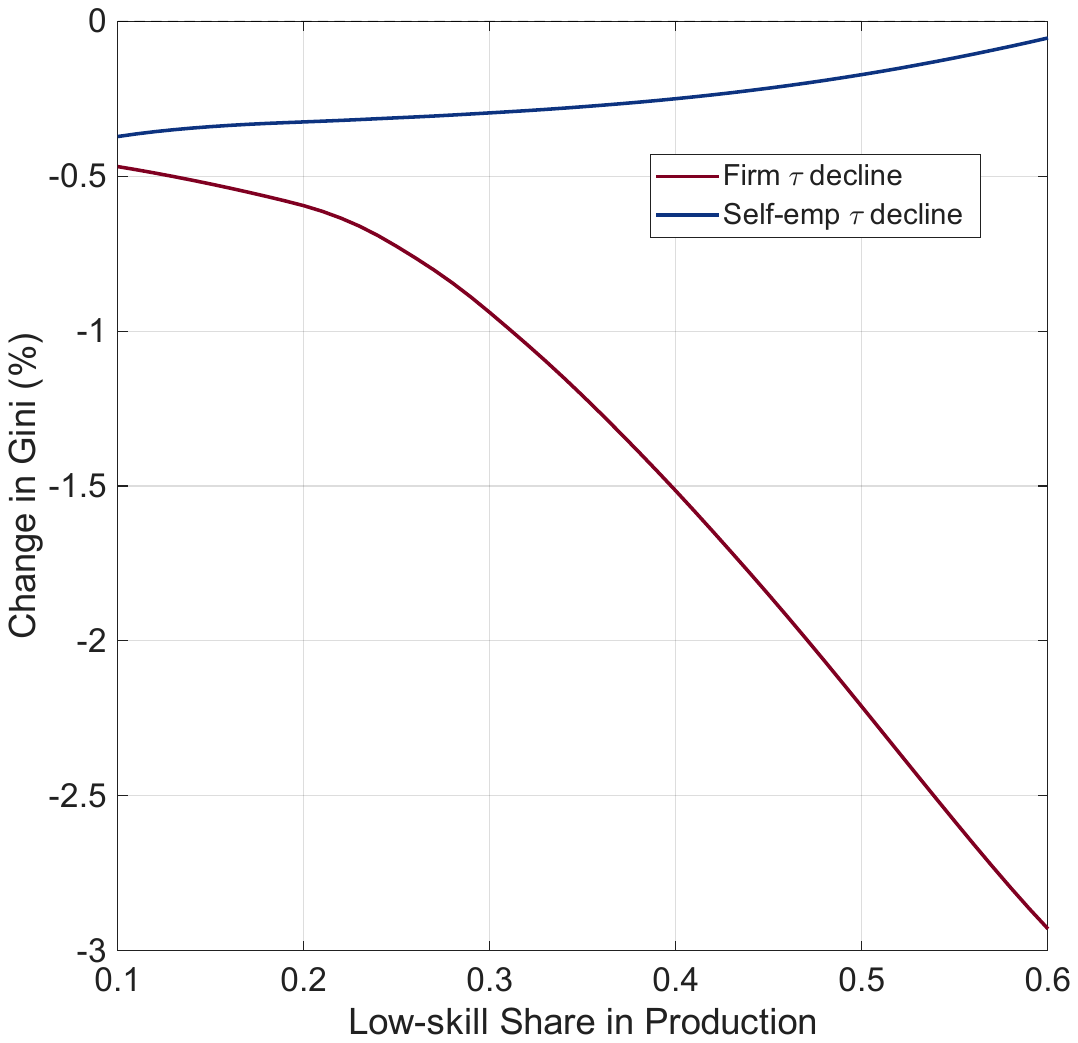}
\end{center}
\vspace{-2cm}

\caption{Effect of Lowering Transaction Cost on Wage Inequality: Decomposition}
\label{fig:cf_inequality_decomp}

\justify
{\footnotesize
This figure displays the percentage change in inequality resulting from reducing transaction costs to zero, using the Gini coefficient as our inequality measure. The navy line corresponds to a scenario where only self-employment experiences a decline in transaction costs. The red line corresponds to a scenario in which only firms benefit from lower transaction costs.
}
\end{figure}

\newpage

\begin{figure}[H]

\begin{center}
\begin{subfigure}{0.55\linewidth}
\includegraphics[width=\textwidth]{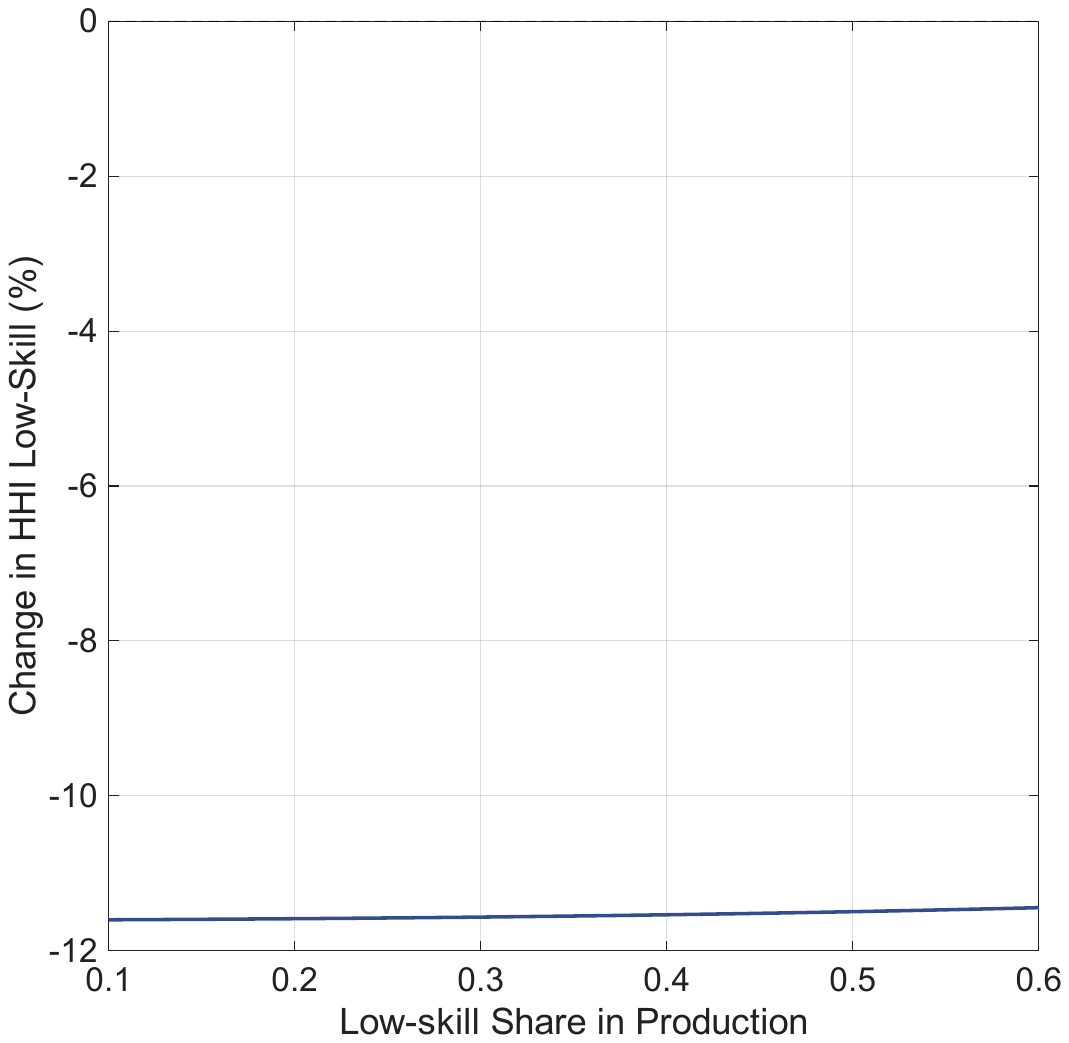}
\vspace{-2.5cm}
\caption*{Panel A. HHI for Low-Skill}
\end{subfigure}

\vspace{-1.5cm}

\begin{subfigure}{0.55\linewidth}
\includegraphics[width=\textwidth]{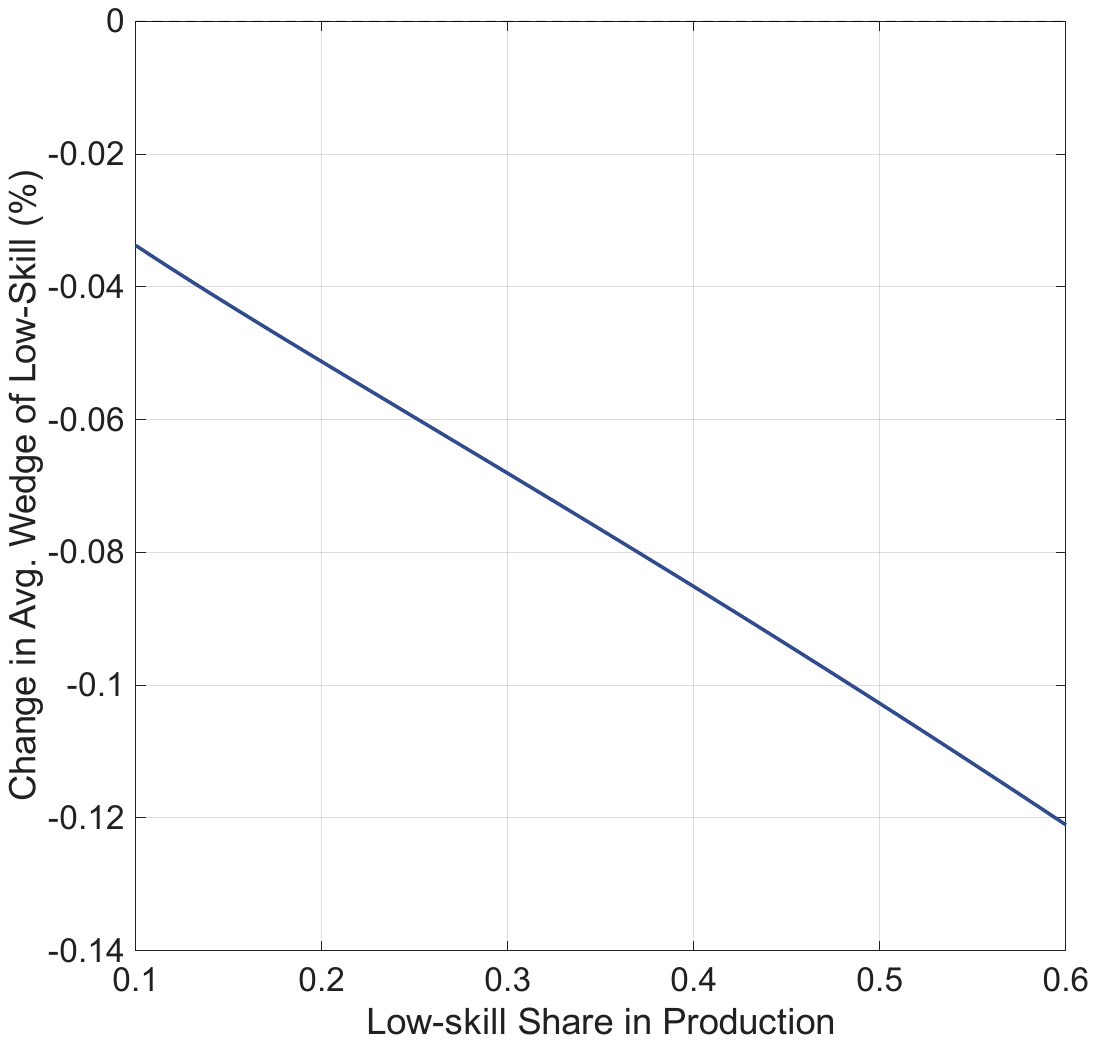}
\vspace{-2.5cm}
\caption*{Panel B. Markdown}
\end{subfigure}
\end{center}

\caption{Effect of lowering transaction costs on HHI and payroll-weighted-average markdown for low-skill}
\label{fig:cf_hhi_markdown}

\justify
{\footnotesize
This figure presents the response of the Herfindahl-Hirschman Index (HHI) and payroll-weighted-average markdown for low-skill workers.
}
\end{figure}

\newpage

\begin{figure}[H]

\begin{center}
\includegraphics[width=0.7\textwidth]{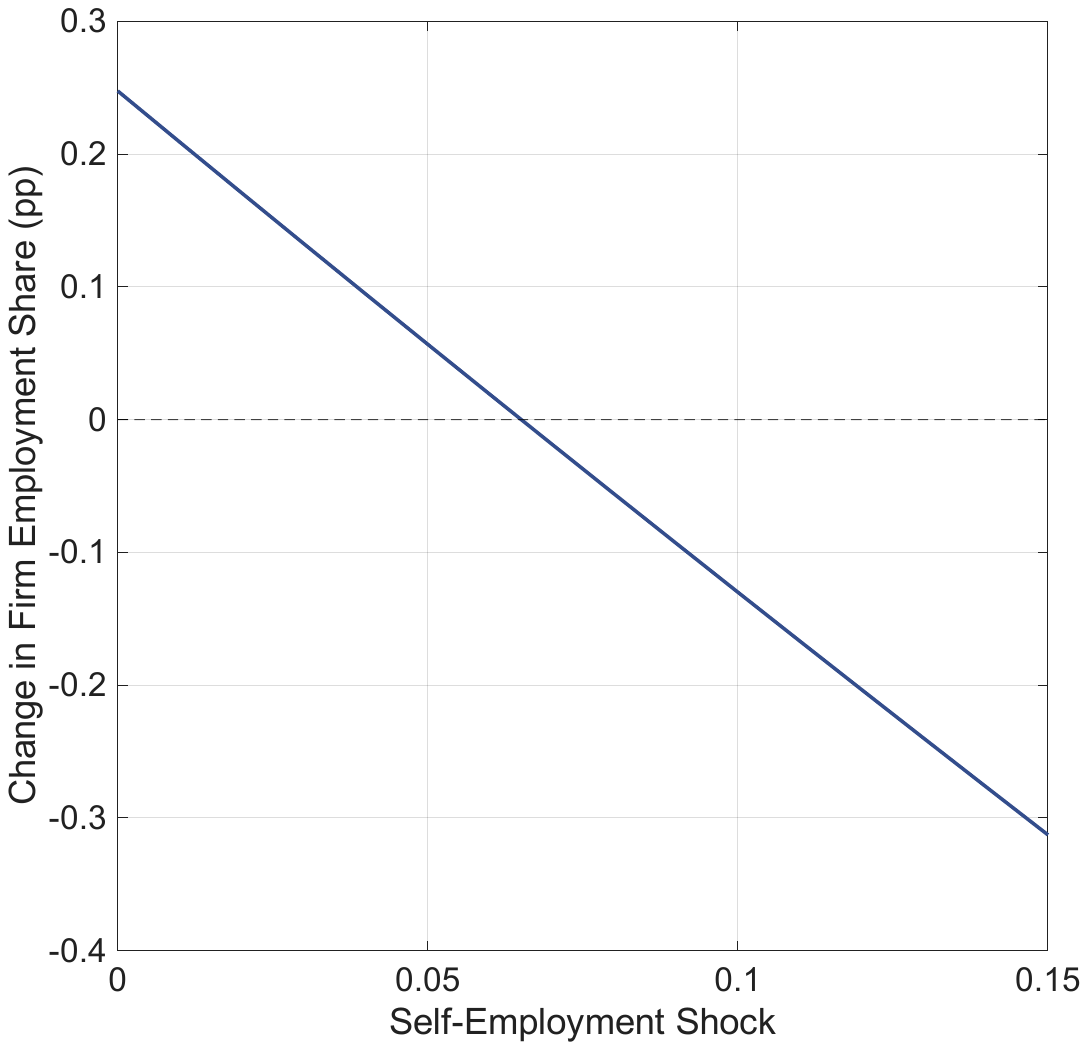}
\end{center}
\vspace{-2cm}

\caption{Effect of Lowering Transaction Cost on Total Employees}
\label{fig:cf_employees}

\justify
{\footnotesize
This figure displays the percentage change in total employees resulting from reducing transaction costs to zero.
}
\end{figure}

\newpage

\begin{figure}[H]

\begin{center}
\includegraphics[width=0.7\textwidth]{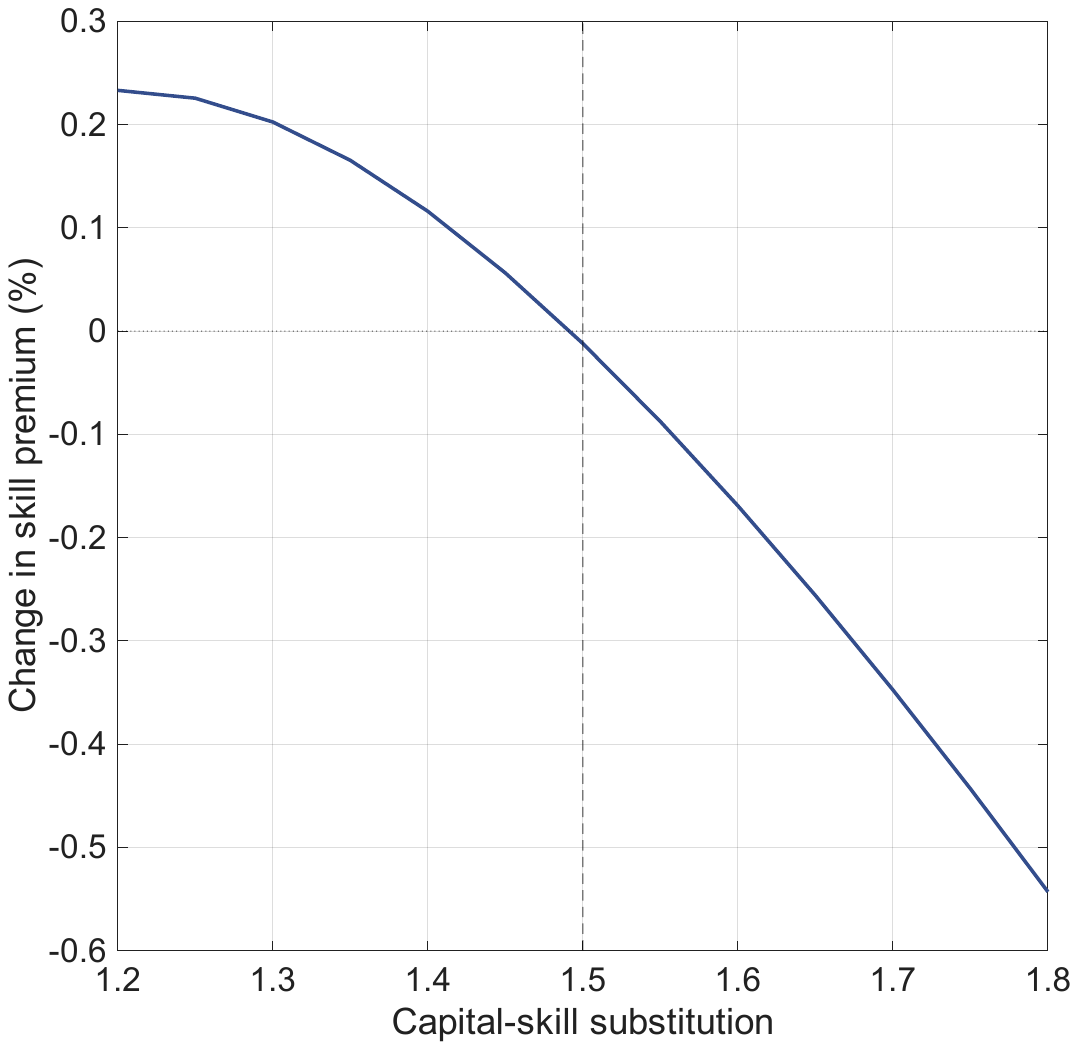}
\end{center}
\vspace{-2cm}

\caption{Effect of Lowering Transaction Cost on Wage Inequality}
\label{fig:cf_inequality_capital}

\justify
{\footnotesize
This figure displays the percentage change in inequality resulting from reducing transaction costs to zero, using the Gini coefficient as our inequality measure.
}
\end{figure}

\newpage

\subsection{Quantitative Framework}

\begin{figure}[H]
\begin{center}
\includegraphics[width=0.7\textwidth]{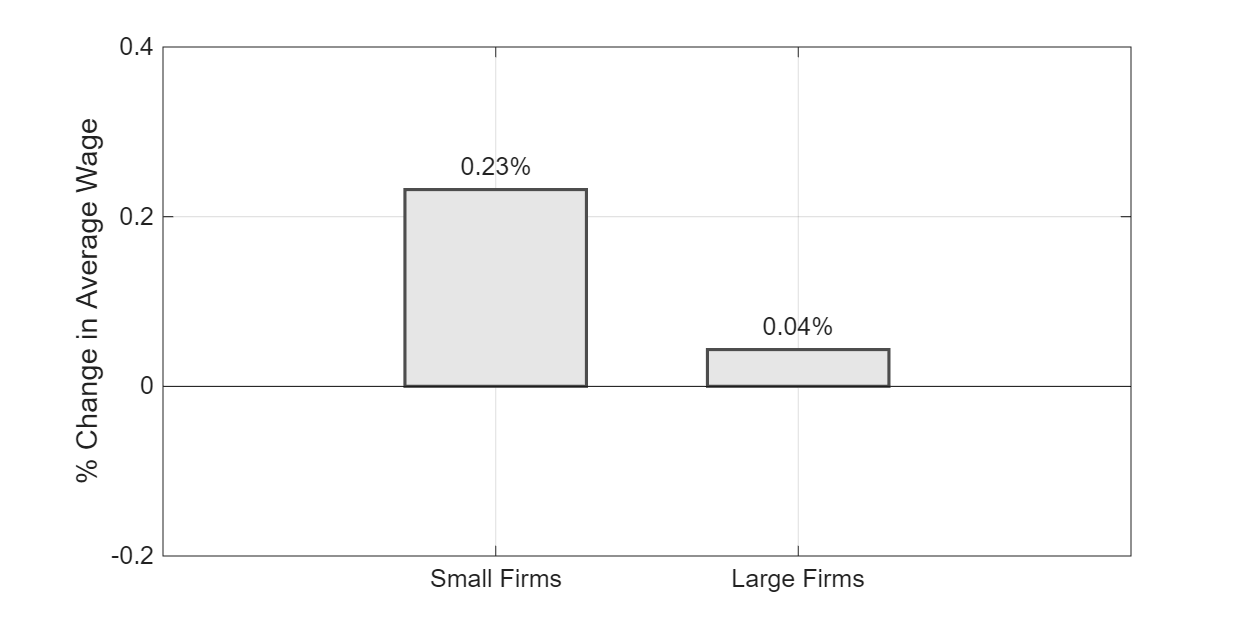}
\end{center}
\caption{Change in Average Wage by Firm Size after Pix Implementation}
\label{fig:wage_change_firmsize}
\justify
{\footnotesize
This figure shows the percentage change in average wages following Pix, comparing small firms directly affected by the transaction cost reduction (left bar) and large firms affected indirectly through increased labor market competition (right bar). Wage change is defined as a percentage relative to the baseline equilibrium with transaction costs.
}
\end{figure}

\newpage
\begin{figure}[H]
\begin{center}
\includegraphics[width=0.7\textwidth]{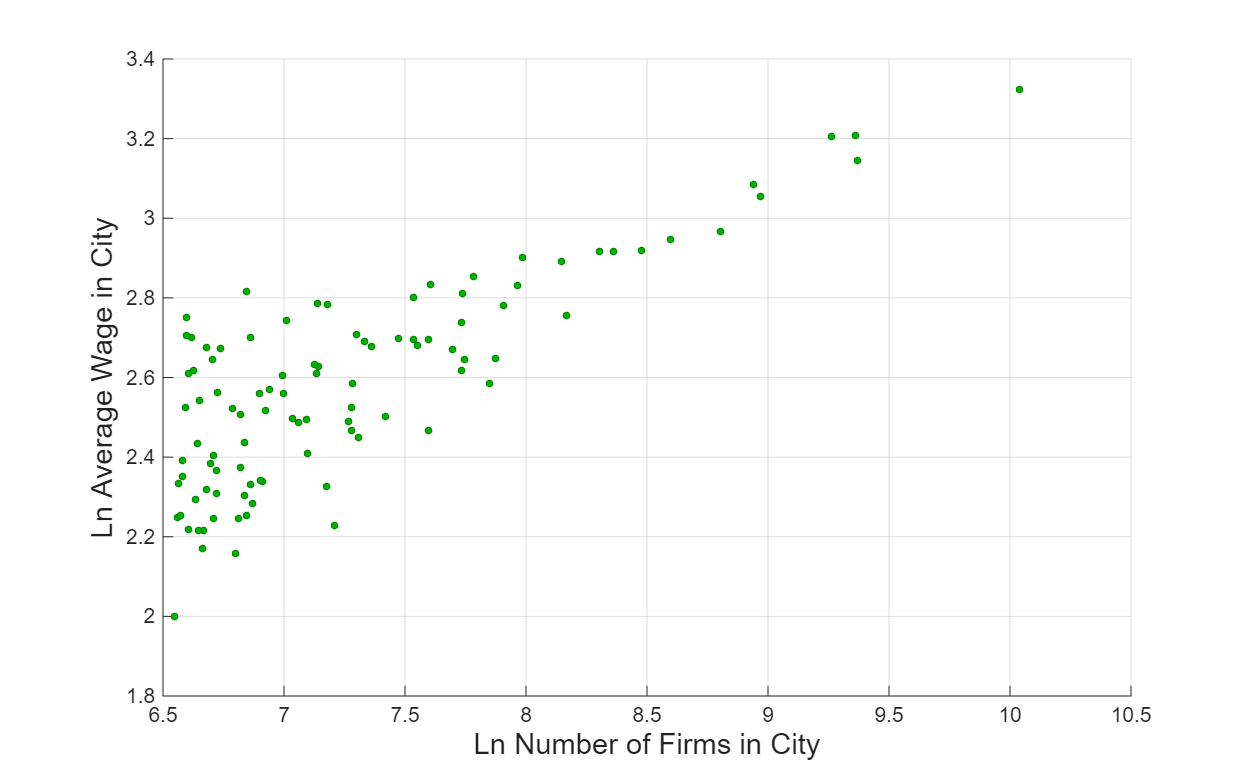}
\end{center}
\caption{Average Wage and Number of Firms across Cities}
\label{fig:city_wage_num_firms}
\justify
{\footnotesize
This figure shows the relationship between the log of number of firms in a city (horizontal axis) and the city's average wage in logs (vertical axis) in the baseline model, prior to the implementation of Pix.
}
\end{figure}

\newpage

\begin{figure}[H]
\begin{center}
\includegraphics[width=0.7\textwidth]{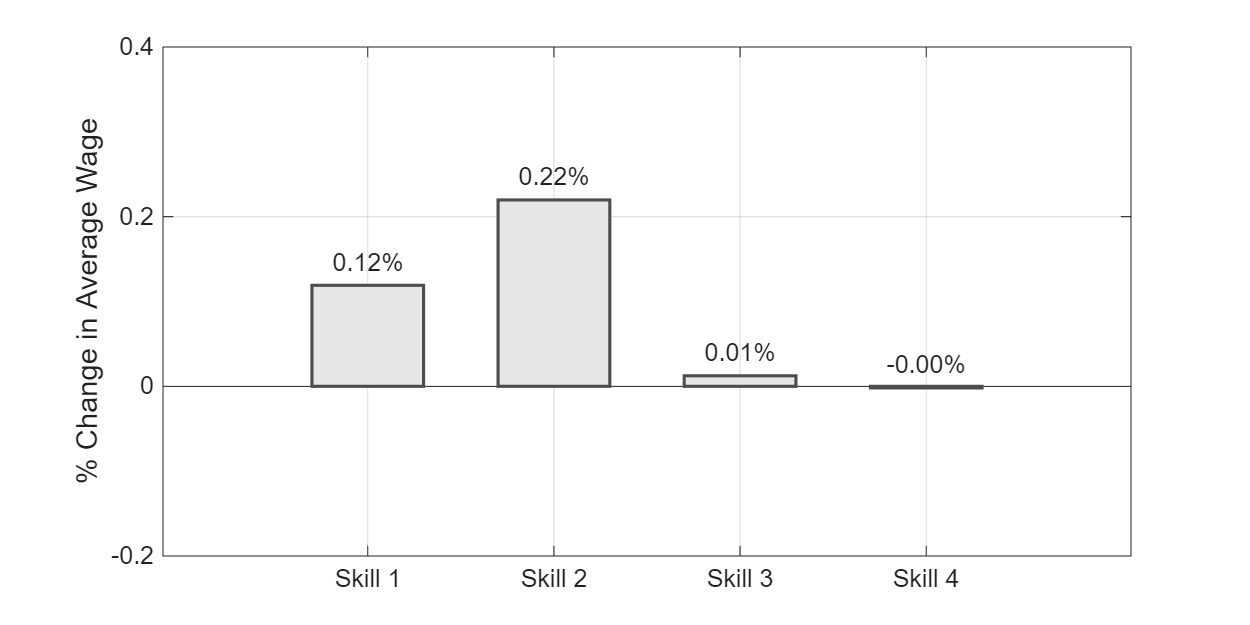}
\end{center}
\caption{Change in Average Wage by Skill Group after Pix Implementation}
\label{fig:wage_change_skill}
\justify
{\footnotesize
This figure shows the percentage change in average wages following Pix for the four skill groups (1: primary education, 2: secondary education, 3: some college, 4: college graduates). Wage change is defined as a percentage with respect to the baseline equilibrium with transaction costs.
}
\end{figure}

\newpage

\newpage

\begin{table}[H]
\centering
\caption{Pre-Determined Model Parameters}
\label{tab:pre_parameters}
\begin{tabular}{l l l}
\toprule
\textbf{Parameter} & \textbf{Description} & \textbf{Value} \\
\midrule
$C$ & Number of cities & 100 \\
$S$ & Number of skill groups & 4 \\
$K$ & Number of sectors & 3 \\
$\alpha_c$ and $x_m$ & Num. cities: shape ($1.24$) and min. ($695$) &  Share of firms in 10\% bigger cities \\
$H_s$ & Skill group masses & $[8.3\%, \: 16.3\%, \: 61.8\%, \: 13.6\%]$ \\
$\gamma_k$ & Sector shares & $[19\%, \: 14\%, \: 67\%]$ \\
$\eta$ and $\theta$ & Within and between-city LS elasticity & 7 and 0.5 (Berger et al., 2022) \\
$\alpha$ & Labor share & 0.7 \\
$\tau$ & Transaction costs & 5\% \\
\bottomrule
\end{tabular}
\end{table}

\newpage

\begin{table}[H]
\centering
\caption{Calibrated Parameters}
\label{tab:cal_parameters}
\begin{tabular}{l l l}
\toprule
\textbf{Parameter} & \textbf{Description and Value} & \textbf{Target} \\
\midrule
$\lambda_{ks}$ & Sector-skill productivity parameters & Skill-specific labor share of: \\
\ \ $k=1$ & $[-0.45, \: -0.63, \: 0.33, \: 0.75]$ & -- Sector 1 \\
\ \ $k=2$ & $[-1.125, \: -1.425, \: 1.02, \: 1.53]$ & -- Sector 2\\
\ \ $k=3$ & $[-2.9, \: -2.1, \: 2.3, \: 2.7]$ & -- Sector 3\\
$\alpha_k$ & Pareto shape by sector $[1.1, \: 1.3, \: 1.6]$ & Sector-specific labor share \\
$A_{s}$ & Skill productivity $[1, \: 1.37, \: 0.208, \: 0.054]$ & Skill-specific average wage \\
\bottomrule
\end{tabular}
\end{table}

\newpage

\begin{table}[H]
\centering
\caption{Model Fit of the Brazilian Labor Market}
\label{tab:model_fit}
\begin{tabular}{lcc}
\toprule
\textbf{Moment} & \textbf{Data} & \textbf{Model} \\
\midrule
I. College Graduate Labor Share by Sector & & \\
\quad Manufacturing & 15.5 & 14.5 \\
\quad Wholesale & 22.2 & 22.0 \\
\quad Retail & 10.1 & 10.1 \\
\addlinespace
II. Sectoral Labor Share & & \\
\quad Manufacturing & 43.9 & 43.0 \\
\quad Wholesale & 11.3 & 13.7 \\
\quad Retail & 44.8 & 43.4 \\
\addlinespace
III. Skill-Specific Average Wages, relative to $s = 1$ & & \\
\quad Skill Group 2 (Secondary) & 1.00 & 0.99 \\
\quad Skill Group 3 (Some College) & 1.12 & 1.12 \\
\quad Skill Group 4 (College Graduates) & 2.99 & 3.01 \\
\bottomrule
\end{tabular}
\end{table}

\newpage

\section{Additional Figures and Tables}

\setcounter{figure}{0}
\renewcommand{\thefigure}{B.\arabic{figure}}
\setcounter{table}{0}
\renewcommand{\thetable}{B.\arabic{table}}

\begin{figure}[h!]
\begin{center}
\begin{subfigure}{0.65\linewidth}
   \includegraphics[width=\textwidth]{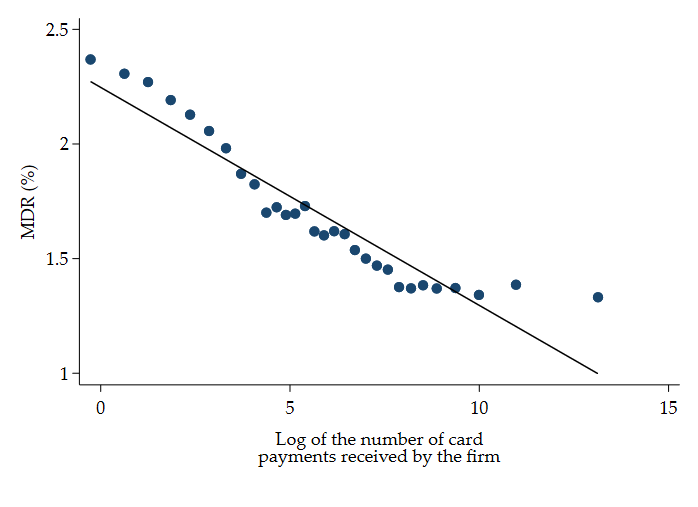}]
   \caption*{Panel A. Debit Cards}
\end{subfigure}
\vspace{0.5cm}
\begin{subfigure}{0.65\linewidth}
   \includegraphics[width=\textwidth]{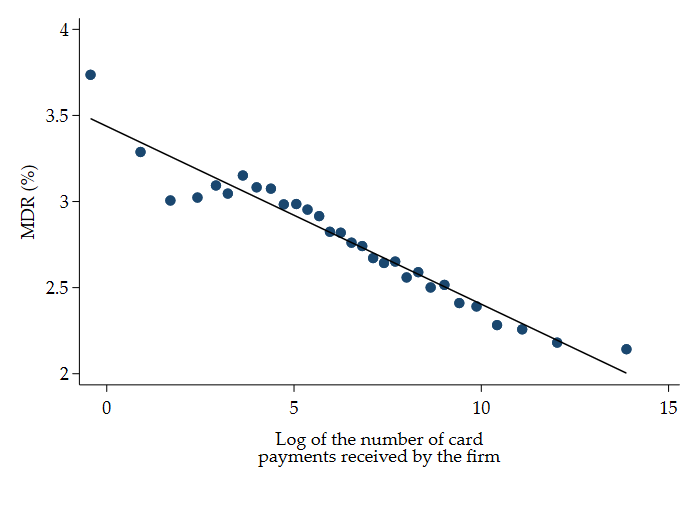}
   \caption*{Panel B. Credit cards}
\end{subfigure}
\end{center}
\caption{Average Merchant Discount Rates (MDRs) by Bins of Card Transactions per Firm}
\label{appendix_fig:MDR}

\justify
{\footnotesize
This figure shows binned scatter plots of the number of quarterly card transactions per firm and merchant discount rates (MDRs). We use quarterly data from the Central Bank of Brazil, controlling for sector and quarter fixed effects. The sample runs from the first quarter of 2022 through the third quarter of 2025.
}
\end{figure}

\newpage
\begin{figure}[H]

\begin{center}
\begin{subfigure}{0.62\linewidth}
   \includegraphics[width=\textwidth]{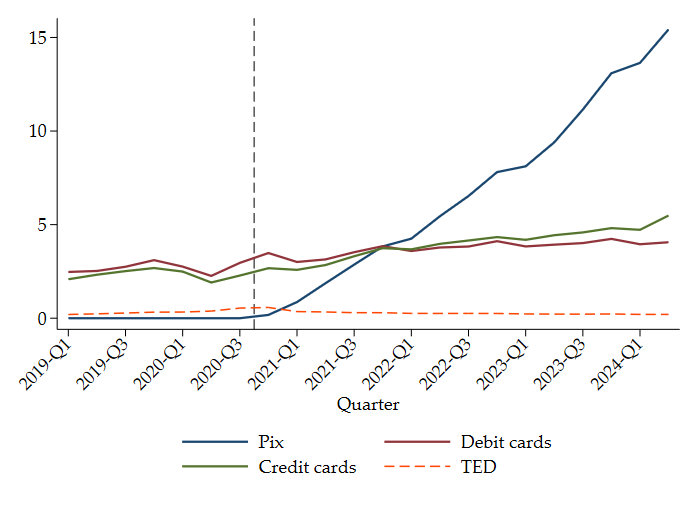}
   \caption*{Panel A. Quantity of Transactions}
\end{subfigure}
\vspace{0.5cm}
\begin{subfigure}{0.62\linewidth}
   \includegraphics[width=\textwidth]{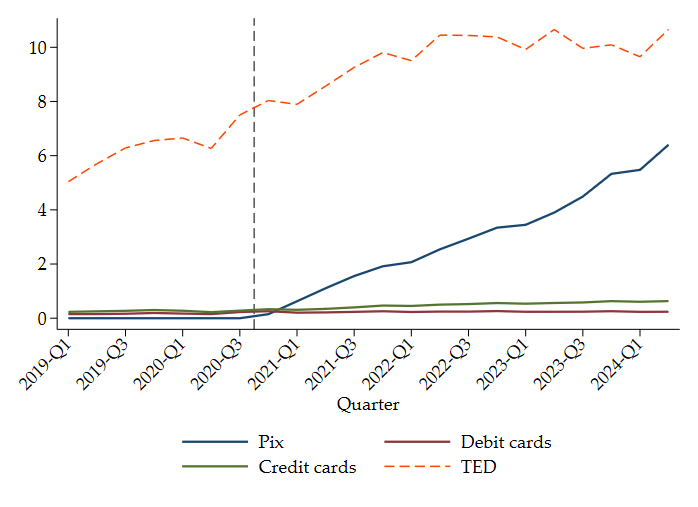}
   \caption*{Panel B. Value of Transactions}
\end{subfigure}
\end{center}
\caption{Evolution of Payment Methods}
\label{appendix_fig:differ_payment}

\justify
{\footnotesize
This figure shows the evolution of different payment methods in Brazil from 2019 to 2024. Panel A plots the quarterly volume of transactions in billions across different payment methods. Panel B displays the quarterly value of transactions in Brazilian Reais (BRL) trillions. The vertical dashed line indicates the introduction of Pix in November 2020. Data are from the Central Bank of Brazil.
}
\end{figure}

\newpage

\begin{figure}[h!]
 \centering
\begin{subfigure}[t]{0.75\textwidth}\includegraphics[scale=.5]{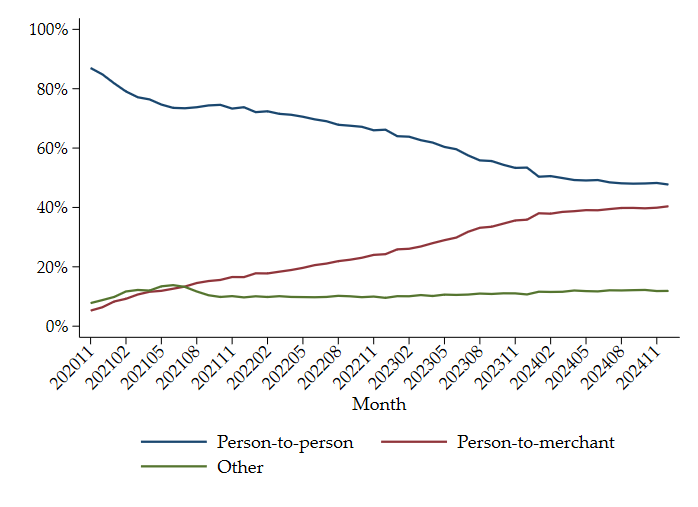}
         \caption*{Panel A. Share of transaction types}
\end{subfigure}    \hfill
\begin{subfigure}[t]{0.7\textwidth}
        \includegraphics[scale=.5]{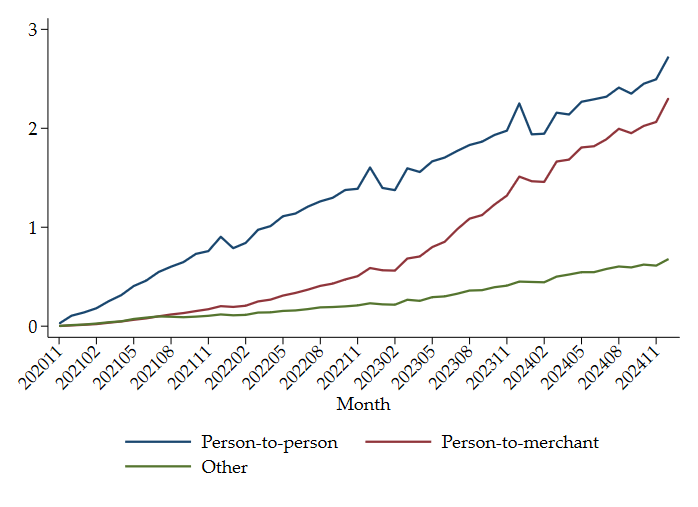}
         \caption*{Panel B. Quantity of transactions (in billions)}
\end{subfigure}
 \caption{Evolution of different types of Pix transactions}
\label{appendix_fig:desc_p2p_p2m}
\footnotesize\justify
This figure displays the evolution of Pix transactions between accounts linked to individuals (person-to-person), transactions from accounts linked to individuals to accounts linked to businesses (person-to-merchant), and others. The data are sourced from the Central Bank of Brazil.
\end{figure}
\newpage

\begin{figure}[H]

\begin{center}
\includegraphics[width=0.7\textwidth]{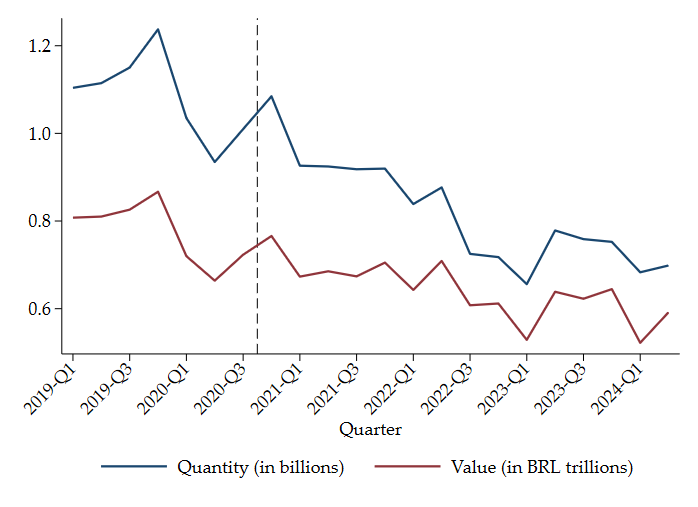}
\end{center}
\caption{Evolution of Cash Withdrawals}
\label{appendix_fig:withdrawal}
\justify
{\footnotesize
This figure shows the evolution of cash withdrawals in Brazil from 2019 to 2024. Data from the Central Bank of Brazil. The dashed line indicates the introduction of Pix. The data are sourced from the Central Bank of Brazil.
}
\end{figure}
\newpage

\begin{figure}[H]

\centering
\includegraphics[width=0.9\textwidth]{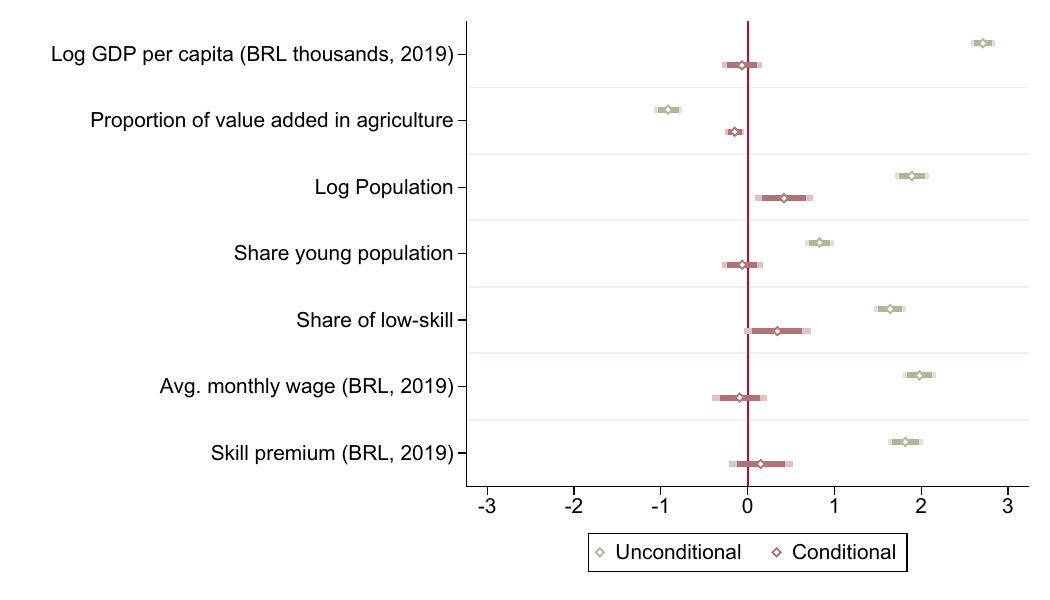}
\caption{Balance of Covariates}
\label{fig:covar_balance}

\justify

{\footnotesize
This figure shows coefficient estimates and 95\% (darker bars) and 99\% (lighter bars) confidence intervals of one standard deviation higher mobile penetration for different variables. All variables are normalized to have a mean of zero and a standard deviation of one. ``Unconditional'' refers to comparing differently treated cities without conditioning on any fixed effects. ``Conditional'' indicates that cities are within the same region, size decile measured by GDP per capita, agriculture share decile measured by the share of GDP in the agriculture sector, and share of young individuals decile, measured by individuals between 20 and 50 years old. Skill premium refers to the difference between wages of college and non-college educated workers.
}
\end{figure}

\newpage
\begin{figure}[H]
\centering
\includegraphics[width=0.9\textwidth]{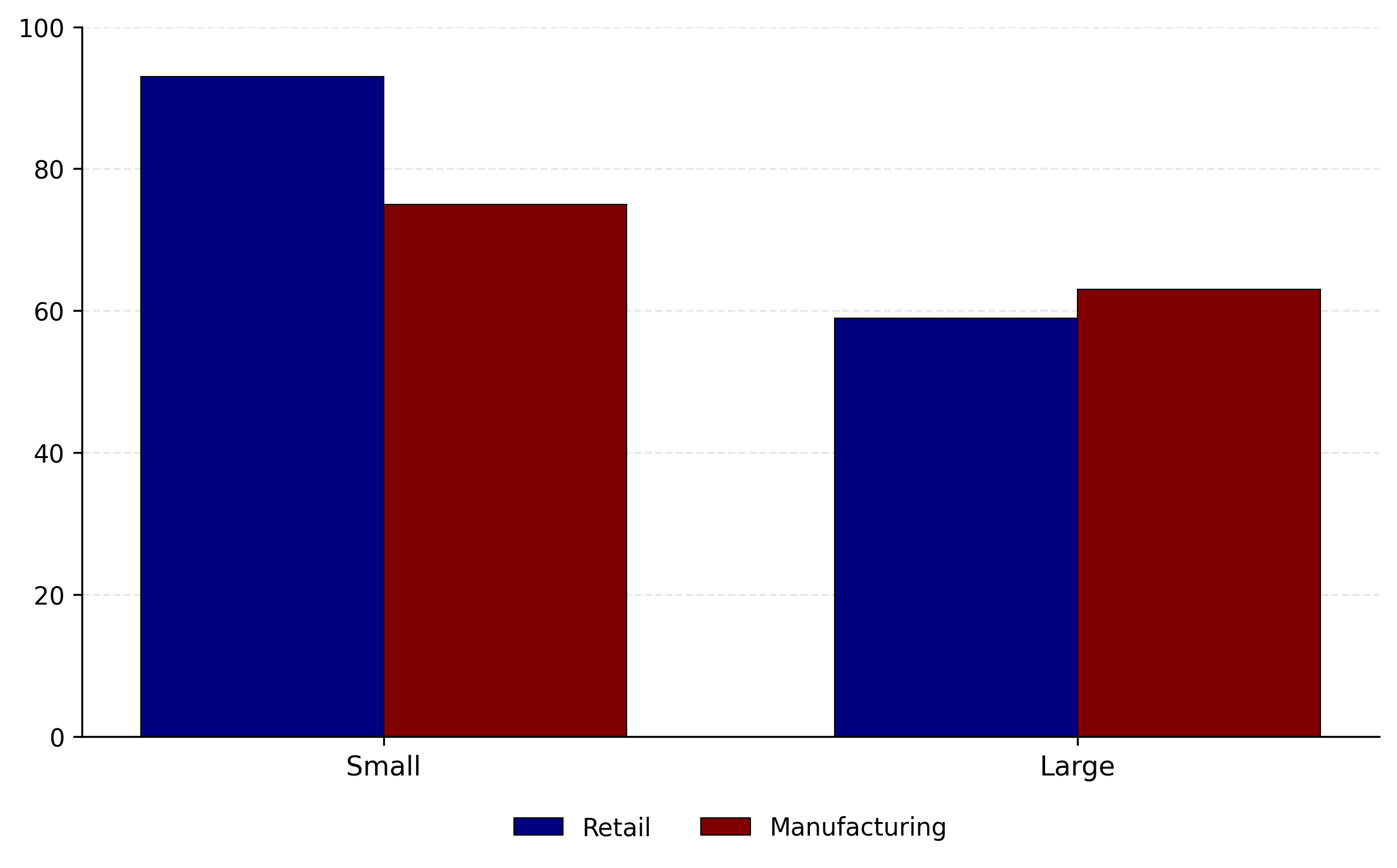}

\caption{Skill Composition by Industry and Firm Size}
\label{fig:skill_composition}

\justify
{\footnotesize
This figure plots the share of low-skill payroll as a percentage of total payroll for small and large establishments in the retail and manufacturing industries. Small firms are defined as having fewer than 20 employees. The data are from RAIS employer-employee matched data for 2019. The bars display results separately for the retail (blue) and manufacturing (maroon) sectors across various firm size categories.
}
\end{figure}

\newpage
\begin{figure}[H]
\centering
\begin{tabular}{c}
    \includegraphics[width=0.6\textwidth]{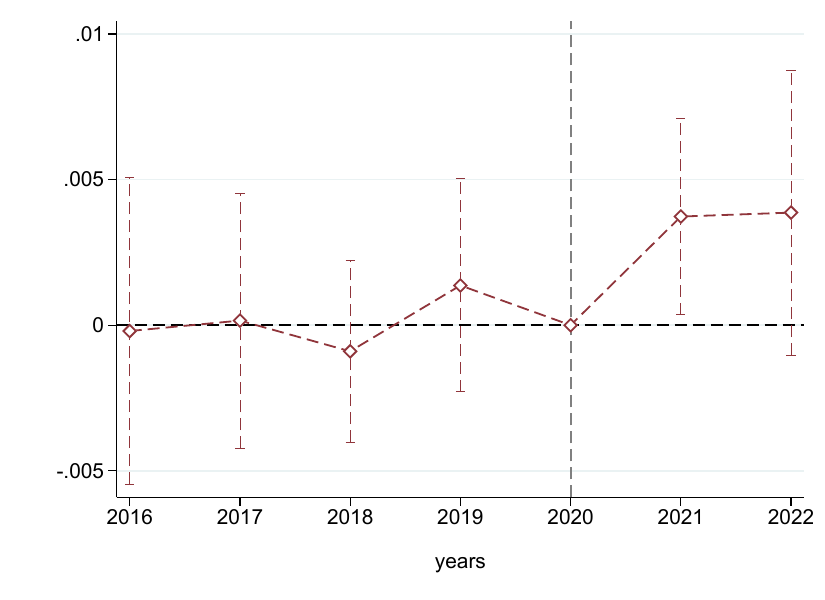} \\
    Panel A. Small Retail \\[0.8cm]
    \includegraphics[width=0.6\textwidth]{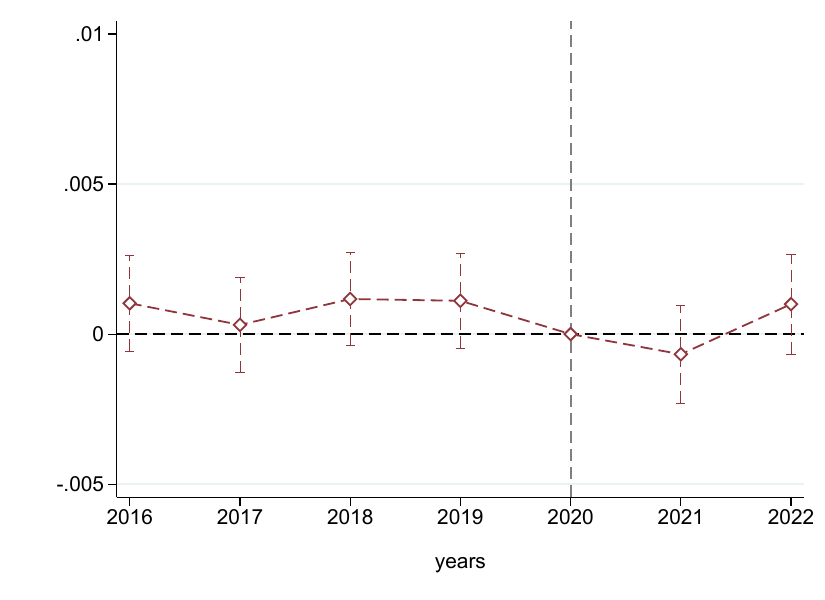} \\
    Panel B. Small Manufacturing
\end{tabular}
\caption{Effect of Mobile Penetration on Firm Entry by Industry and Size}
\label{fig:firm_entry}
\justify
{\footnotesize
This figure presents the estimated yearly coefficients from municipality-level difference-in-differences regressions examining the effect of standardized mobile penetration on the number of firm entrants per 1,000 population around the introduction of Pix in November 2020, estimated separately for small retail and small manufacturing firms. Small firms are defined as those with annual sales between USD 70,000 and USD 970,000. The regression specification includes municipality fixed effects, time fixed effects, and municipality characteristics interacted with time fixed effects. Municipality characteristics include state membership, deciles of GDP per capita, percentage of young population, and percentage of value added in agriculture. The dashed lines represent 95\% confidence intervals constructed using robust standard errors clustered at the municipality level. A vertical dashed line marks the introduction of Pix in 2020. Mobile penetration is defined as the ratio of mobile devices with 3G or higher capability to total municipal population, standardized to have unit variance and winsorized at the 1\% level.
}
\end{figure}

\newpage
\begin{figure}[H]
\centering

    \includegraphics[width=0.6\textwidth]{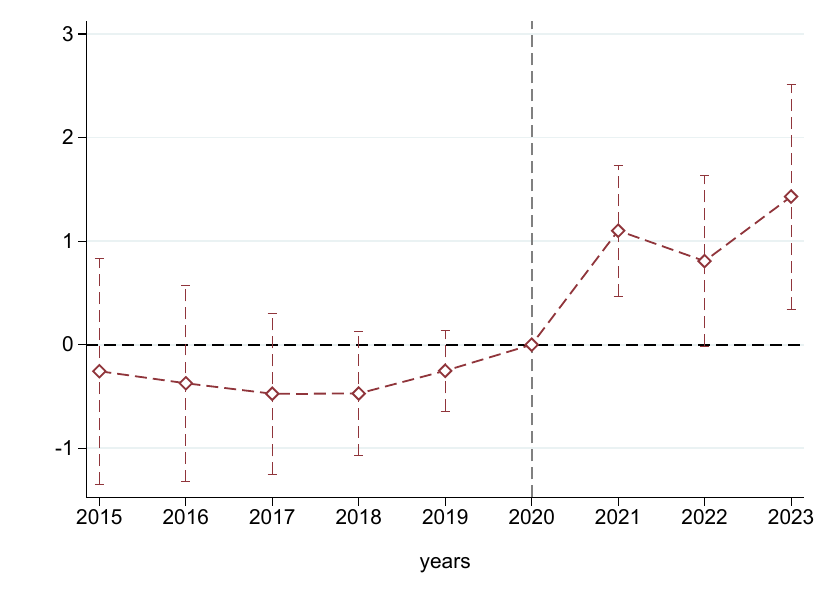}

\caption{Effect of Mobile Penetration on Micro-entrepreneur Entry}
\label{fig:taxes}

\justify
{\footnotesize
This figure presents the estimated yearly coefficients from municipality-level difference-in-differences regressions examining the effect of standardized mobile penetration on the entry of individual micro-entrepreneurs (MEI) around the introduction of Pix in November 2020. The dependent variable is the annual municipal tax revenue from the Tax on the Circulation of Goods and Services (Imposto sobre Circulação de Mercadorias e Serviços, ICMS) per 1,000 population paid by MEI firms. The ICMS fee for MEIs is a fixed amount of one Brazilian real (BRL 1.00), predominantly paid by retailers. The regression specification includes municipality fixed effects, region-by-year fixed effects, and municipality characteristics interacted with year fixed effects. Municipality characteristics include deciles of GDP per capita, percentage of young population, and percentage of value added in agriculture. The dashed lines represent 95\% confidence intervals constructed using robust standard errors clustered at the municipality level. A vertical dashed line marks the introduction of Pix in 2020. Mobile penetration is defined as the ratio of mobile devices with 3G or higher capability to total municipal population, standardized to have unit variance and winsorized at the 1\% level.
}

\end{figure}

\newpage
\begin{figure}[H]
\centering
    \includegraphics[width=0.6\textwidth]{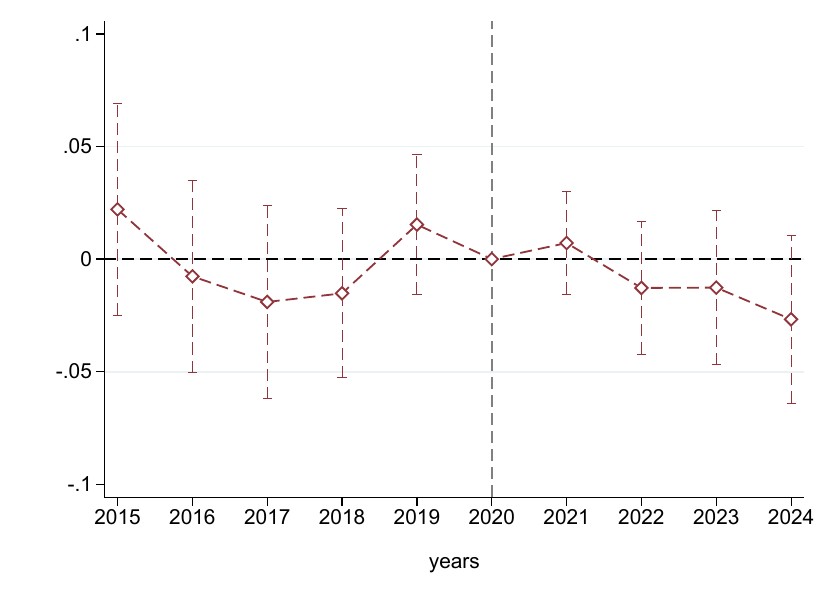} \\

\caption{The Effect of Mobile Penetration on Business Loans}
\label{fig:business_loans}

\justify
{\footnotesize
This figure presents the estimated yearly coefficients from municipality-level difference-in-differences regressions examining the effect of standardized mobile penetration on business loans per capita around the introduction of Pix in November 2020. The dependent variable is business loans, defined as all formal lending to business entities at the municipality level. The regression specification includes municipality fixed effects, region-by-year fixed effects, and municipality characteristics interacted with year fixed effects. Municipality characteristics include deciles of GDP per capita, percentage of young population, and percentage of value added in agriculture. The dashed lines represent 95\% confidence intervals constructed using robust standard errors clustered at the municipality level. A vertical dashed line marks the introduction of Pix in 2020. Mobile penetration is defined as the ratio of mobile devices with 3G or higher capability to total municipal population, standardized to have unit variance and winsorized at the 1\% level.
}
\end{figure}

\newpage 

\begin{table}[htbp]
  \centering\caption{\label{appendix_tab:desc_pix_sector} Share of Pix Transactions and GDP by Industry (\%)}
  \begin{singlespace}
  \justify
  {\footnotesize
This table presents the participation of selected industries in the total quantity and value of Pix transactions from 2021 to 2023. The GDP figures correspond to the year 2019. The Pix data come from the Central Bank of Brazil and the GDP data come from the Brazilian Institute of Geography and Statistics (IBGE).
}
\end{singlespace}
\vspace{0.5cm}
    \small\begin{tabular}{lccc}
    \toprule
          & Quantity & Value & GDP \\
    \midrule
    Retail & 32.8  & 37.1  & 10.1 \\
    Services (excluding financial, health, education, food) & 31.6  & 14.7  & 8.4 \\
    Financial services & 10.9  & 10.6  & 5.4 \\
    Accommodation and Food Services & 7.3   & 1.8   & 2.6 \\
    Information & 6.7   & 3.1   & 3.3 \\
    Manufacturing & 3.1   & 16.0  & 27.7 \\
    Arts, Entertainment, and Recreation & 1.6   & 0.4   & 0.4 \\
    Education & 1.1   & 1.0   & 4.1 \\
    Health & 0.9   & 1.9   & 4.2 \\
    Other & 4.1   & 13.3  & 33.8 \\
    \bottomrule
    \end{tabular}%
\end{table}%

\newpage 

\begin{table}[H]
\setlength{\tabcolsep}{0pt}
  \centering\caption{\label{appendix_tab:penetration_pix_adoption} Effect of Mobile Penetration on Pix Transactions (\%)}
  \begin{singlespace}
  \justify
  {\footnotesize
This table presents regression estimates of the impact of mobile penetration on Pix adoption following its introduction in late 2020. The dependent variables are the number of Pix transactions per capita (Column (1)) and the value of Pix transactions per capita (Column (2)). The interaction term Mobile penetration × Post captures the differential effect of mobile penetration on Pix adoption after its introduction, representing the average quarterly effect across the post-period. The specification includes municipality fixed effects, region-by-year fixed effects, and municipality characteristics interacted with year fixed effects. Standard errors clustered at the municipality level are reported in parentheses. Mobile penetration is defined as the ratio of mobile devices with 3G or higher capability to total municipal population, standardized to have unit variance and winsorized at the 1\% level. ***p$<$0.01, **p$<$0.05, *p$<$0.1.
}
\end{singlespace}
\vspace{0.5cm}
{\small
\begin{tabular*}{\linewidth}{l@{\extracolsep{\fill}}*{2}{c}}
\toprule

& (1) & (2) \\ 
\midrule
& Number of Transactions per Capita & Value of Transactions per Capita \\ 
\midrule
Mobile penetration  $\times$ Post & 5.78\sym{***} & 2,072.7\sym{***} \\
& (0.64) & (265.2) \\[0.5em]
\midrule
Observations & 72,371 & 72,371 \\
\midrule
Region $\times$ Year  & $\checkmark$ & $\checkmark$ \\
Controls $\times$ Year & $\checkmark$ & $\checkmark$ \\
\bottomrule
\end{tabular*}
}
\end{table}%

\newpage
\begin{table}[H]
\caption{Effect of Mobile Penetration on Wages in Small Establishments: Excluding Owners}
\label{tab:triple_diff_size_noceo}
\begin{singlespace}
{\footnotesize
This table replicates Panel A of Table \ref{tab:triple_diff_size} after excluding CEOs as a proxy for establishment owners. The dependent variable is the logarithm of average monthly wages. Small establishments are defined as those with fewer than 20 employees. The key coefficient of interest, Mobile Penetration $\times$ Small $\times$ Post, measures how the effect of mobile penetration on wages varies with establishment size after Pix introduction. Columns vary the fixed effects specification following Table \ref{tab:triple_diff_size}. Standard errors clustered at the municipality level are reported in parentheses. Mobile penetration is defined as the ratio of mobile devices with 3G or higher capability to total municipal population, standardized to have unit variance and winsorized at the 1\% level. ***p$<$0.01, **p$<$0.05, *p$<$0.1.
}
\end{singlespace}

\vspace{0.5cm}
{\small
\begin{tabular*}{\linewidth}{l@{\extracolsep{\fill}}*{3}{c}}
\toprule
& (1) & (2) & (3) \\ 
\midrule
Mobile Penetration $\times$ Small $\times$ Post & 0.003** & 0.003** & 0.005*** \\
 & (0.001) & (0.001) & (0.001) \\
\midrule
Observations & 1,529,548 & 1,529,548 & 1,529,548 \\
\midrule
Municipality $\times$ Size $\times$ Industry & $\checkmark$ & $\checkmark$ & $\checkmark$ \\
Municipality $\times$ Industry $\times$ Year & \xmark & $\checkmark$ & $\checkmark$ \\
Size $\times$ Year & $\checkmark$ & $\checkmark$ & \xmark \\
Industry $\times$ Year & $\checkmark$ & \xmark & \xmark \\
Municipality $\times$ Year & $\checkmark$ & \xmark & \xmark \\
Size $\times$ Industry $\times$ Year & \xmark & \xmark & $\checkmark$ \\
\bottomrule
\end{tabular*}
}
\end{table}

\begin{table}[H]
\setlength{\tabcolsep}{0pt}
\caption{Effect of Mobile Penetration on Average Wages: Continuing Establishments}
\label{tab:continuing_estabs}
\justify
{\footnotesize
This table presents triple difference-in-differences estimates examining the differential effect of standardized mobile penetration on average wages in small establishments relative to large establishments following the introduction of Pix in November 2020, restricting the sample to establishments in every year from 2016 to 2024. The dependent variable is the logarithm of average monthly wages. Small establishments are defined as those with fewer than 20 employees. The key coefficient of interest, Mobile Penetration $\times$ Small $\times$ Post, measures how the effect of mobile penetration on wages varies with establishment size after Pix introduction. Columns present the alternative fixed effects specifications of Table \ref{tab:triple_diff_size}. Standard errors clustered at the municipality level are reported in parentheses. Mobile penetration is defined as the ratio of mobile devices with 3G or higher capability to total municipal population, standardized to have unit variance and winsorized at the 1\% level. ***p<0.01, **p<0.05, *p<0.1.
}

\vspace{0.5cm}

\noindent
{\small
\makebox[\textwidth]{
\begin{tabular*}{\linewidth}{l@{\extracolsep{\fill}}*{3}{c}}
\toprule
& (1) & (2) & (3) \\ 
\midrule
Mobile Penetration $\times$ Small $\times$ Post & 0.016*** & 0.015*** & 0.020***\\
& (0.002) & (0.002) & (0.003)  \\
\midrule
Observations & 964,774 & 964,774 & 964,774  \\
\midrule
Municipality $\times$ Size $\times$ Industry & $\checkmark$ & $\checkmark$ & $\checkmark$ \\
Municipality $\times$ Industry $\times$ Year & \xmark & $\checkmark$ & $\checkmark$ \\
Size $\times$ Year & $\checkmark$ & $\checkmark$ & \xmark \\
Industry $\times$ Year & $\checkmark$ & \xmark & \xmark \\
Municipality $\times$ Year & $\checkmark$ & \xmark & \xmark \\
Size $\times$ Industry $\times$ Year & \xmark & \xmark & $\checkmark$ \\
\bottomrule
\end{tabular*}}
}
\end{table}

\newpage
\begin{table}[H]
\setlength{\tabcolsep}{0pt}
\caption{Effect of Mobile Penetration on Average Wages: Commuting-Zone Level}
\label{tab:commuting_zone}
\justify
{\footnotesize
This table presents triple difference-in-differences estimates examining the differential effect of standardized mobile penetration on average wages in small establishments relative to large establishments following the introduction of Pix in November 2020, defining local labor markets as commuting zones rather than municipalities. The dependent variable is the logarithm of average monthly wages. Small establishments are defined as those with fewer than 20 employees. The key coefficient of interest, Mobile Penetration $\times$ Small $\times$ Post, measures how the effect of mobile penetration on wages varies with establishment size after Pix introduction. Columns present the alternative fixed effects specifications of Table \ref{tab:triple_diff_size}, replacing municipality with commuting zone. Standard errors clustered at the commuting-zone level are reported in parentheses. Mobile penetration is defined as the ratio of mobile devices with 3G or higher capability to total municipal population, standardized to have unit variance and winsorized at the 1\% level, and aggregated to the commuting zone as the population-weighted average across member municipalities. ***p<0.01, **p<0.05, *p<0.1.
}

\vspace{0.5cm}

\noindent
{\small
\makebox[\textwidth]{
\begin{tabular*}{\linewidth}{l@{\extracolsep{\fill}}*{3}{c}}
\toprule
& (1) & (2) & (3) \\ 
\midrule
Mobile Penetration $\times$ Small $\times$ Post & 0.002 & 0.004** & 0.004** \\
 & (0.001) & (0.002) & (0.002) \\
\midrule
Observations & 779,550 & 779,550 & 779,550 \\
\midrule
Commuting $\times$ Size $\times$ Industry & $\checkmark$ & $\checkmark$ & $\checkmark$ \\
Commuting $\times$ Industry $\times$ Year & \xmark & $\checkmark$ & $\checkmark$ \\
Size $\times$ Year & $\checkmark$ & $\checkmark$ & \xmark \\
Industry $\times$ Year & $\checkmark$ & \xmark & \xmark \\
Commuting $\times$ Year & $\checkmark$ & \xmark & \xmark \\
Size $\times$ Industry $\times$ Year & \xmark & \xmark & $\checkmark$ \\
\bottomrule
\end{tabular*}}
}
\end{table}
\newpage
\begin{table}[H]
\caption{Effect of Mobile Penetration on Wages in Small Establishments: Excluding 2020}
\label{tab:triple_diff_size_donut}

{\footnotesize
This table replicates Panel A of Table \ref{tab:triple_diff_size} after dropping the year 2020 from the sample. The pre-period covers 2016--2019 and the post-period covers 2021--2024. The dependent variable is the logarithm of average monthly wages. Small establishments are defined as those with fewer than 20 employees. The key coefficient of interest, Mobile Penetration $\times$ Small $\times$ Post, measures how the effect of mobile penetration on wages varies with establishment size after Pix introduction. Columns vary the fixed effects specification following Table \ref{tab:triple_diff_size}. Standard errors clustered at the municipality level are reported in parentheses. Mobile penetration is defined as the ratio of mobile devices with 3G or higher capability to total municipal population, standardized to have unit variance and winsorized at the 1\% level. ***p$<$0.01, **p$<$0.05, *p$<$0.1.
}

\vspace{0.5cm}
{\small
\begin{tabular*}{\linewidth}{l@{\extracolsep{\fill}}*{3}{c}}
\toprule
& (1) & (2) & (3) \\ 
\midrule
Mobile Penetration $\times$ Small $\times$ Post & 0.003* & 0.003** & 0.005*** \\
 & (0.001) & (0.001) & (0.002) \\
\midrule
Observations & 1,379,830 & 1,379,830 & 1,379,830 \\
\midrule
Municipality $\times$ Size $\times$ Industry & $\checkmark$ & $\checkmark$ & $\checkmark$ \\
Municipality $\times$ Industry $\times$ Year & \xmark & $\checkmark$ & $\checkmark$ \\
Size $\times$ Year & $\checkmark$ & $\checkmark$ & \xmark \\
Industry $\times$ Year & $\checkmark$ & \xmark & \xmark \\
Municipality $\times$ Year & $\checkmark$ & \xmark & \xmark \\
Size $\times$ Industry $\times$ Year & \xmark & \xmark & $\checkmark$ \\
\bottomrule
\end{tabular*}
}
\end{table}

\newpage
\begin{table}[H]
\setlength{\tabcolsep}{0pt}
\caption{\label{tab:delivery_split} Effect of Mobile Penetration on Wages in Physical Services and Delivery-Capable Services}

\justify
{\footnotesize
This table presents triple difference-in-differences estimates examining the differential effect of standardized mobile penetration on average wages in small establishments relative to large establishments following the introduction of Pix in November 2020, estimated separately for physical services and delivery-capable services. The dependent variable is the logarithm of average monthly wages. Small establishments are defined as those with fewer than 20 employees. Delivery-capable services include restaurants. Physical services include repair of vehicles, sport activities, repair of computers, and personal services. Column 1 restricts the sample to physical services, and Column 2 to delivery-capable services. All specifications include municipality-by-size-by-industry, municipality-by-industry-by-year, and size-by-industry-by-year fixed effects. Standard errors clustered at the municipality level are reported in parentheses. Mobile penetration is defined as the ratio of mobile devices with 3G or higher capability to total municipal population, standardized to have unit variance and winsorized at the 1\% level. ***p$<$0.01, **p$<$0.05, *p$<$0.1.
}

\vspace{0.5cm}

\noindent
{\small
\makebox[\textwidth]{
\begin{tabular*}{\textwidth}{@{\extracolsep{\fill}}l*{2}{c}@{}}
\toprule
& (1) & (2) \\
\midrule
& Physical Services & Delivery Capable \\
\midrule
Mobile Penetration $\times$ Small $\times$ Post & 0.019*** & 0.002 \\
 & (0.005) & (0.004) \\
\midrule
Observations & 133,506 & 64,406 \\
\midrule
Municipality $\times$ Size $\times$ Industry & $\checkmark$ & $\checkmark$ \\
Municipality $\times$ Industry $\times$ Year & $\checkmark$ & $\checkmark$ \\
Size $\times$ Industry $\times$ Year & $\checkmark$ & $\checkmark$ \\
\bottomrule
\end{tabular*}}
}
\end{table}

\newpage

\begin{table}[H]
\caption{\label{tab:mei_entry} Effect of Mobile Penetration on Micro-entrepreneur Entry}

{\footnotesize
This table presents difference-in-differences estimates examining the effect of standardized mobile penetration on the entry of individual micro-entrepreneurs (MEI) following the introduction of Pix in November 2020. The dependent variable is the annual municipal tax revenue from the Tax on the Circulation of Goods and Services (Imposto sobre Circulação de Mercadorias e Serviços, ICMS) per 1,000 population paid by MEI firms. The ICMS fee for MEIs is a fixed amount of one Brazilian real (BRL 1.00), predominantly paid by retailers. The key coefficient of interest, Mobile Penetration $\times$ Post, measures how mobile penetration affects MEI entry after Pix introduction. The specification includes municipality fixed effects, region-by-year fixed effects, municipality characteristics interacted with year fixed effects, and pre-Pix MEI levels interacted with year fixed effects. Standard errors clustered at the municipality level are reported in parentheses. Mobile penetration is defined as the ratio of mobile devices with 3G or higher capability to total municipal population, standardized to have unit variance and winsorized at the 1\% level. **p$<$0.05, *p$<$0.1.
}

\vspace{0.5cm}
{\small
\centering
\begin{tabular*}{\linewidth}{l@{\extracolsep{\fill}}*{1}{c}}
\toprule
& (1) \\ 
\midrule
Mobile penetration $\times$ Post & 1.417** \\
 & (0.571) \\
\midrule
Observations & 23,562 \\
\midrule
Municipality  & $\checkmark$  \\
Region $\times$ Year  & $\checkmark$ \\
Controls $\times$ Year & $\checkmark$ \\
Pre-MEI $\times$ Year  & $\checkmark$ \\
\bottomrule
\end{tabular*}
}
\end{table}

\newpage

\begin{table}[H]
\caption{Effect of Mobile Penetration on Average Wages of Small Establishments by Industry Tradability}
\label{tab:triple_diff_T_vs_NT}
\begin{singlespace}
{\footnotesize
This table presents triple difference-in-differences estimates examining the differential effect of standardized mobile penetration on average wages in small establishments relative to large establishments across different groups of industries following the introduction of Pix in November 2020. We define an industry-specific measure of tradability equal to the geographical concentration of wage-bill defined by the Herfindahl Hirschman Index (HHI). We exclude low-cash sectors and classify industries as non-tradable if they are in the bottom half of the HHI distribution, and as tradable if they are in the top half. The dependent variable is the logarithm of average monthly wages. Small establishments are defined as those with fewer than 20 employees. The key coefficient of interest, Mobile Penetration $\times$ Small $\times$ Post, measures how the effect of mobile penetration on wages varies with establishment size after Pix introduction, estimated separately for each group of industries. Column (1) presents the estimate for non-tradable industries and column (2) for tradable industries. All specifications include municipality-by-size-by-industry fixed effects, municipality-by-industry-by-year fixed effects, and size-by-industry-by-year fixed effects. Standard errors clustered at the municipality level are reported in parentheses. Mobile penetration is defined as the ratio of mobile devices with 3G or higher capability to total municipal population, standardized to have unit variance and winsorized at the 1\% level. ***p$<$0.01, **p$<$0.05, *p$<$0.1.
}
\end{singlespace}
\vspace{0.5cm}
{\small
\begin{tabular*}{\linewidth}{l@{\extracolsep{\fill}}*{2}{c}}
\toprule
 & (1) & (2) \\ 
\midrule
 & Non-Tradable & Tradable \\
\midrule
Mobile Penetration $\times$ Small $\times$ Post & 0.007*** & 0.007** \\
 & (0.002) & (0.003) \\
\midrule
Observations & 436,015 & 412,851 \\
\midrule
Municipality $\times$ Size $\times$ Industry & $\checkmark$ & $\checkmark$ \\
Municipality $\times$ Industry $\times$ Year & $\checkmark$ & $\checkmark$ \\
Size $\times$ Industry $\times$ Year & $\checkmark$ & $\checkmark$ \\
\bottomrule
\end{tabular*}
}
\end{table}

\newpage 
\begin{table}[H]
\setlength{\tabcolsep}{0pt}
\caption{\label{tab:ddd_wage_informal} Effect of Mobile Penetration on Average Wages in Small Establishments by Municipal Informal Taxpayer Rate}

\justify
{\footnotesize
This table presents triple difference-in-differences estimates examining the differential effect of standardized mobile penetration on average wages in small establishments relative to large establishments following the introduction of Pix in November 2020, separately for municipalities with above-median and below-median informal taxpayer rates. The dependent variable is the logarithm of average monthly wages. Small establishments are defined as those with fewer than 20 employees. Column 1 restricts the sample to municipalities with above-median informal taxpayer rates, and Column 2 to municipalities with below-median informal taxpayer rates. All specifications include municipality-by-size-by-industry, municipality-by-industry-by-year, and size-by-industry-by-year fixed effects. Standard errors clustered at the municipality level are reported in parentheses. Mobile penetration is defined as the ratio of mobile devices with 3G or higher capability to total municipal population, standardized to have unit variance and winsorized at the 1\% level. ***p$<$0.01, **p$<$0.05, *p$<$0.1.
}

\vspace{0.5cm}

\noindent
{\small
\makebox[\textwidth]{
\begin{tabular*}{\textwidth}{@{\extracolsep{\fill}}l*{2}{c}@{}}
\toprule
& (1) & (2) \\
\midrule
& Above-Median informality & Below-Median informality \\

\midrule
Mobile Penetration $\times$ Small $\times$ Post & 0.005** & 0.005** \\
 & (0.002) & (0.002) \\
\midrule
Observations & 626,518 & 747,103 \\
\midrule
Municipality $\times$ Size $\times$ Industry & $\checkmark$ & $\checkmark$ \\
Municipality $\times$ Industry $\times$ Year & $\checkmark$ & $\checkmark$ \\
Size $\times$ Industry $\times$ Year & $\checkmark$ & $\checkmark$ \\
\bottomrule
\end{tabular*}}
}
\end{table}

\newpage
\begin{table}[H]
\setlength{\tabcolsep}{0pt}
\caption{\label{tab:composition} Effect of Mobile Penetration on Workforce Composition in Small Establishments}

\justify
{\footnotesize
This table presents triple difference-in-differences estimates examining the differential effect of standardized mobile penetration on workforce composition in small establishments relative to large establishments following the introduction of Pix in November 2020. Each column reports the coefficient on Mobile Penetration $\times$ Small $\times$ Post for a different dependent variable. Female, High Skill, and Age $\geq$ 40 are indicators equal to one if the worker is female, has at least a college degree, or is 40 years or older, respectively. Log of Tenure is the logarithm of months employed at the current establishment. Age is measured in years. Small establishments are defined as those with fewer than 20 employees. All specifications include municipality-by-size-by-industry, municipality-by-industry-by-year, and size-by-industry-by-year fixed effects. Standard errors clustered at the municipality level are reported in parentheses. Mobile penetration is defined as the ratio of mobile devices with 3G or higher capability to total municipal population, standardized to have unit variance and winsorized at the 1\% level. ***p$<$0.01, **p$<$0.05, *p$<$0.1.
}

\vspace{0.5cm}

\noindent
{\small
\makebox[\textwidth]{
\begin{tabular*}{\textwidth}{@{\extracolsep{\fill}}l*{5}{c}@{}}
\toprule
& (1) & (2) & (3) & (4) & (5) \\
\midrule
& Female & High Skill & Log Tenure & Age & Age $\geq$ 40 \\
\midrule
Mobile Penetration $\times$ Small $\times$ Post & 0.001 & 0.001 & $-$0.001 & 0.022 & $-$0.000 \\
 & (0.001) & (0.001) & (0.005) & (0.021) & (0.001) \\
\midrule
Observations & 1,299,277 & 1,299,277 & 1,299,250 & 1,299,277 & 1,299,277 \\
\midrule
Municipality $\times$ Size $\times$ Industry & $\checkmark$ & $\checkmark$ & $\checkmark$ & $\checkmark$ & $\checkmark$ \\
Municipality $\times$ Industry $\times$ Year & $\checkmark$ & $\checkmark$ & $\checkmark$ & $\checkmark$ & $\checkmark$ \\
Size $\times$ Industry $\times$ Year & $\checkmark$ & $\checkmark$ & $\checkmark$ & $\checkmark$ & $\checkmark$ \\
\bottomrule
\end{tabular*}}
}
\end{table}

\newpage
\begin{table}[H]
\caption{Effect of Mobile Penetration on Average Wages in Small Establishments by Worker Tenure}
\label{tab:tenure_monotonicity}
\begin{singlespace}
{\footnotesize
This table presents triple difference-in-differences estimates examining whether the differential effect of standardized mobile penetration on average wages in small establishments relative to large establishments varies with worker tenure following the introduction of Pix in November 2020. The dependent variable is the logarithm of average monthly wages. The unit of observation is a municipality $\times$ education $\times$ establishment-size $\times$ industry $\times$ tenure-group $\times$ year cell. Worker tenure is divided into three mutually exclusive groups: fewer than 12 months, 12 to 24 months, and at least 24 months. The key coefficients of interest interact Mobile Penetration $\times$ Small $\times$ Post with indicators for each tenure group, with workers with fewer than 12 months of tenure as the omitted category. Small establishments are defined as those with fewer than 20 employees. The reported $p$-value tests equality of the coefficients for the 12 to 24 months and at least 24 months tenure groups. All specifications include municipality-by-size-by-industry fixed effects, municipality-by-industry-by-year fixed effects, and size-by-industry-by-year fixed effects. Standard errors clustered at the municipality level are reported in parentheses. Mobile penetration is defined as the ratio of mobile devices with 3G or higher capability to total municipal population, standardized to have unit variance and winsorized at the 1\% level. ***p$<$0.01, **p$<$0.05, *p$<$0.1.
}
\end{singlespace}
\vspace{0.5cm}
{\small
\begin{tabular*}{\linewidth}{l@{\extracolsep{\fill}}c}
\toprule
 & (1) \\
\midrule
Mobile Penetration $\times$ Small $\times$ Post $\times$ 12--24 Months
    & 0.0165*** \\
    & (0.0003) \\
Mobile Penetration $\times$ Small $\times$ Post $\times$ $\geq$ 24 Months
    & 0.0365*** \\
    & (0.0014) \\
\midrule
$p$-value: 12--24 Months $=$ $\geq$ 24 Months & $<$0.001 \\
Observations & 15,893,472 \\
\midrule
Municipality $\times$ Size $\times$ Industry & $\checkmark$ \\
Municipality $\times$ Industry $\times$ Year & $\checkmark$ \\
Size $\times$ Industry $\times$ Year & $\checkmark$ \\
\bottomrule
\end{tabular*}
}
\end{table}

\newpage
\begin{table}[H]
\caption{\label{tab:pnad_informality} Effect of Mobile Penetration on Informal Employees}
\justify
{\footnotesize
This table presents difference-in-differences estimates examining the effect of standardized mobile penetration on informal employment following the introduction of Pix in November 2020. The outcome is measured from the Brazilian National Household Sample Survey (PNAD Cont\'{\i}nua) microdata aggregated to 146 survey strata. The dependent variable is the number of private-sector employees without a signed work contract divided by all employed workers in the stratum and year. The key coefficient of interest is Mobile Penetration $\times$ Post. All specifications include stratum fixed effects, state-by-year fixed effects, and stratum characteristics interacted with year. Standard errors clustered at the stratum level are reported in parentheses. Mobile penetration is the municipal ratio of mobile devices with 3G or higher capability to population before Pix, aggregated to the stratum as a population-weighted average, standardized to have unit variance. ***p$<$0.01, **p$<$0.05, *p$<$0.1.
}
\vspace{0.5cm}
\noindent
{\small
\makebox[\textwidth]{
\begin{tabular*}{\textwidth}{@{\extracolsep{\fill}}l*{1}{c}@{}}
\toprule
& (1) \\
\midrule
& Employees without contract/total employed \\

\midrule
Mobile Penetration $\times$ Post & 0.0026 \\
 & (0.0016) \\
\midrule
Observations & 1,305 \\
\midrule
Stratum & $\checkmark$ \\
State $\times$ Year & $\checkmark$ \\
Controls $\times$ Year & $\checkmark$ \\
\bottomrule
\end{tabular*}}
}
\end{table}

\newpage

\begin{table}[H]
\setlength{\tabcolsep}{0pt}
\caption{\label{tab:formal_informal} Differences Between Formal and Informal Workers}

\justify
{\footnotesize
This table examines differences in observable characteristics between formal and informal workers using individual-level microdata from the 2010 Brazilian Census. The dependent variable is an indicator equal to one if the worker is informally employed. Each column reports the coefficient from a separate regression of informal employment status on a single worker characteristic. Female is an indicator for female workers. High Skill is an indicator for workers with at least a college degree. Age is measured in years. The sample includes all individuals who report being actively employed at the time of the census. All specifications include municipality fixed effects. Standard errors are reported in parentheses. ***p$<$0.01, **p$<$0.05, *p$<$0.1.
}

\vspace{0.5cm}

\noindent
{\small
\makebox[\textwidth]{
\begin{tabular*}{\textwidth}{@{\extracolsep{\fill}}l*{3}{c}@{}}
\toprule
& (1) & (2) & (3) \\
\midrule
& \multicolumn{3}{c}{Informal Worker Indicator} \\
\midrule
Female & 0.018*** &  &  \\
       & (0.0004) &  &  \\
High Skill &  & $-$0.208*** &  \\
           &  & (0.0004) &  \\
Age (years) &  &  & 0.001*** \\
            &  &  & (0.00002) \\
\midrule
Observations & 8,382,159 & 8,382,159 & 8,382,094 \\
\midrule
Municipality & $\checkmark$ & $\checkmark$ & $\checkmark$ \\
\bottomrule
\end{tabular*}}
}
\end{table}

\end{appendices}
\end{document}